\documentclass[preprint,12pt]{elsarticle}

\usepackage{caption}
\usepackage{amsmath}
\usepackage{amssymb}
\usepackage{amsthm}
\usepackage{mathrsfs}

\usepackage{epstopdf}
\usepackage{makecell}

\usepackage{subfig}

\usepackage{booktabs} 
\usepackage{multirow} 

\usepackage{url}
\usepackage{hyperref}
\usepackage{xcolor}
\usepackage{cleveref} 

\usepackage{lineno}

\usepackage{threeparttable}
\usepackage{color}
\usepackage{epstopdf}
\usepackage{float}
\usepackage[top=2.5cm, bottom=2.5cm, left=2cm, right=2cm]{geometry}
\renewcommand{\vec}[1]{\boldsymbol{#1}}

\biboptions{numbers,sort&compress}

\journal{Journal of Computational Physics}

\begin{document}


\begin{frontmatter}



\title{An implicit adaptive unified gas-kinetic scheme for steady-state solutions of non-equilibrium flows}


\author[a]{Wenpei Long}
\author[a]{Yufeng Wei}
\author[a,b,c]{Kun Xu\corref{cor1}}

\cortext[cor1] {Corresponding author.}
\ead{makxu@ust.hk}

\address[a]{Department of Mathematics, Hong Kong University of Science and Technology, Hong Kong, China}
\address[b]{Department of Mechanical and Aerospace Engineering, Hong Kong University of Science and Technology, Hong Kong, China}
\address[c]{HKUST Shenzhen Research Institute, Shenzhen, 518057, China}
\begin{abstract}

In recent years, non-equilibrium flows have been frequently encountered in various aerospace engineering and micro-electro-mechanical systems applications. Numerical simulations enable us to understand non-equilibrium physics, multiscale effects, and the dynamics in these applications, which require an effective and reliable multiscale scheme for all flow regimes.
Following the direct modeling methodology, the adaptive unified gas-kinetic scheme employs discrete velocity space to accurately capture the non-equilibrium physics, recovering the original unified gas-kinetic scheme (UGKS). Additionally, it adaptively employs continuous distribution functions based on the Chapman--Enskog expansion to efficiently achieve near-equilibrium flow regions with the gas-kinetic scheme (GKS). The UGKS and GKS are dynamically coupled at the cell interface through the fluxes from the discrete and continuous gas distribution functions, thereby avoiding any buffer zone between them. In the current study, an implicit adaptive unified gas-kinetic scheme (IAUGKS) is constructed to further enhance the efficiency of steady-state solutions. To preserve the capability for all flow regimes, a physical time step is introduced to determine the flow scale, while a numerical time step controls the time iteration. The current scheme employs implicit macroscopic governing equations and couples them with implicit microscopic governing equations within the non-equilibrium region, resulting in high convergence efficiency in all flow regimes. The point relaxation scheme is used to solve the implicit governing equations. To validate the efficiency and robustness of the IAUGKS, a series of numerical tests were conducted for high Mach number flows around diverse geometries such as a two-dimensional cylinder, a three-dimensional sphere, an X-38-like vehicle, and a space station. The current scheme can capture the non-equilibrium physics and provide accurate predictions of surface quantities. In comparison with the original UGKS, the velocity space adaptation, unstructured discrete velocity space (DVS), and implicit iteration significantly improve the efficiency by one or two orders of magnitude. Given its exceptional efficiency and accuracy, the IAUGKS serves as an effective tool for non-equilibrium flow simulations.
\end{abstract}


\end{frontmatter}



\section{Introduction}\label{sec:intro}

Non-equilibrium flow plays a critical role in aerospace engineering~\cite{xu2021unified}. During the re-entry process of spacecraft, the atmospheric environment undergoes dramatic changes across multiple flow regimes from rarefied regime to continuum flow regime~\cite{votta_hypersonic_2013}. In hypersonic flows, non-equilibrium flow exists in different parts of the spacecraft~\cite{ivanov_computational_1998}, even though the Knudsen number of the incoming flow is relatively small. Traditional computational fluid dynamics methods, based on the continuity assumption, fail to capture these non-equilibrium flow physics. Consequently, the development of efficient and accurate multiscale simulation methods is of great importance.

To solve non-equilibrium flows, mainstream computational methods can be divided into two categories: stochastic and deterministic methods. Among stochastic methods, the direct simulation Monte Carlo (DSMC)~\cite{bird1963approach,bird_recent_1998,fan2001statistical} method is the most popular, which uses probabilistic Monte Carlo simulation to solve the Boltzmann equation, providing efficient and accurate results in rarefied flow regimes. Deterministic methods like the discrete velocity method (DVM)~\cite{chu_kinetic-theoretic_1965, JCHuang1995, Mieussens2000, tcheremissine2005direct} or the discrete ordinate method (DOM) use the discrete velocity distribution function to solve the Boltzmann equation with different models. However, the computational efficiency of the DSMC method and the DVM decreases in continuum flow regimes due to the decoupling of particle collision and transport processes, limiting the time step and grid size to the scale of mean collision time and mean free path scale respectively.

Several explorations were made for deterministic methods to overcome the efficiency shortage of the conventional DVM in multiscale flows. Based on the direct modeling methodology~\cite{directmodeling}, the unified gas-kinetic scheme (UGKS)~\cite{ugks2010} was proposed as a genuine multiscale method for all flow regimes. Under the framework of the finite volume method, the UGKS locally solves the Boltzmann equation with the Bhatnagar--Gross--Krook (BGK)~\cite{bhatnagar1954model} model when constructing the cell interface flux, thereby coupling the collision and free transport processes and breaking the time step limitation of the traditional DVM. In recent years, a series of multiscale methods based on the methodology of the DVM have emerged. Among them, the discrete unified gas-kinetic scheme (DUGKS)~\cite{guo_discrete_2013,guo_progress_2021} overcomes many shortages of the lattice Boltzmann equation method~\cite{succi2001lattice}, and simplifies the UGKS, making the scheme easy to be implemented. Based on the traditional DVM, the improved DVM~\cite{yang_improved_2018} and a variant of the improved DVM~\cite{yuan_variant_2024} were further developed. These methods use weighted coefficients to couple the near-continuous distribution function and discrete distribution function, breaking the limitation on time step and grid size of the original DVM. A rapidly converging general synthetic iterative scheme (GSIS) was proposed~\cite{su_multiscale_2021,zhang2023efficient, zeng_general_2023} by coupling the kinetic equation and its moment equations, while maintaining the ability to capture non-equilibrium physics. However, all methods mentioned above require discretizing the gas distribution function on a velocity space mesh, thus dramatically increasing the computational cost and memory consumption. In three-dimensional hypersonic flows, a large discrete velocity space (DVS) is needed to obtain accurate results, which consumes a significant amount of memory resources and slows down computational efficiency.

Multiscale stochastic methods that use statistical particles to describe distribution functions have also made significant progress.  A unified gas-kinetic wave-particle (UGKWP) method~\cite{liu2020unified,zhu2019unified} was developed. This method replaces the DVS in the UGKS with stochastic particles, overcoming the substantial computational load and memory consumption of the UGKS, while maintaining its advantage of large time steps and grid sizes in the continuum flow regime. By using a wave-particle formulation, the UGKWP method substantially reduces the number of simulation particles in near-continuum flow regime. The method has shown great performance in hypersonic non-equilibrium flows~\cite{chen_three-dimensional_2020,long2024nonequilibrium}, and has been extended to diatomic gas with
rotational and vibrational non-equilibrium~\cite{wei_unified_2024}, plasma~\cite{liu_unified_2021}, photon transport~\cite{li_unified_2020}, and multiphase flow problems~\cite{yang_unified_2022,yang_unified_2022-1,yang_unified_2023,yang_unified_2024}. In addition, based on the stochastic particle BGK method~\cite{Macrossan2001426}, the unified stochastic particle method~\cite{fei_unified_2020,fei_hybrid_2021} is proposed. This method couples molecular motion and collision effects, overcoming the shortcoming of short time steps. A multi-scale Boltzmann equation method~\cite{liu2024multi} was developed by coupling the DSMC method and NS solver directly and simply, and is validated in varies cases. However, all stochastic methods face several challenges: In low-speed flows, the statistical noise contaminates the simulation result, leading to the use of a great number of particles and a long averaging process to get smooth results. Moreover, stochastic methods have difficulties when implementing acceleration techniques such as implicit iteration, resulting in slow convergence in steady-state solutions.

Given the above issues, research into acceleration and reducing memory consumption of deterministic methods is urgently needed. The implicit UGKS was proposed for both steady~\cite{zhu_implicit_2016} and unsteady~\cite{zhu_implicit_2019} solutions. By combining macroscopic prediction and microscopic implicit iteration, it achieves convergence 1 to 2 orders of magnitude faster than explicit UGKS in all flow regimes. The IUGKS was extended to three-dimensional thermal non-equilibrium flows~\cite{zhang_conservative_2024}, involving rotational and vibrational degrees of freedom. An implicit DUGKS~\cite{pan_implicit_2019} was also developed. Based on the traditional semi-implicit DVM, a fully implicit improved DVM~\cite{yang_improved_2018-1} was proposed. In addition, A new implicit scheme~\cite{peng_implicit_2016} covering various flow regimes that directly solves the unified Boltzmann model equation was developed based on the gas-kinetic unified algorithm~\cite{li2004study,li2019gas}. However, due to the explicit handling of the equilibrium distribution function, the convergence efficiency is reduced. In terms of memory reduction techniques of the DVM, an unstructured DVS~\cite{yuan_conservative_2020} was introduced, which significantly reduces the velocity space mesh in three-dimensional problems without affecting the accuracy of the results. The adaptive UGKS (AUGKS)~\cite{xiao_velocity-space_2020} adopts a adaptive velocity space decomposition to capture the non-equilibrium part of the flow field, and use the discrete distribution function in the corresponding area and the continuous distribution function in other areas, saving memory consumption in the near-continuous flow regime. The effectiveness of the AUGKS in three-dimensional flows with rotational and vibrational non-equilibrium~\cite{wei_adaptive_2023} was further proved. In addition, an adaptive DUGKS~\cite{yang_adaptive_2023} and an adaptive UGKWP method~\cite{wei_adaptivewp} were proposed.

Considering the efficiency requirement in multiscale flows and the accuracy to capture non-equilibrium physics in engineering problems, an implicit adaptive unified gas-kinetic scheme is developed. This scheme significantly reduces memory consumption compared to the original UGKS and improves the convergence speed by one to two orders of magnitude in all flow regimes. Simultaneously, the IAUGKS retains the ability of the UGKS to capture non-equilibrium physics. It can effectively predict the heat flux, pressure, and shear stress on the surface of spacecraft, with results consistent with the UGKS or the DSMC data. Combining these properties, this scheme can provide effective and rapid simulation results in aerospace engineering, MEMS systems, and various other practical application fields.

The organization of this paper is as follows: in section \ref{sec:method}, we introduce the general framework of the IAUGKS, along with its implicit iteration and velocity space adaptation. Section \ref{sec:case} validates the efficiency and accuracy of the current scheme by applying it to non-equilibrium flows, specifically, hypersonic flow around a cylinder, a sphere, an X-38 like vehicle, and a space station. The efficiency of the IAUGKS is compared with the orginal UGKS, the UGKWP method, and the GSIS, while the accuracy of surface quantities is validated by the UGKS and DSMC data. Finally, Section \ref{sec:conclusion} concludes the paper and outlines potential future work.

\section{Numerical method}\label{sec:method}
The IAUGKS adopts discrete and continuous velocity space adaptively for each region, thereby maintaining the ability to capture non-equilibrium physics and perform efficient computation simultaneously. The macroscopic implicit prediction is applied to the whole domain, and the microscopic implicit iterative update is used for the non-equilibrium region. In this section, the IUGKS is introduced first, including the macroscopic implicit prediction and microscopic implicit evolution. The IAUGKS is further demonstrated by introducing the treatment in near-continuum flow region and the velocity space adaptation at the interface of the discrete and continuous velocity space regions.

\subsection{Implicit unified gas-kinetic scheme for steady-state solutions}
In the current study, the BGK model is considered for the collision term in the Boltzmann Equation 
\begin{equation}
	\label{eq:Boltzmann Equation}
	\frac{\partial f}{\partial t}+u\frac{\partial f}{\partial x}=\frac{g-f}{\tau}.
\end{equation}
Under the framework of the finite volume method, the implicit evolution of gas distribution function $f_{i,k}$ at the discrete velocity $\vec{u}_{k}$ with backward Euler method is  

\begin{equation}
	\label{eq:implicit-f}
	f_{i,k}^{n+1} = f_{i,k}^n 
	- \frac{\Delta t}{\Omega_i} \sum_{j \in N(i)} \mathcal{F}_{ij,k}^{n+1} \mathcal{A}_{ij}
	+ \frac{\Delta t}{\tau_i^{n+1}}\left( g_{i,k}^{n+1} - f_{i,k}^{n+1} \right),
\end{equation}
where the equilibrium gas distribution function $g_{i,k}^{n+1}$ and relaxation time $\tau_i^{n+1}$ are calculated from the macroscopic variables $\vec{\tilde{W}}_i^{n+1}$ given by macroscopic implicit prediction as 

\begin{equation}
	\label{eq:implicit-W}
	\vec{\tilde{W}}_i^{n+1}=\vec{W}_i^n
	-\frac{\Delta t}{\Omega_i} \sum_{j \in N(i)} \vec{F}^{n+1}_{i j} \mathcal{A}_{i j}.
\end{equation}
The macroscopic variables at step $n+1$ are provided by taking moments of the gas distribution function 
\begin{equation}
	\vec{W}_i^{n+1} = \sum_k f_{i,k} \vec{\psi}_k \mathcal{V}_k.
\end{equation}
Therefore, the IUGKS mainly consists of two steps: macroscopic prediction and microscopic implicit iterative update.
\subsection{Macroscopic implicit prediction}

By subtracting the time-averaged flux $\vec{F}^n_{ij}$ corresponding to the macroscopic variables at time step $n$ from both sides of the macroscopic implicit prediction formula~\eqref{eq:implicit-W}, we can rewrite it in the $\Delta$ form

\begin{equation}
	\label{eq:delta-W}
	\frac{\Omega_i}{\Delta t} \Delta \vec{W}_i
	+\sum_{j \in N(i)} \Delta \vec{F}_{i j} \mathcal{A}_{i j}
	= \vec{R}_i,
\end{equation}
where $\Delta \vec{W}_i = \vec{\tilde{W}}_i^{n+1} - \vec{W}_i^{n}$ and  
$\Delta \vec{F}_{ij} = \vec{F}_{ij}^{n+1} - \vec{F}_{ij}^n$. $\vec{R}_i$ is the residual 

\begin{equation}
	\vec{R}_i = -\sum_{j \in N(i)} \vec{F}_{i j}^n \mathcal{A}_{i j}.
\end{equation}
For the UGKS, the macroscopic flux $\vec{F}^n_{ij}$ is provided by taking moments of the time-averaged flux of the discrete gas distribution function

\begin{equation}
	\vec{F}_{ij}^n = \sum_k \mathcal{F}_{ij,k}^n \vec{\psi}_k \mathcal{V}_k.
\end{equation}
For steady-state solutions, the local time step $\Delta t_{ij}$ at the interface is 
\begin{equation}
	\Delta t_{ij} = \min(\Delta t_{i},\Delta t_{j}).
\end{equation}
The time-averaged flux is expressed as 

\begin{equation}
	\mathcal{F}_{ij,k}^n = \frac{1}{\Delta t_{ij}} 
	\int_0^{\Delta t_{ij}} \vec{u}_k \cdot \vec{n}_{ij} f_{ij,k}(t) {\rm d} t,
\end{equation}
where the time-dependent gas distribution function $f_{ij,k}(t)$ at the interface is modeled by the integral solution of the BGK Boltzmann equation along the characteristic line  
\begin{equation}
	\label{eq:integral solution}
f(\boldsymbol{r},t)=\frac1\tau\int_0^te^{-(t-t')/\tau}g(\boldsymbol{r}',t')\mathrm{d}t'+e^{-t/\tau}f_0(\boldsymbol{r}-\boldsymbol{u}t),
\end{equation}
where $\boldsymbol{r}-\boldsymbol{u}t$ denotes the trajectory of particles. The integral solution gives a multiscale modeling of an evolution process from an initial non-equilibrium distribution $f$ to an equilibrium distribution $g$ by collision. 
To achieve second-order accuracy, the initial and equilibrium distribution functions are constructed as
\begin{equation}
	\begin{aligned}
		&g_{k}(\boldsymbol{r},t)=g_{0,k}+\boldsymbol{r}\cdot\frac{\partial g_{0,k}}{\partial\boldsymbol{r}}+\frac{\partial g_{0,k}}{\partial t}t,\\&f_{0,k}(\boldsymbol{r})=f_k^{l,r}+\boldsymbol{r}\cdot\frac{\partial f_k^{l,r}}{\partial\boldsymbol{r}},
	\end{aligned}
\end{equation}
where $f_k^{l,r}$ is constructed by the initial distribution function interpolated to the left and right side of the interface
\begin{equation}
	f_k^{l,r}=f_k^lH(\bar{u}_{ij,k})+f_k^r\left[1-H(\bar{u}_{ij,k})\right],
\end{equation}
$H(\bar{u}_{ij,k})$ denotes the Heaviside step function, and $\bar{u}_{ij,k}$ is the particle velocity perpendicular to the interface. $g_{0}$ is the Maxwellian distribution function derived from the conservative variables contributed by all particles transported from both sides of the interface. Details of the distribution functions and derivatives are demonstrated in earlier work~\cite{ugks2010,directmodeling}.

By performing time integration, we have 

\begin{equation}
	\begin{aligned}
		\mathcal{F}^n_{ij,k}
		&= \vec{u}_k \cdot \vec{n}_{ij}
		\left(
		L_1^u g^0_k
		+ L_2^u \vec{u}_k \cdot \frac{\partial g^0_k}{\partial \vec{x}}
		+ L_3^u \frac{\partial g^0_k}{\partial t} 
		+ L_4^u f^{l,r}_{k}
		+ L_5^u \vec{u}_k \cdot \frac{\partial f^{l,r}_k}{\partial \vec{x}} 
		\right),
	\end{aligned}
\end{equation}
where $L_1^u$ to $L_5^u$ are time coefficients considering the local time step

\begin{equation}
	\begin{aligned}
		L_1^u &= 1 - \frac{\tau}{\Delta t_{ij}} \left( 1 - e^{-\Delta t_{ij} / \tau} \right) , \\
		L_2^u &= -\tau + \frac{2\tau^2}{\Delta t_{ij}} - e^{-\Delta t_{ij} / \tau} \left( \frac{2\tau^2}{\Delta t_{ij}} + \tau\right) ,\\
		L_3^u &=  \frac12 \Delta t_{ij} - \tau + \frac{\tau^2}{\Delta t_{ij}} \left( 1 - e^{-\Delta t_{ij} / \tau} \right) , \\
		L_4^u &= \frac{\tau}{\Delta t_{ij}} \left(1 - e^{-\Delta t_{ij} / \tau}\right), \\
		L_5^u & = \tau  e^{-\Delta t_{ij} / \tau} - \frac{\tau^2}{\Delta t_{ij}}(1 -  e^{-\Delta t_{ij} / \tau}) . 
	\end{aligned}
\end{equation}
In addition, the macroscopic numerical flux corresponding to $g^0_k$ is directly given by the integration of the Maxwellian distribution function, which reduces the computational cost by summation.

The $\Delta \vec{F}_{ij}$ is approximated by flux vector splitting method as   

\begin{equation}
	\label{eq:delta-F}
	\Delta \vec{F}_{i j}=\frac{1}{2}
	\left[
	\Delta \vec{T}_i
	+ \Delta \vec{T}_j
	+ \Gamma_{i j}\left(\Delta \vec{W}_i
	-\Delta \vec{W}_j\right)\right],
\end{equation}
where~$\Delta \vec{T}_i = \vec{T}_i^{n+1} - \vec{T}_i^{n}$.
$\vec{T}$ represents the Euler flux calculated from macroscopic variables by					

\begin{equation}
	\vec{T} = \left( \rho U_n, \rho \vec{U}U_n  + p \vec{n}_{ij}, (\rho E + p) U_n\right)^T,
\end{equation}
where $U_n = \vec{U} \cdot \vec{n}_{ij}$ denotes the component of macroscopic velocity $\vec{U}$ in the interface normal direction $\vec{n}_{ij}$. $\Gamma_{ij}$ represents the spectral radius of the Euler flux Jacobian matrix, and an additional stabilizing term related to kinetic viscosity is added considering the viscous effects

\begin{equation}
	\Gamma_{ij} = \mid U_n \mid + a_s + \frac{2\mu}{\rho \mid \vec{n}_{ij} \cdot (\vec{x}_j - \vec{x}_i) \mid},
\end{equation}
where $a_s$ is sound speed. By substituting the expression of $\Delta \vec{F}_{ij}$ in ~\eqref{eq:delta-F} into Eq.~\eqref{eq:delta-W}, and combining the fact that within a closed finite volume element $\vec{T}_i$ satisfying

\begin{equation}
	\sum_{j \in N(i)} \vec{T}_i \mathcal{A}_{ij} = \vec{0},
\end{equation}
the macroscopic governing equation is
\begin{equation}
	\label{eq:delta-W-final}
	\left(
	\frac{\Omega_i}{\Delta t}
	+\frac{1}{2} \sum_{j \in N(i)} \Gamma_{i j} \mathcal{A}_{i j}
	\right) 
	\Delta \vec{W}_i
	+ \frac{1}{2} \sum_{j \in N(i)}
	\left(
	\Delta \vec{T}_j
	-\Gamma_{i j} \Delta \vec{W}_j
	\right) \mathcal{A}_{i j}
	=-\sum_{j \in N(i)} \vec{F}_{i j}^n \mathcal{A}_{i j},
\end{equation}
where $\Delta \vec{T}_j$ can be obtained by the Jacobian matrix

\begin{equation}
	\Delta \vec{T}_j = \left( 
	\frac{\partial \vec{T}}{\partial \vec{W}} 
	\right)_j
	\Delta \vec{W}_j.
\end{equation}
In the current study, to avoid matrix operations, $\Delta \vec{T}_j$ can also use direct differential calculation

\begin{equation}
	\Delta \vec{T}_j = \vec{T}\left(\vec{W}^n_j + \Delta \vec{W}_j\right) 
	- \vec{T}\left(\vec{W}_j^n\right).
\end{equation}
The point relaxation scheme is adopted to solve the macroscopic variables prediction equation. Instead of one forward-backward scanning process of the traditional lower–upper symmetric Gauss--Seidel (LU-SGS) method, the point relaxation scheme adopts multiple iteration steps. For the $m$th forward scanning process 

\begin{equation}
	\begin{aligned}
		\left(\frac{\Omega_i}{\Delta t}+\right. & \left.\frac{1}{2} \sum_{j \in N(i)} \Gamma_{ij} \mathcal{A}_{ij}\right) \Delta \vec{W}_i^*+\frac{1}{2} \sum_{j \in L(i)} \mathcal{A}_{ij}\left[\vec{T}\left(\vec{W}_j^n+\Delta \vec{W}_j^*\right)-\vec{T}\left(\vec{W}_j^n\right)-\Gamma_{i j} \Delta \vec{W}_j^*\right] \\
		     & +\frac{1}{2} \sum_{j \in U(i)} \mathcal{A}_{ij}\left[\vec{T}\left(\vec{W}_j^n+\Delta \vec{W}_j^{m-1}\right)-\vec{T}\left(\vec{W}_j^n\right)-\Gamma_{i j} \Delta \vec{W}_j^{m-1}\right]= \vec{R}_i,
	\end{aligned}
\end{equation}
and the $m$th backward scanning process is
\begin{equation}
	\begin{aligned}
		\left(\frac{\Omega_i}{\Delta t}+\right. & \left.\frac{1}{2} \sum_{j \in N(i)} \Gamma_{ij} \mathcal{A}_{ij}\right) \Delta \vec{W}_i^{m}+\frac{1}{2} \sum_{j \in L(i)} \mathcal{A}_{ij}\left[\vec{T}\left(\vec{W}_j^n+\Delta \vec{W}_j^*\right)-\vec{T}\left(\vec{W}_j^n\right)-\Gamma_{ij} \Delta \vec{W}_j^*\right] \\
		& +\frac{1}{2} \sum_{j \in U(i)} \mathcal{A}_{ij}\left[\vec{T}\left(\vec{W}_j^n+\Delta \vec{W}_j^m\right)-\vec{T}\left(\vec{W}_j^n\right)-\Gamma_{i j} \Delta \vec{W}_j^m\right]=\vec{R}_i ,
	\end{aligned}
\end{equation}
where $\vec{W}_j^{m-1}$ is the result of $m-1$th iteration and $\Delta \vec{W}_i^*$ is the result of $m$th forward scanning process. 

\subsubsection{Microscopic implicit iterative update}
After the macroscopic implicit prediction, the equilibrium gas distribution function $\tilde{g}^{n+1}$ and relaxation time $\tilde{\tau}^{n+1}$ at $n+1$ step can be directly calculated from the predicted $\vec{\tilde{W}}^{n+1}$. By substituting $\tilde{g}^{n+1}$ and $\tilde{\tau}^{n+1}$, and subtracting the time-averaged microscopic flux at $n$ step, the delta-form of the microscopic implicit iteration equation is

\begin{equation}
	\label{eq:delta-f}
	\left(\frac{\Omega_i}{\Delta t}
	+ \frac{\Omega_i}{\tilde{\tau}_i^{n+1}}\right) \Delta f_{i,k}
	+ \sum_{j \in N(i)} \Delta \mathcal{F}_{ij,k} \mathcal{A}_{ij}
	= r_{i,k},
\end{equation}
where $r_{i,k}$ is the residual of discretized gas distribution function at $\vec{u}_{k}$

\begin{equation}
	r_{i,k}
	= \frac{\Omega_i}{\tilde{\tau}_i^{n+1}}\left(
	\tilde{g}_{i,k}^{n+1}-f_{i,k}^n
	\right)
	- \sum_{j \in N(i)} \mathcal{F}_{ij,k}^n \mathcal{A}_{ij}.
\end{equation}
For microscopic distribution function flux, a single first-order upwind scheme is adopted

\begin{equation}
	\label{eq:delta-micro-F}
	\Delta \mathcal{F}_{ij,k} = u_{k,n}
	\left\{
	H(u_{k,n}) \Delta f_{i,k} +   [1-H(u_{k,n})] \Delta f_{j,k}
	\right\}.
\end{equation}
By substituting the microscopic distribution function flux~\eqref{eq:delta-micro-F} into the implicit governing equation~\eqref{eq:delta-f}, we have

\begin{equation}
	D_{i,k} \Delta f_{i,k} + \sum_{j \in N(i) } D_{j,k} \Delta f_{j,k} = r_{i,k},
\end{equation}
where  
\begin{equation}
	\begin{aligned}
		D_{i,k} & = \frac{\Omega_i}{\Delta t} 
		+ \frac{\Omega_i}{\tilde{\tau}_i^{n+1}}
		+ \sum_{j \in N(i)} u_{k,n}\mathcal{A}_{ij} H(u_{k,n}),
		\\
		D_{j,k} & = u_{k,n}\mathcal{A}_{ij} \left[ 1-H(u_{k,n})\right] .
	\end{aligned}
\end{equation}
The microscopic implicit equation adopts traditional the LU-SGS method to decompose the process into a forward scanning process

\begin{equation}
	D_{i,k} \Delta f_{i,k}^{\star}
	+\sum_{j \in L(i)} D_{j,k} \Delta f_{j,k}^{\star}
	+\sum_{j \in U(i)} D_{j,k} \Delta f_{j,k}^{n}=r_{i,k},
\end{equation}
and a backward scanning process
\begin{equation}
	D_{i,k} \Delta f_{i,k}^{n+1}
	+\sum_{j \in L(i)} D_{j,k} \Delta f_{j,k}^{\star}
	+\sum_{j \in U(i)} D_{j,k} \Delta f_{j,k}^{n+1}=r_{i,k}.
\end{equation}

\subsection{Implicit adaptive unified gas-kinetic scheme}

In the near-continuum flow region, giving the validity of the Chapman--Enskog expansion, the IAUGKS adopts continuous gas distribution functions to reduce memory consumption and computational cost. The criterion for velocity space adaptation is given by the gradient-length local Knudsen number 
\begin{equation}
	\mathrm{Kn}_{Gll}=\frac l{\rho/|\nabla\rho|},
\end{equation}
where $l$ is the local mean free path of gas molecules. When the $\mathrm{Kn}_{Gll}$ is less than the given switching criterion $C_t$, the flow region is considered as near-continuum, thus the continuous gas distribution function is adopted; otherwise, the discretized gas distribution function is employed.

The treatment of the evolution in near-continuum flow regions is simplified as 

\begin{equation}
	\vec{W}^{n+1}_i = \vec{\tilde{W}}^{n+1}_i.
\end{equation}
Instead of the summation over the whole discrete velocity space, the flux at the interface can be directly constructed from the continuous distribution function

\begin{equation}
	\vec{F}_{ij}^n = \int \mathcal{F}_{ij} \vec{\psi} {\rm d} \vec{u},
\end{equation}
where $\mathcal{F}_{ij}$ is a time-averaged flux constructed by a continuous distribution function the same as Eq~(\ref{eq:integral solution}). To capture the near-continuum flow physics and achieve second-order accuracy, the distribution functions are obtained by the Chapman--Enskog and Taylor expansion as
\begin{equation}
	\begin{aligned}
		&g(\boldsymbol{r},t)=g_0+\boldsymbol{r}\cdot\frac{\partial g_0}{\partial\boldsymbol{r}}+\frac{\partial g_0}{\partial t}t,\\&f_0(\boldsymbol{r})=g^{l,r}-\tau\left(\boldsymbol{u}\cdot\frac{\partial g^{l,r}}{\partial\boldsymbol{r}}+\frac{\partial g^{l,r}}{\partial t}\right)+\boldsymbol{r}\cdot\frac{\partial g^{l,r}}{\partial\boldsymbol{r}}.
	\end{aligned}
\end{equation}
The time integration of the distribution function as the same as the GKS at the interface 

\begin{equation}
	\begin{aligned}
		\mathcal{F}_{ij}^n & = \frac{1}{\Delta t_{ij}} 
		\int_0^{\Delta t_{ij}} \vec{u} \cdot \vec{n}_{ij} f_{ij}(t) {\rm d} t \\
		&= \vec{u} \cdot \vec{n}_{ij}
		\left(
		L_1^g g^0
		+ L_2^g \vec{u} \cdot \frac{\partial g^0}{\partial \vec{x}}
		+ L_3^g \frac{\partial g^0}{\partial t} 
		+ L_4^g g^{l,r}
		+ L_5^g \vec{u} \cdot \frac{\partial g^{l,r}}{\partial \vec{x}} 
		+ L_6^g \frac{\partial g^{l,r}}{\partial \vec{t}} 
		\right),
	\end{aligned}
\end{equation}
where $L_1^g$ to $L_6^g$ are time coefficients

\begin{equation}
	\label{eq:gks-tCoef}
	\begin{aligned}
		L_1^g &= 1 - \frac{\tau}{\Delta t_{ij}} \left( 1 - e^{-\Delta t_{ij} / \tau} \right) , \\
		L_2^g &= -\tau + \frac{2\tau^2}{\Delta t_{ij}} - e^{-\Delta t_{ij} / \tau} \left( \frac{2\tau^2}{\Delta t_{ij}} + \tau\right) ,\\
		L_3^g &=  \frac12 \Delta t_{ij} - \tau + \frac{\tau^2}{\Delta t_{ij}} \left( 1 - e^{-\Delta t_{ij} / \tau} \right) , \\
		L_4^g &= \frac{\tau}{\Delta t_{ij}} \left(1 - e^{-\Delta t_{ij} / \tau}\right), \\
		L_5^g &= \tau  e^{-\Delta t_{ij} / \tau} - \frac{2 \tau^2}{\Delta t_{ij}}(1 -  e^{-\Delta t_{ij} / \tau}) , \\
		L_6^g &= - \frac{\tau^2}{\Delta t_{ij}}\left( 1- e^{-\Delta t_{ij} / \tau} \right).
	\end{aligned}
\end{equation}
The details and validity of the GKS are demonstrated in the previous work~\cite{xu2001}.

At the interface of the discrete and continuous velocity space regions, the gas distribution function is a combination of continuous and discrete distribution functions in each cell. By performing integration on both sides of the interface, we have

\begin{equation} 
	\begin{aligned}
		\mathcal{F}_{ij,k} &
		= u_{k,n}
		\left(L_1^g g_{0,k}
		+ L_2^g \vec{u}_k \cdot \frac{\partial g_{t,k}}{\partial \vec{x}}
		+ L_3^g \frac{\partial g_{t,k}}{\partial t} \right) \\
		&+
		u_{k,n}
		\left\{
		\left(
		L_4^g g_k^l
		+ L_5^g \vec{u}_k \cdot \frac{\partial g^l_k}{\partial \vec{x}}
		+ L_6^g \frac{\partial g^l_k}{\partial t}
		\right) H\left(u_{k,n}\right) 
		+
		\left(
		L_4^u f_k^r
		+ L_5^u \vec{u}_k \cdot \frac{\partial f^r_k}{\partial \vec{x}}
		\right) \left[ 1-H(u_{k,n})\right] \right\}.
	\end{aligned}
\end{equation}
In addition, the $\Delta f_{j,k}$ used in the microscopic implicit iteration process is required at the region interface. For simplicity, the value is approximated by Maxwellian distribution instead of Champann--Enskog expansion 

\begin{equation}
	\Delta f_{j,k}=g_{j,k}^{n+1} - g_{j,k}^{n}.
\end{equation}
\subsection{Summary of algorithm}
\begin{description}
	\item[Step 1] Calculate gradients of macroscopic variables, and obtain ${\rm Kn}_{Gll}$.
	\item[Step 2] Perform velocity adaptation. Decompose the physical domain into continuous (GKS) and discrete (UGKS) distribution function regions by comparing ${\rm Kn}_{Gll}$ and given criterion $C_t$. 
	\item[Step 3] Calculate the microscopic and macroscopic numerical flux on the cell interface.  
	\item[Step 4] Evaluate the macroscopic residual $\vec{R}_i$.
	\item[Step 5] Perform macroscopic implicit prediction using $\vec{R}_i$ with the point relaxation scheme.
	\item[Step 6] Evaluate the microscopic residual $r_{i,k}$ in the discretized (UGKS) distribution function region using $\tilde{g}^{n+1}$ and $\tilde{\tau}^{n+1}$ obtained in step 4.
	\item[Step 7] Perform microscopic implicit iterative update using $r_{i,k}$ with the LUSGS method. Use the Maxwellian distribution function to obtain $\Delta f_{j,k}$ in adjacent GKS cells of UGKS cells.
	\item[Step 8] Update macroscopic conservative variables. In near-continuum flow regions, the macroscopic variables are directly updated by implicit prediction; otherwise, the macroscopic variables are calculated by the summation of updated discretized distribution function $f_{j,k}^{n+1}$.
\end{description}

\begin{figure}[H]
	\centering
	\includegraphics[width=0.8\textwidth]
	{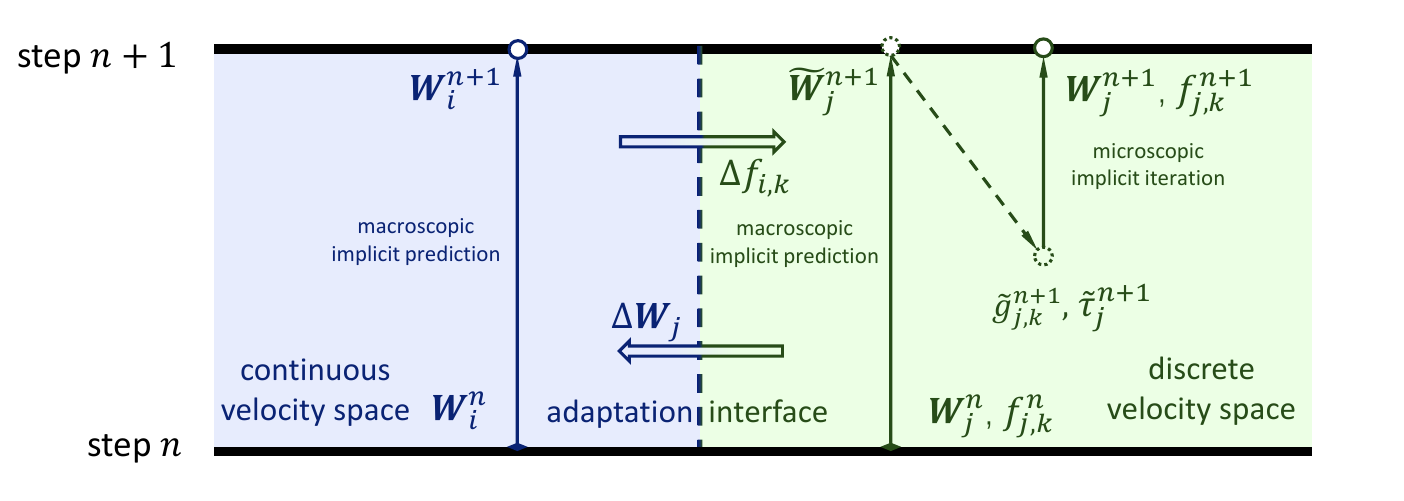}
	\caption{A schematic diagram of the IAUGKS. The figure demonstrates the iteration of the IAUGKS in different regions and the treatment at their interfaces. }
	\label{fig:IAUGKSdiagram}
\end{figure}

A concise schematic diagram of the velocity space adaptation and the iteration of the IAUGKS is plotted in Fig.~\ref{fig:IAUGKSdiagram}.

\section{Numerical Validation}\label{sec:case}

\subsection{Hypersonic flow around a circular cylinder}
\begin{figure}[H]
	\centering
	\includegraphics[width=0.4\textwidth]
		{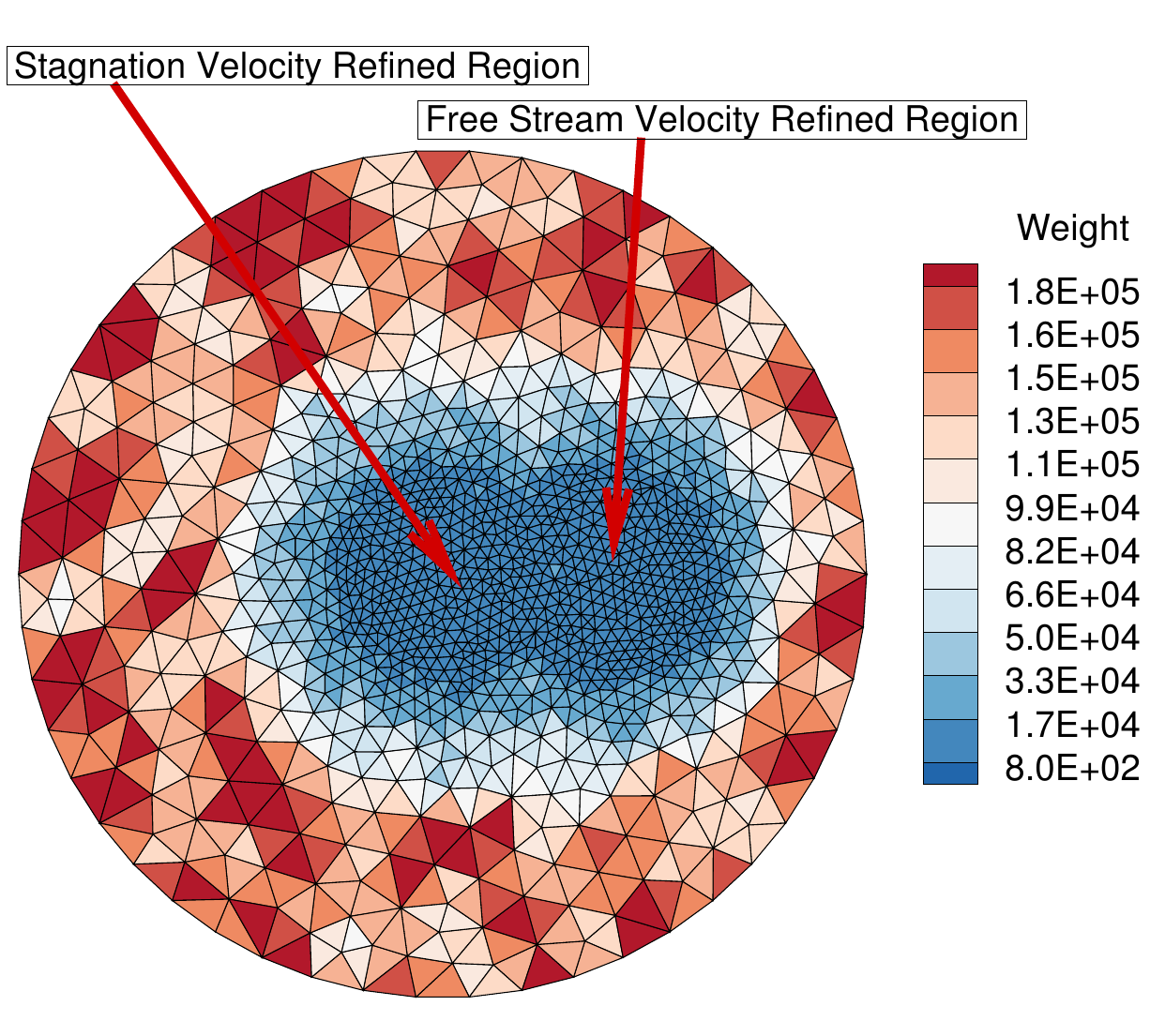}
	\caption{Unstructured DVS mesh used for hypersonic flow at ${\rm Kn} = 0.1$ and ${\rm Ma} = 5$ passing over a cylinder by the IAUGKS.}
	\label{fig:cylinder-Ma5-DVS}
\end{figure}

Hypersonic gas flow of nitrogen around a circular cylinder has been simulated at ${\rm Ma}_\infty = 5$ and ${\rm Kn}_\infty = 0.1$. The characteristic length used to define the Knudsen number is the cylinder diameter $D = 1$ m. The temperature of free stream flow is $T_\infty = 273$ K, and an isothermal wall with a fixed temperature of $T_w = 273$ K is applied for the cylinder surface. The physical domain is discretized by $5000$ quadrilateral cells, while the unstructured discrete velocity space (DVS) mesh consists of 2,060 cells as shown in Fig.~\ref{fig:cylinder-Ma5-DVS}. The DVS is discretized in a circle region with the center $0.4\times(U_\infty,V_\infty,W_\infty)$ with a total radius of $6\sqrt{R T_s}$ where $T_s$ is the stagnation temperature of the free stream flow. The unstructured DVS mesh is refined at zero velocity point with a radius of $3\sqrt{R T_w}$, and the free stream velocity point with a radius of $3\sqrt{R T_\infty}$. 

\begin{figure}[H]
	\centering
	\subfloat[]{\includegraphics[width=0.3\textwidth]
	{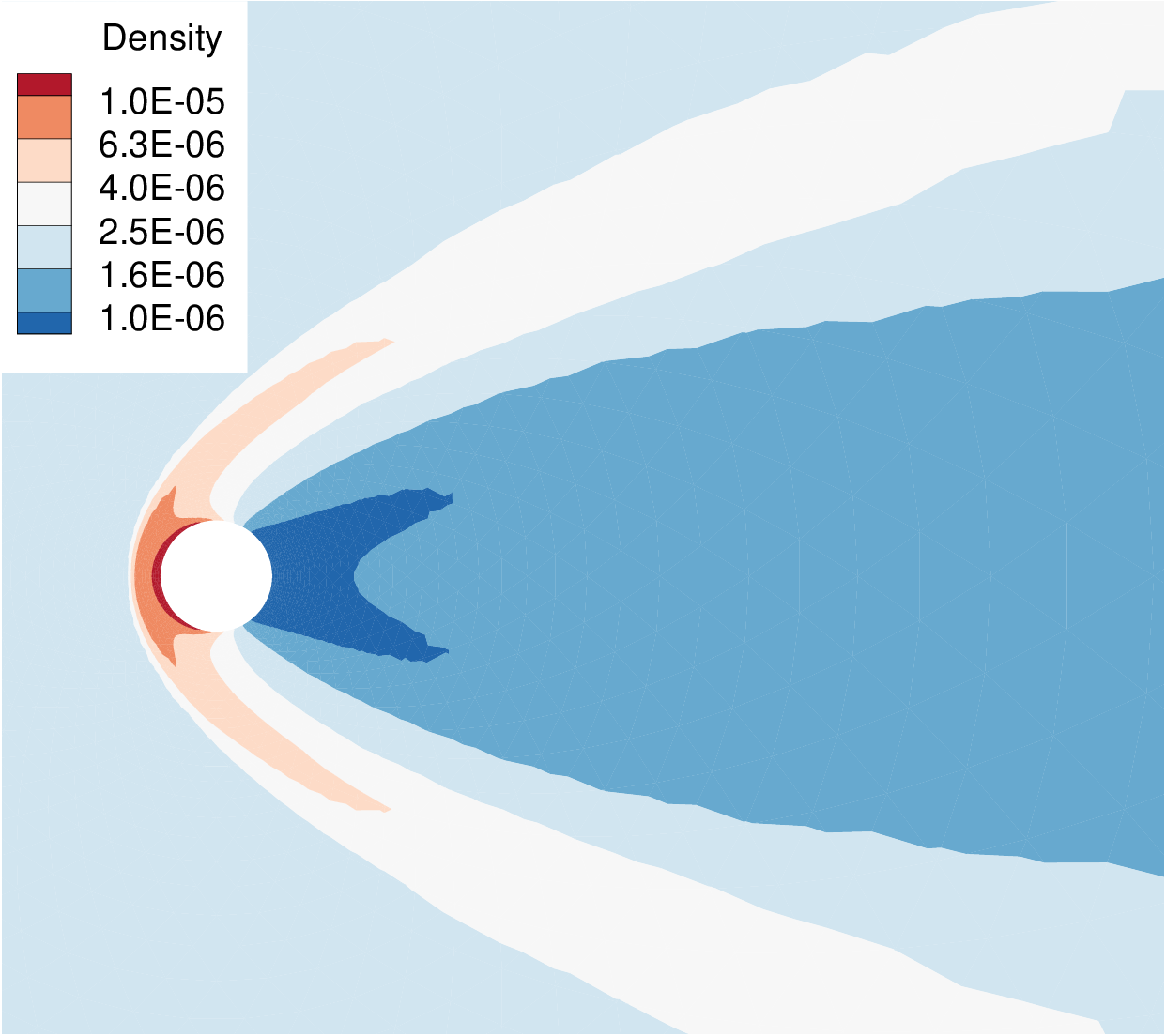}}~
	\subfloat[]{\includegraphics[width=0.3\textwidth]
	{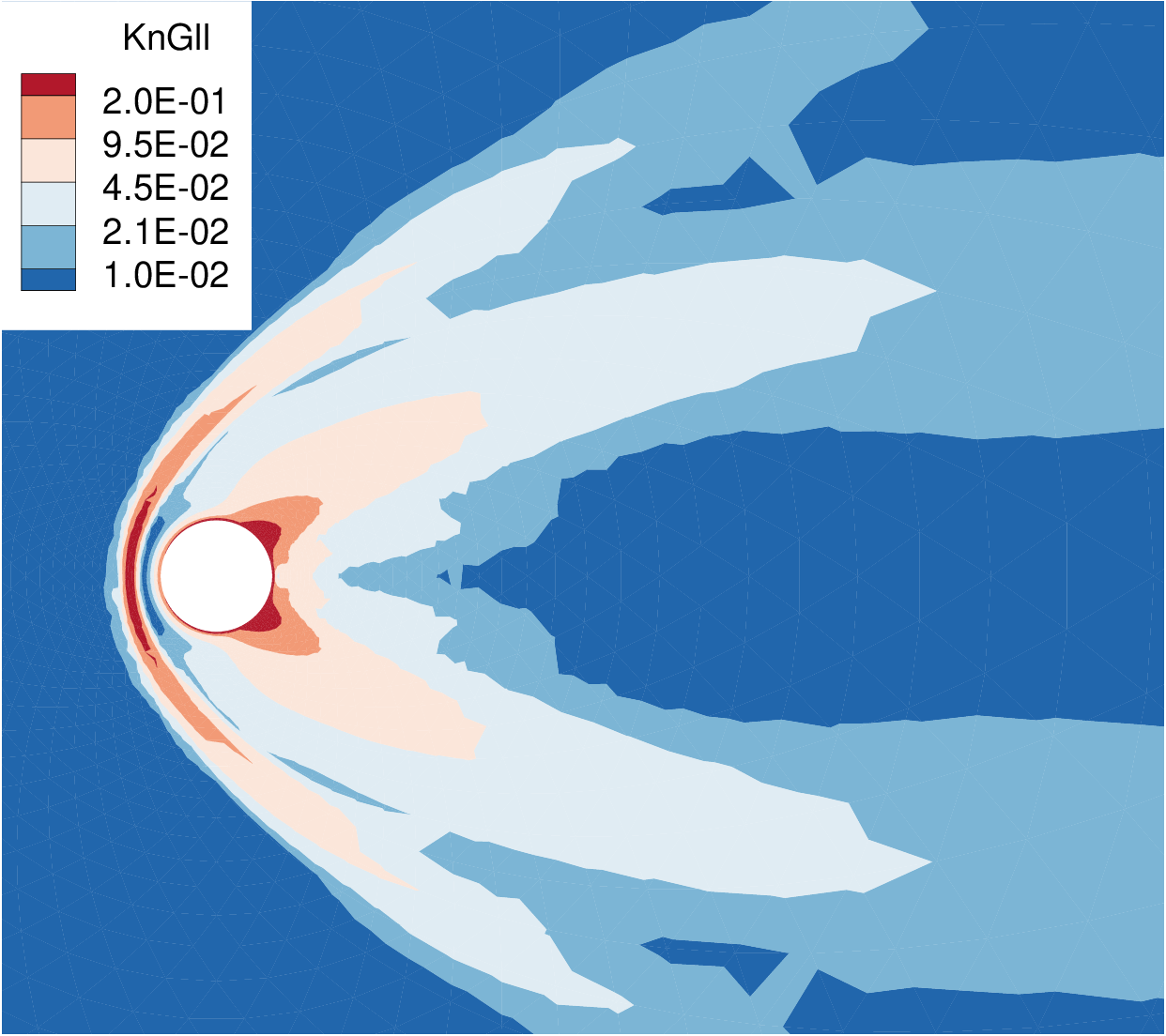}}\\
	\subfloat[]{\includegraphics[width=0.3\textwidth]
	{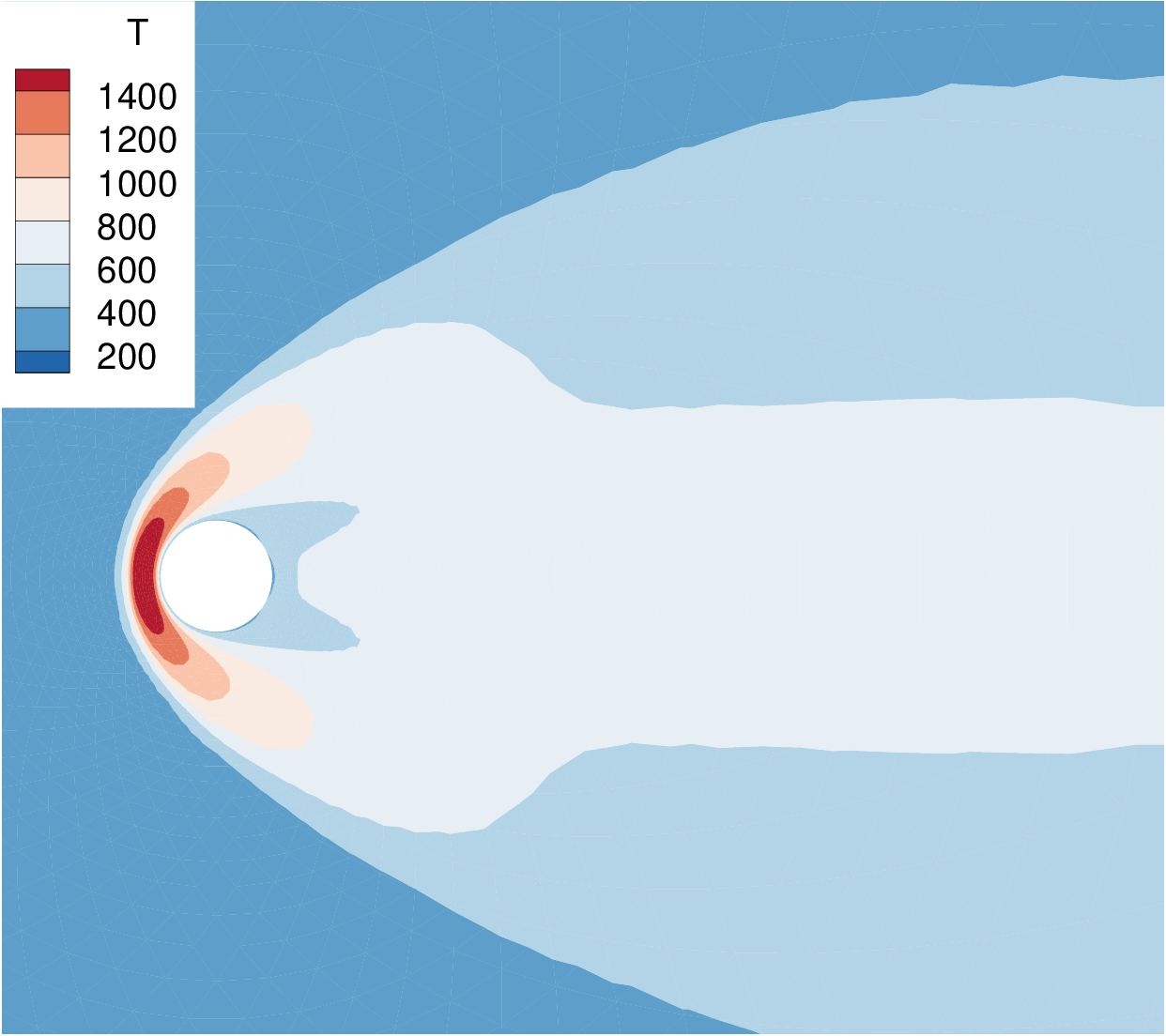}} ~
	\subfloat[]{\includegraphics[width=0.3\textwidth]
	{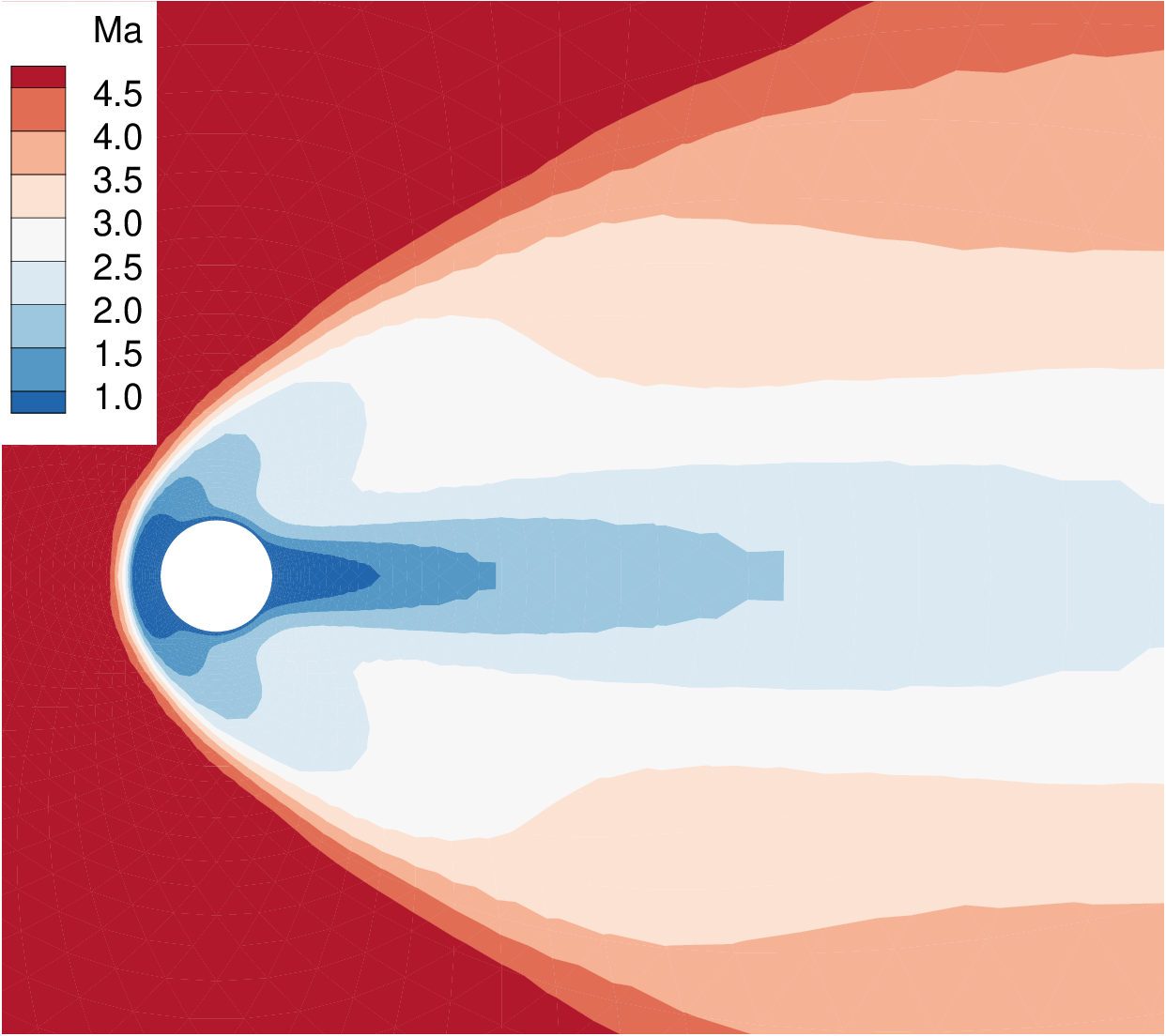}}\\
	\caption{Hypersonic flow at ${\rm Kn} = 0.1$ and ${\rm Ma} = 5$ passing over a circular cylinder with $C_t = 0.005$ by the IAUGKS. (a) Density, (b) $\rm{Kn}_{Gll}$,
	(c) temperature, and (d) Mach number contours.}
	\label{fig:cylinder-Ma5-0.005}
\end{figure}

\begin{figure}[H]
	\centering
	\subfloat[]{\includegraphics[width=0.3\textwidth]
		{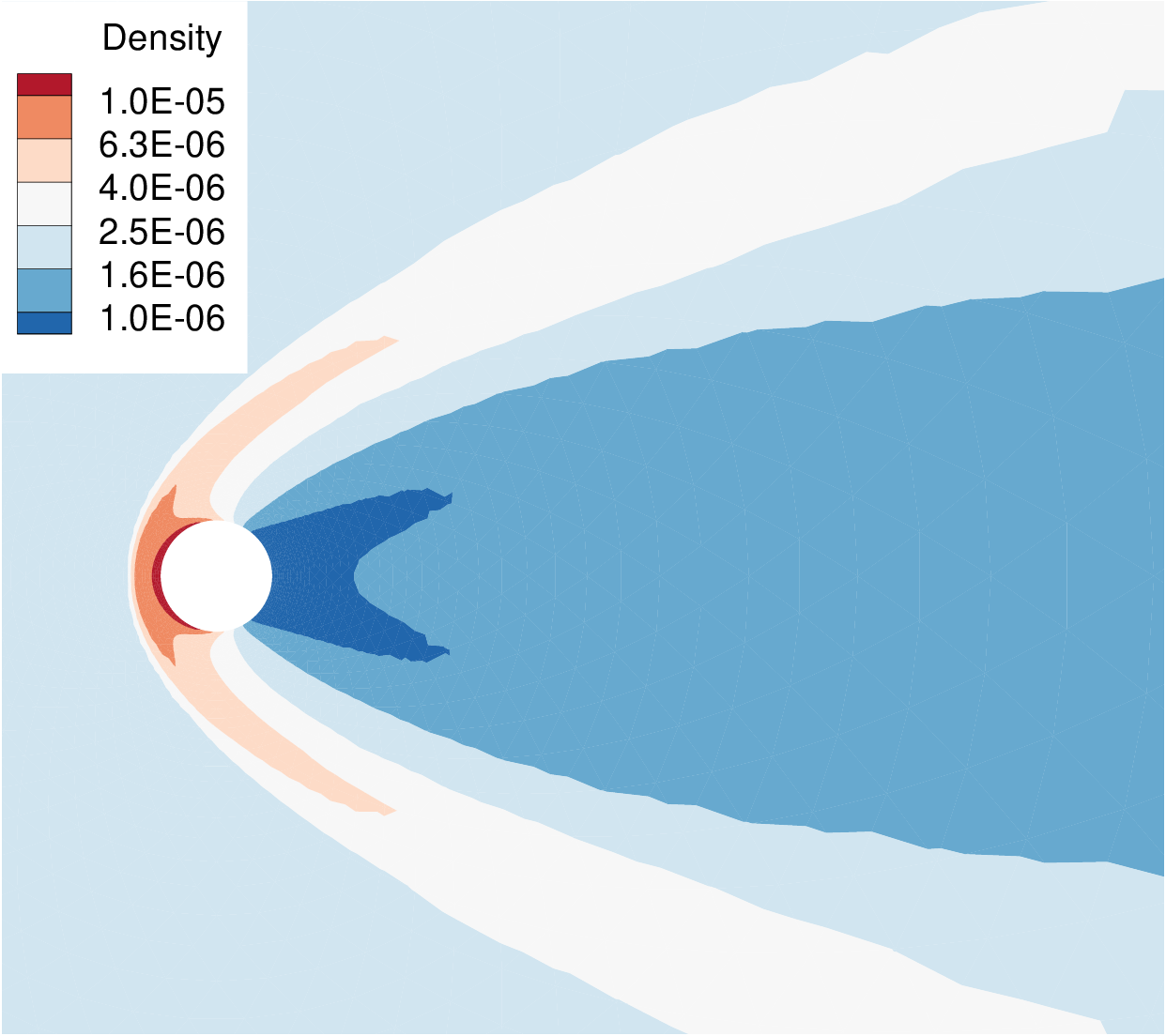}}~
	\subfloat[]{\includegraphics[width=0.3\textwidth]
		{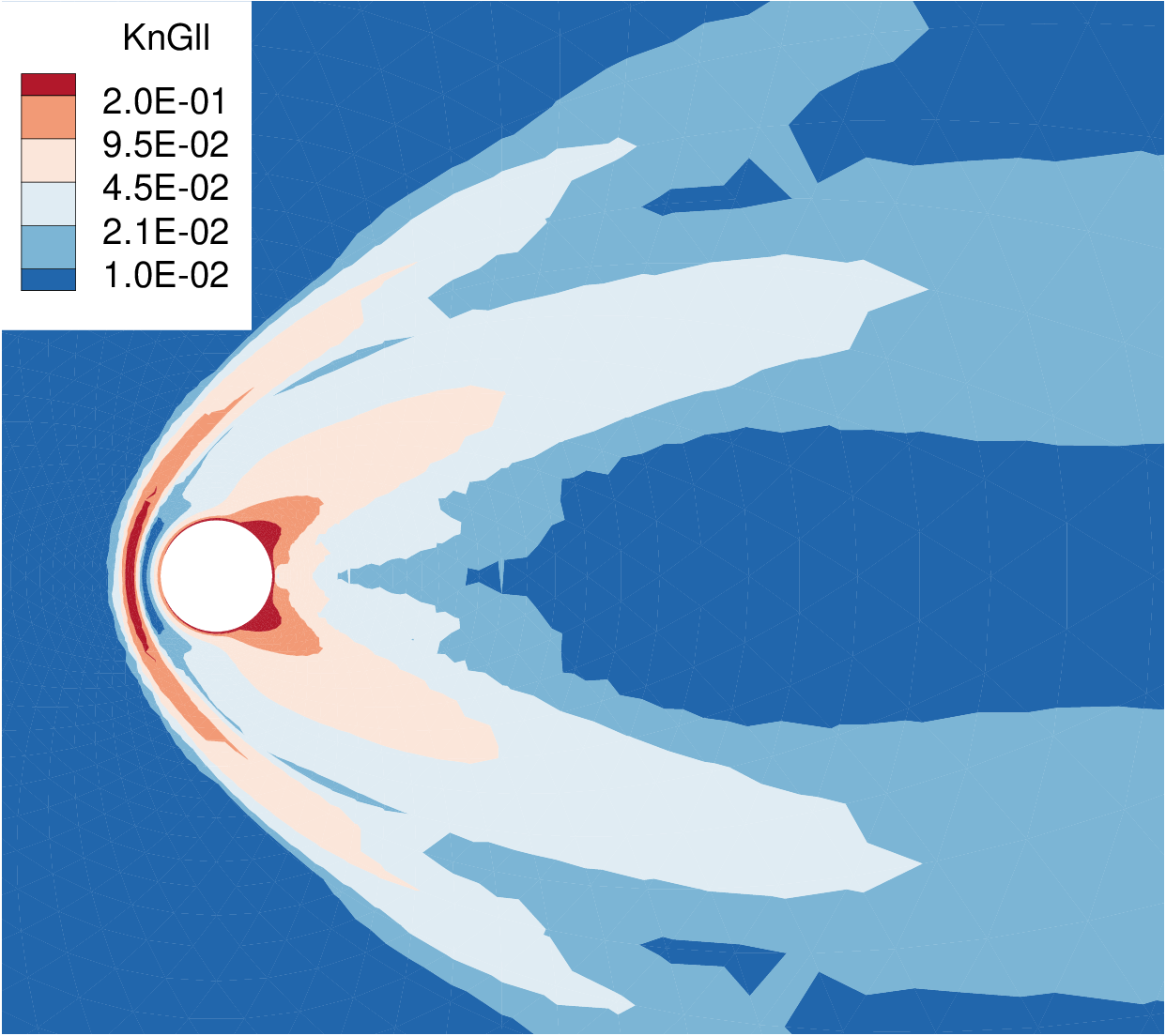}}\\
	\subfloat[]{\includegraphics[width=0.3\textwidth]
		{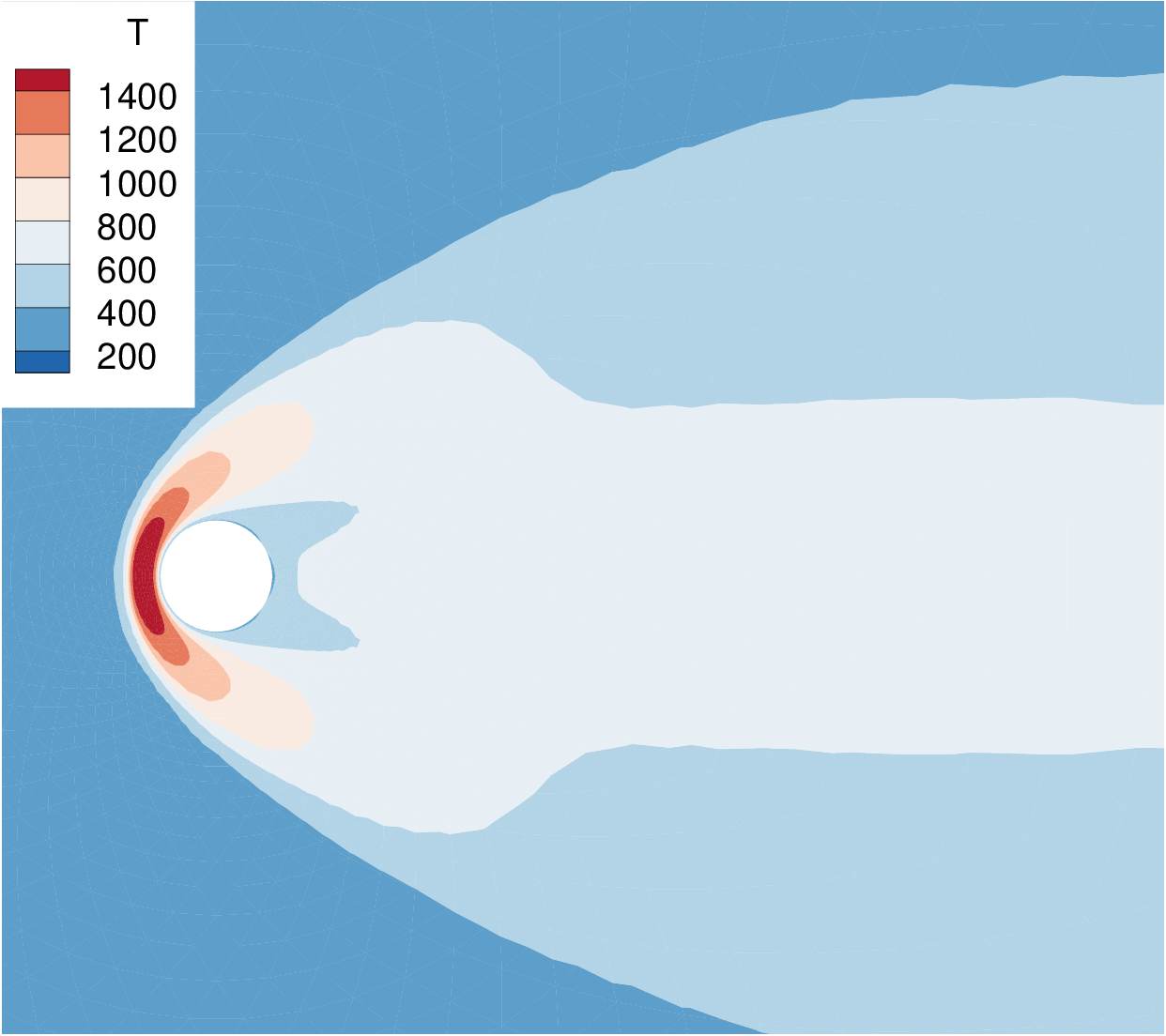}} ~
	\subfloat[]{\includegraphics[width=0.3\textwidth]
		{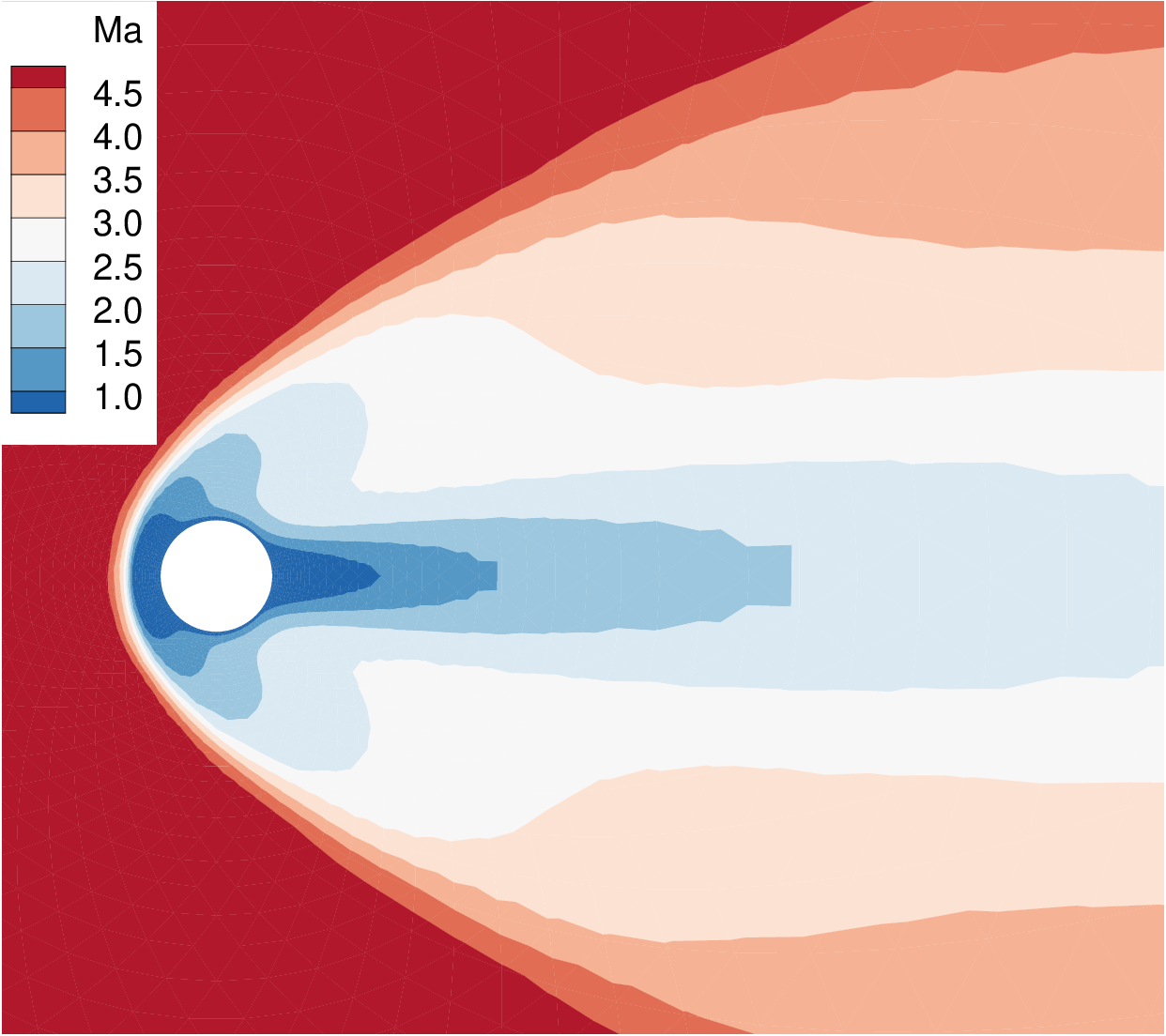}}\\
	\caption{Hypersonic flow at ${\rm Kn} = 0.1$ and ${\rm Ma} = 5$ passing over a circular cylinder with $C_t = 0.01$ by the IAUGKS. (a) Density, (b) $\rm{Kn}_{Gll}$,
		(c) temperature, and (d) Mach number contours.}
	\label{fig:cylinder-Ma5-0.01}
\end{figure}

\begin{figure}[H]
	\centering
	\subfloat[]{\includegraphics[width=0.3\textwidth]
		{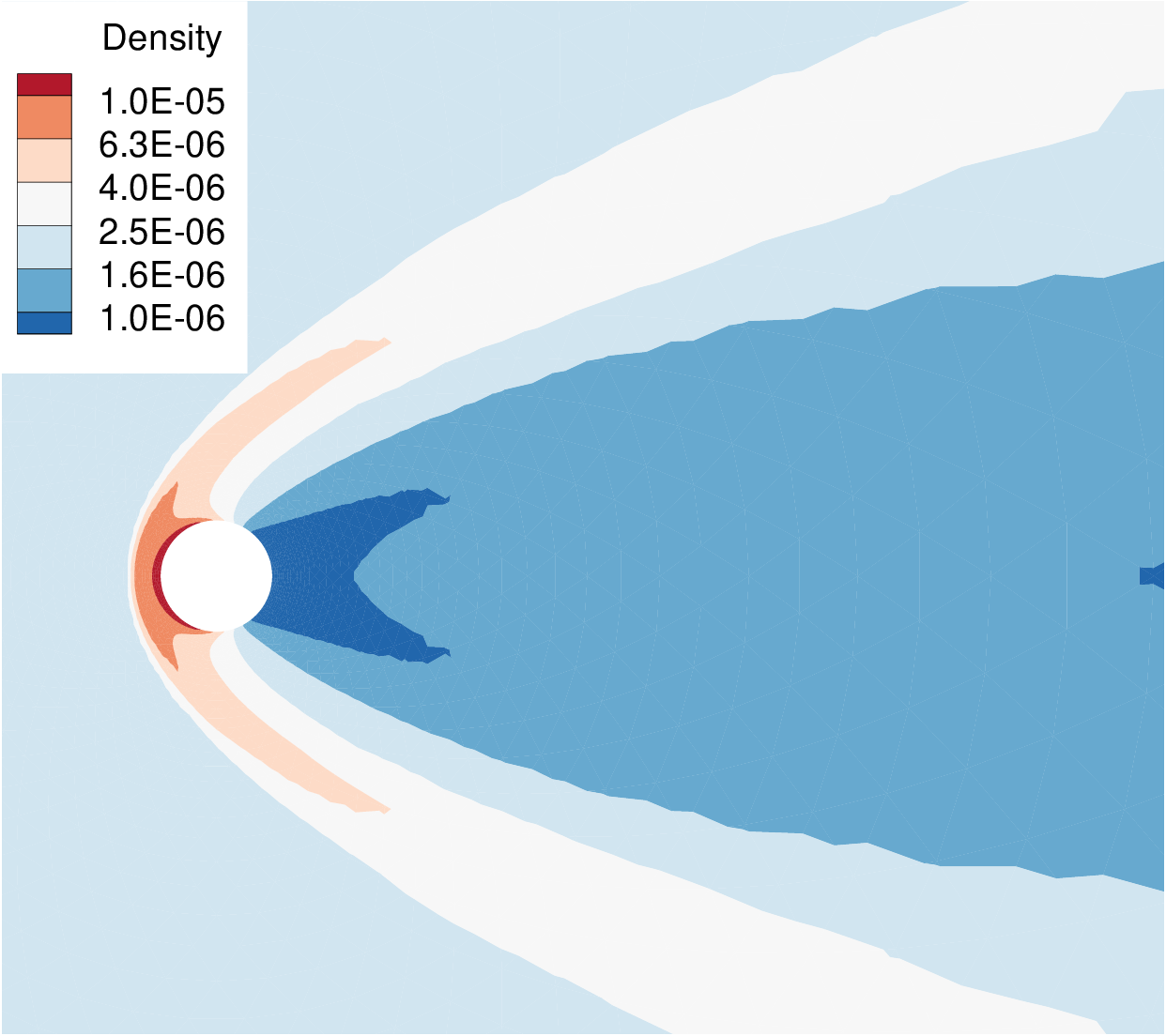}}~
	\subfloat[]{\includegraphics[width=0.3\textwidth]
		{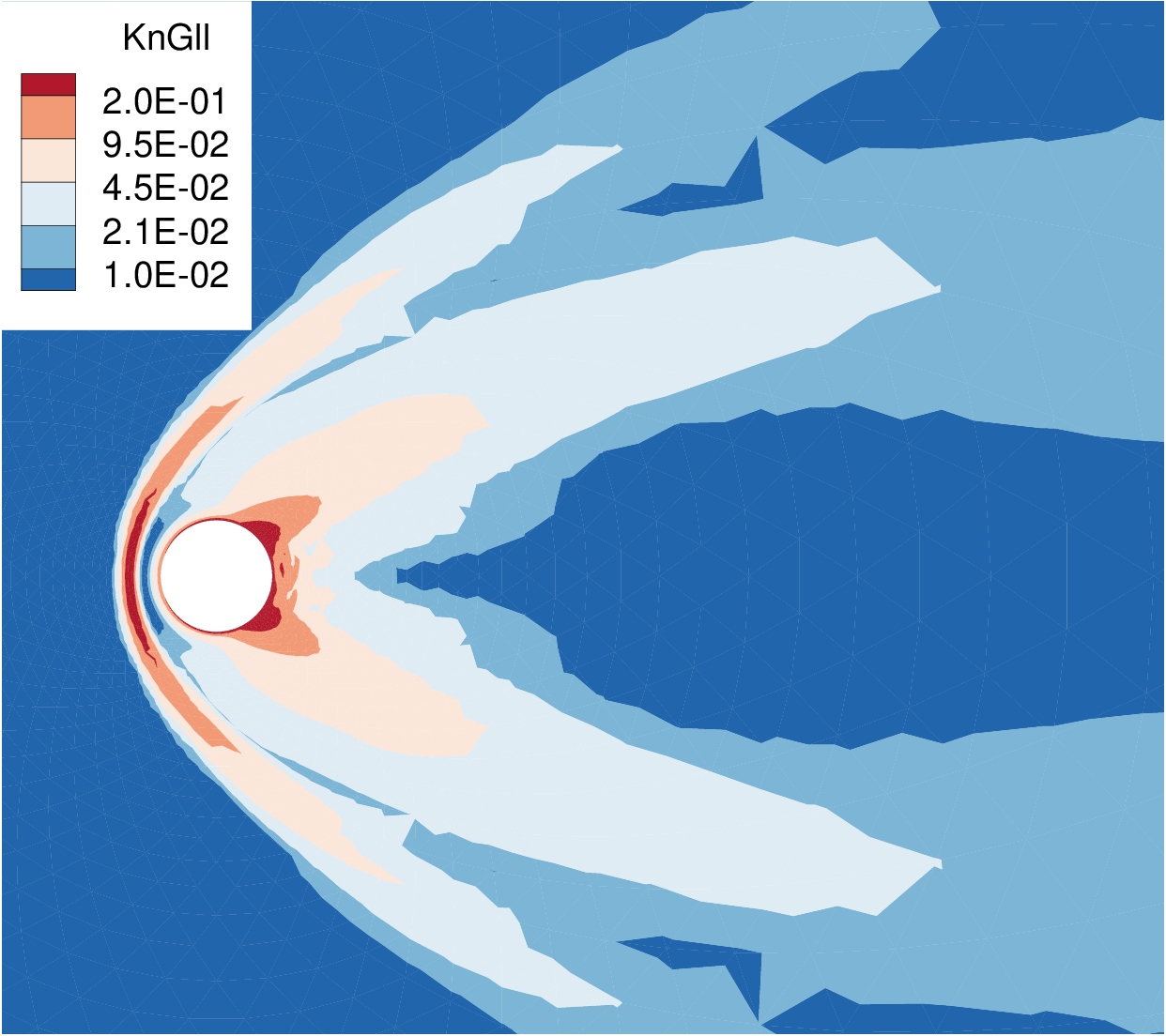}}\\
	\subfloat[]{\includegraphics[width=0.3\textwidth]
		{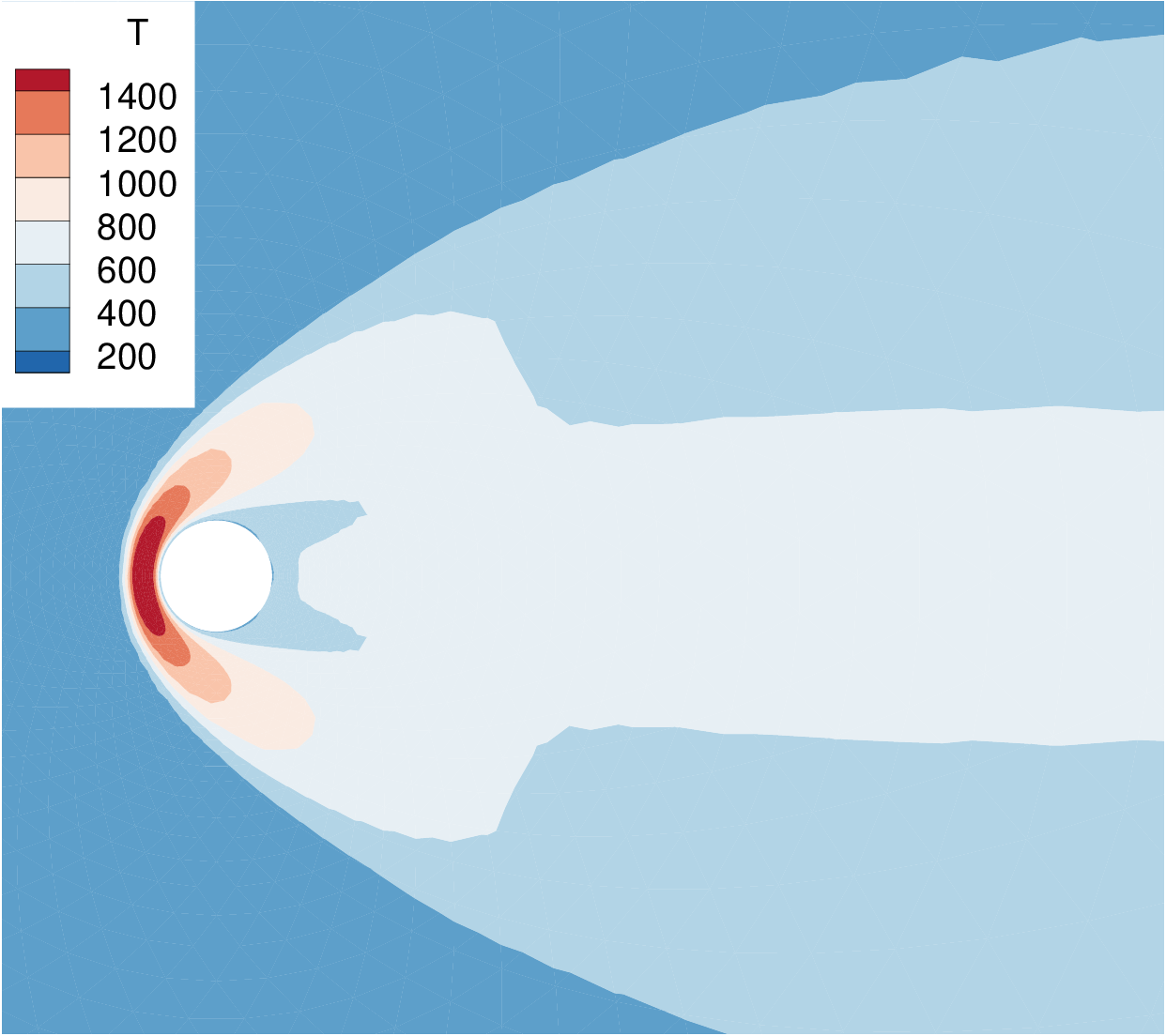}} ~
	\subfloat[]{\includegraphics[width=0.3\textwidth]
		{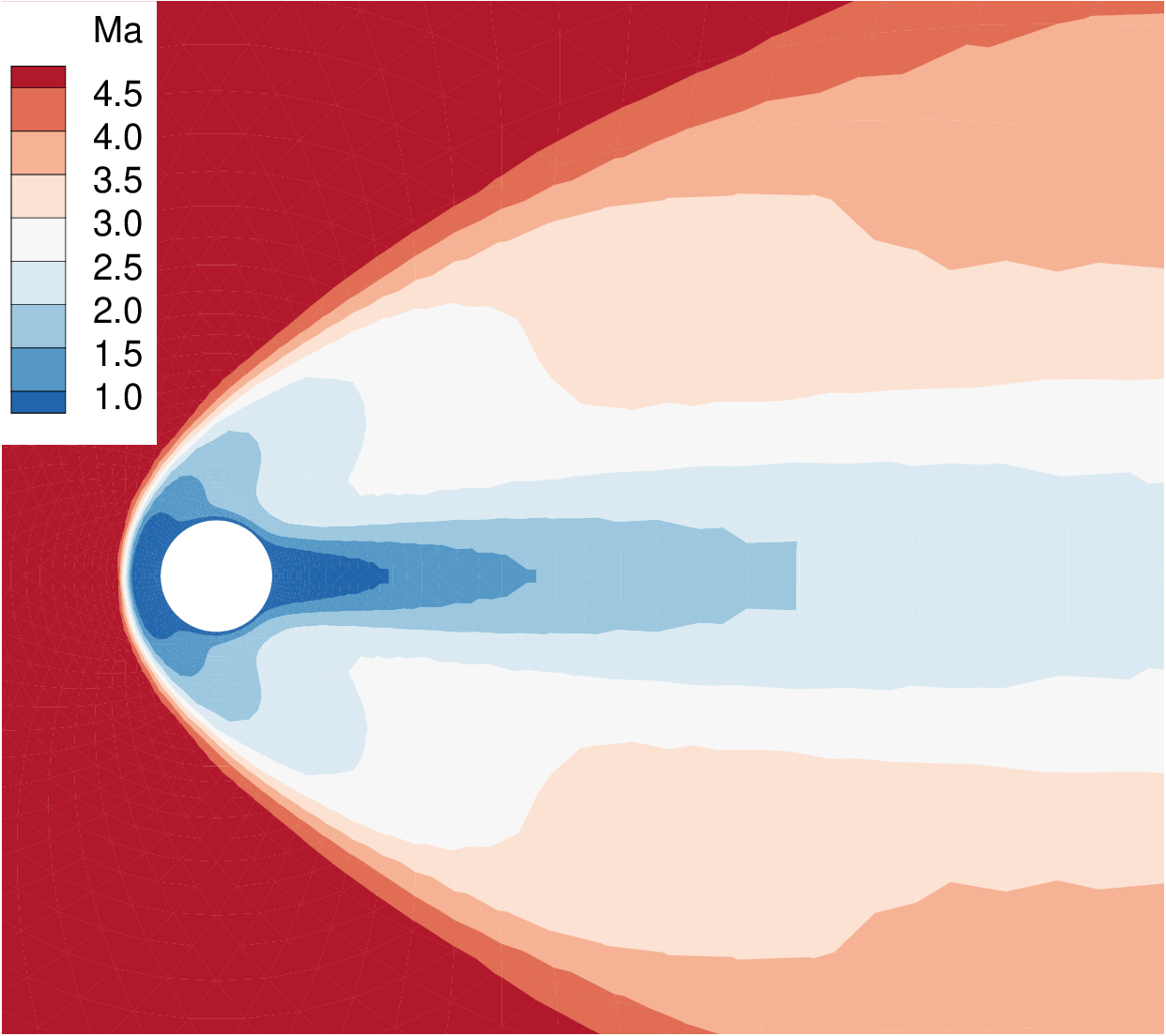}}\\
	\caption{Hypersonic flow at ${\rm Kn} = 0.1$ and ${\rm Ma} = 5$ passing over a circular cylinder with $C_t = 0.05$ by the IAUGKS. (a) Density, (b) $\rm{Kn}_{Gll}$,
		(c) temperature, and (d) Mach number contours.}
	\label{fig:cylinder-Ma5-0.05}
\end{figure}

\begin{figure}[H]
	\centering
	\subfloat[]{\includegraphics[width=0.3\textwidth]
		{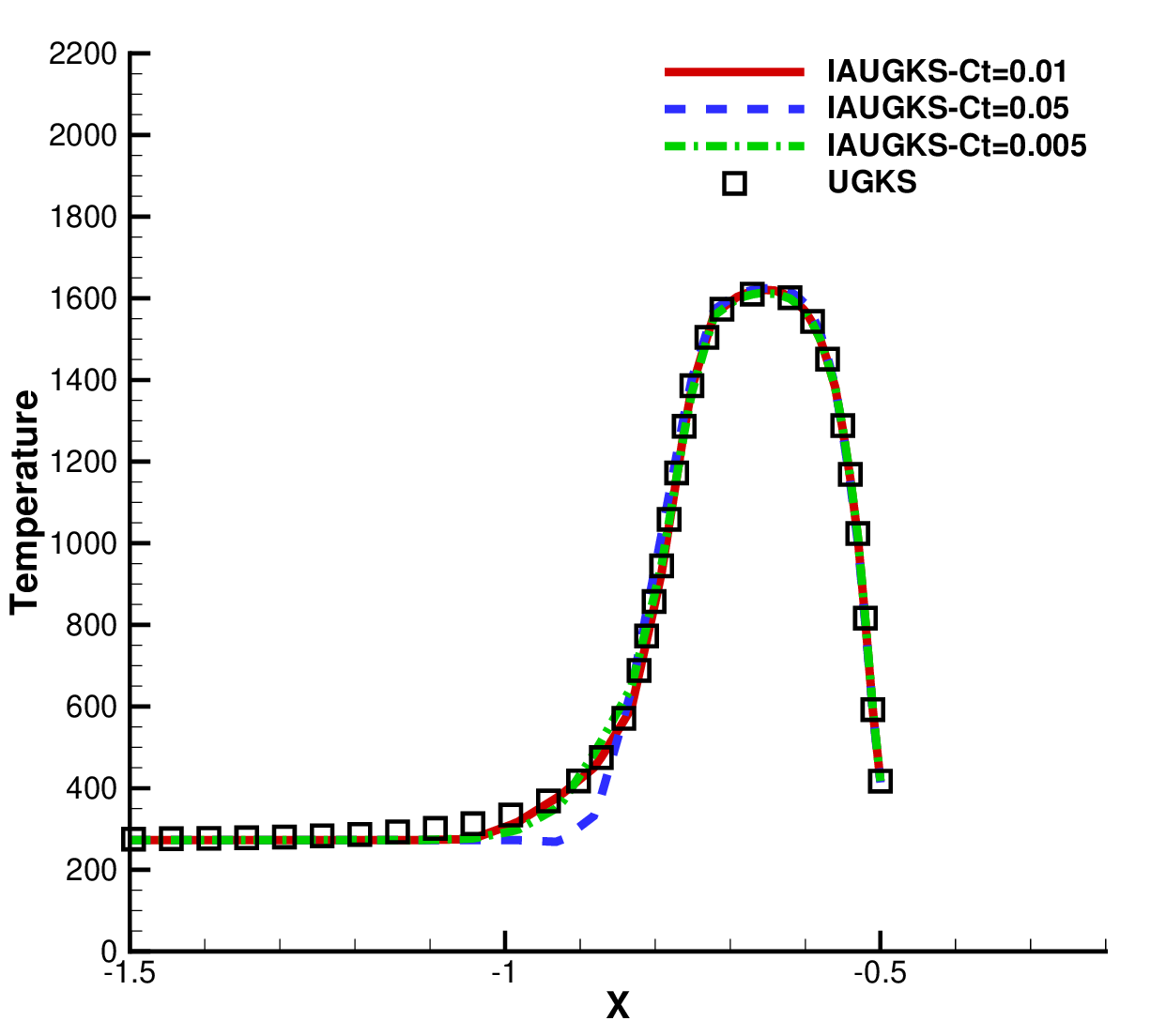}}\\
	\caption{Hypersonic flow at ${\rm Kn}_\infty = 0.1$ and ${\rm Ma} = 5$ passing over a circular cylinder by the IAUGKS. The temperature distribution along the stagnation line.}
	\label{fig:cylinder-Ma5-Tline}
\end{figure}

\begin{figure}[H]
	\centering
	\subfloat[]{\includegraphics[width=0.3\textwidth]
		{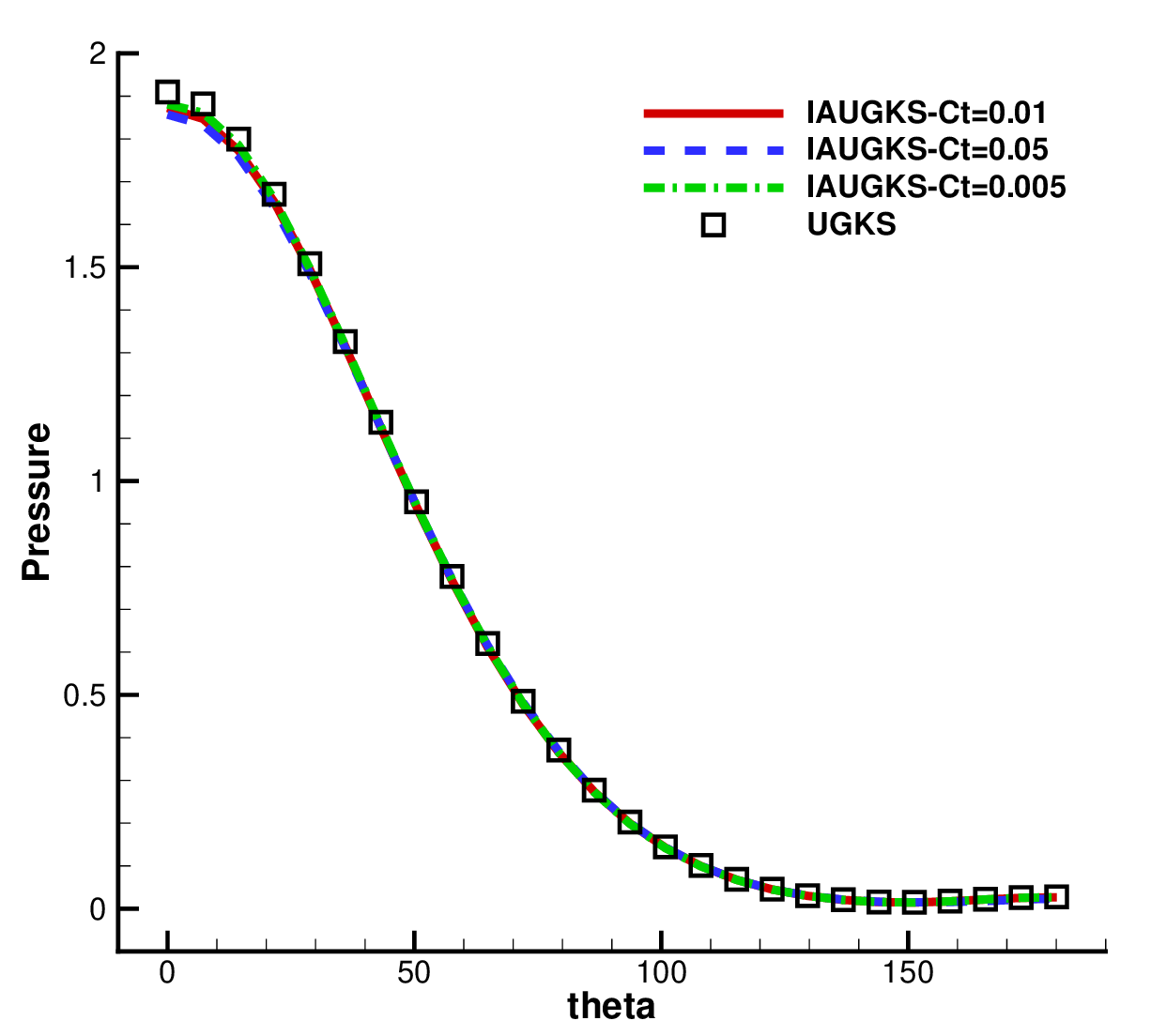}}
	\subfloat[]{\includegraphics[width=0.3\textwidth]
		{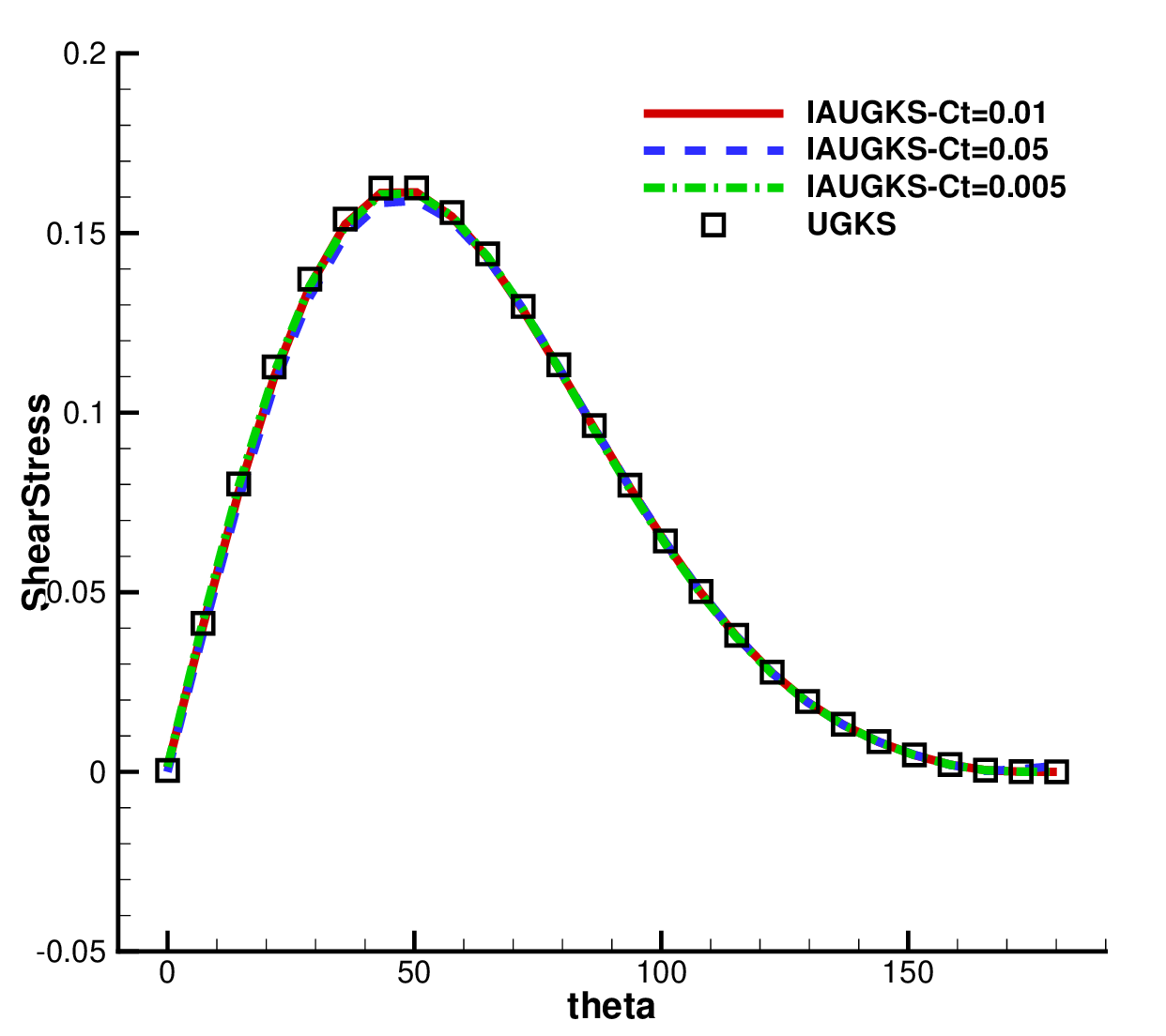}}
	\subfloat[]{\includegraphics[width=0.3\textwidth]
		{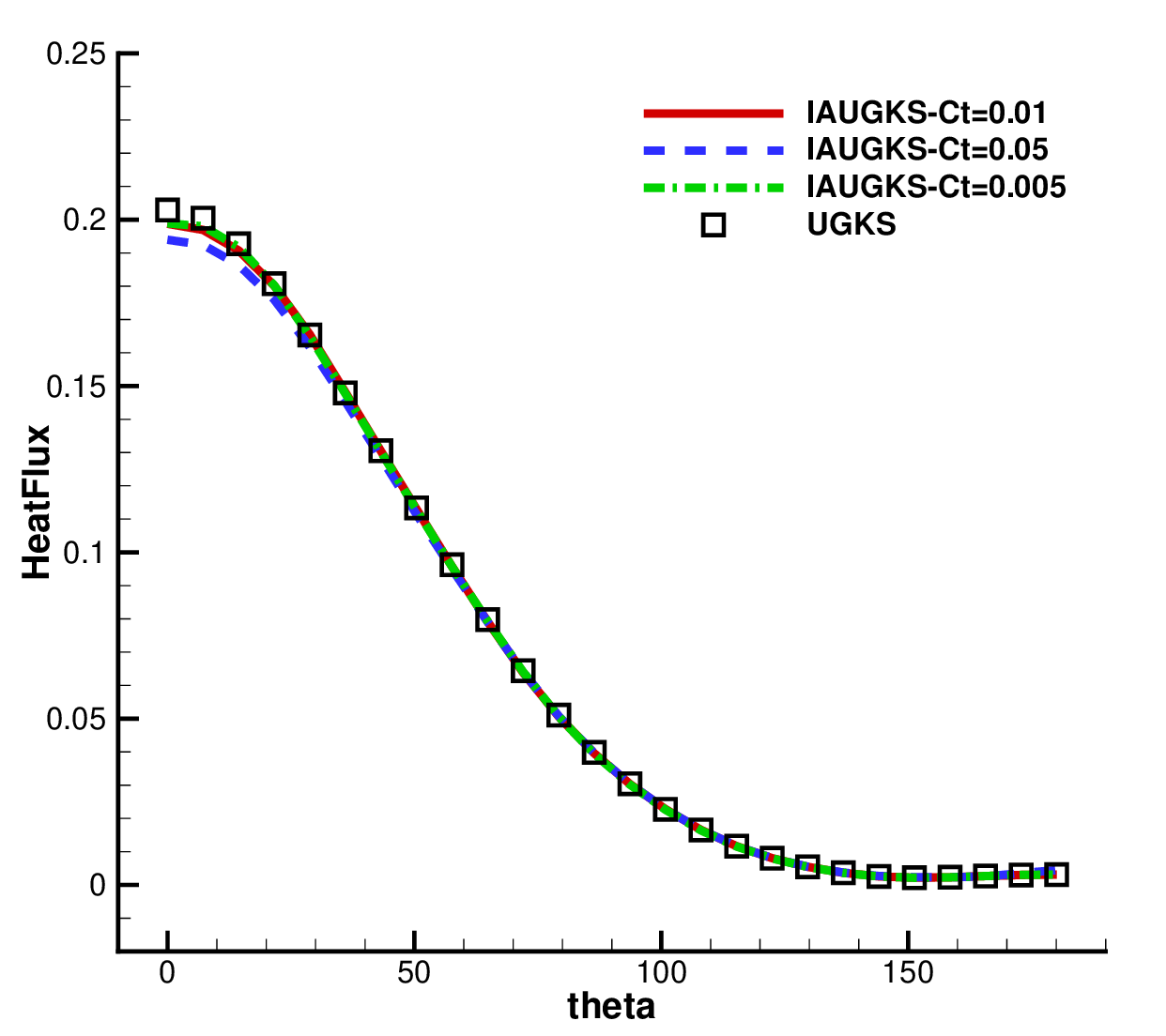}}\\
	\caption{Hypersonic flow at ${\rm Kn}_\infty = 0.1$ and ${\rm Ma} = 5$ passing over a circular cylinder by the IAUGKS. Surface quantities distributions: (a) pressure, (b) shear stress, and (c) heat flux.}
	\label{fig:cylinder-Ma5-surface}
\end{figure}

To validate the velocity space adaptation, cases with criteria of $C_t = 0.005$, $C_t = 0.01$, and $C_t = 0.05$ are simulated by the IAUGKS. The flow contours are plotted in Figures~\ref{fig:cylinder-Ma5-0.005} to \ref{fig:cylinder-Ma5-0.05}, where a tiny difference can be observed in the Mach number and temperature contour at the leeward region. Figure~\ref{fig:cylinder-Ma5-Tline} shows the temperature distribution along the stagnation line. Compared with the explicit UGKS data, a deviation at the front of the shock region appears in $C_t = 0.05$ while other data all fit well. As depicted in Figure ~\ref{fig:cylinder-Ma5-surface}, the pressure and shear stress coefficients on the cylinder surface are consistent for all criteria, but for $C_t = 0.05$ the heat flux coefficient at the windward region slightly deviates from the reference.

\begin{figure}[H]
	\centering
	\subfloat[]{\includegraphics[width=0.3\textwidth]
		{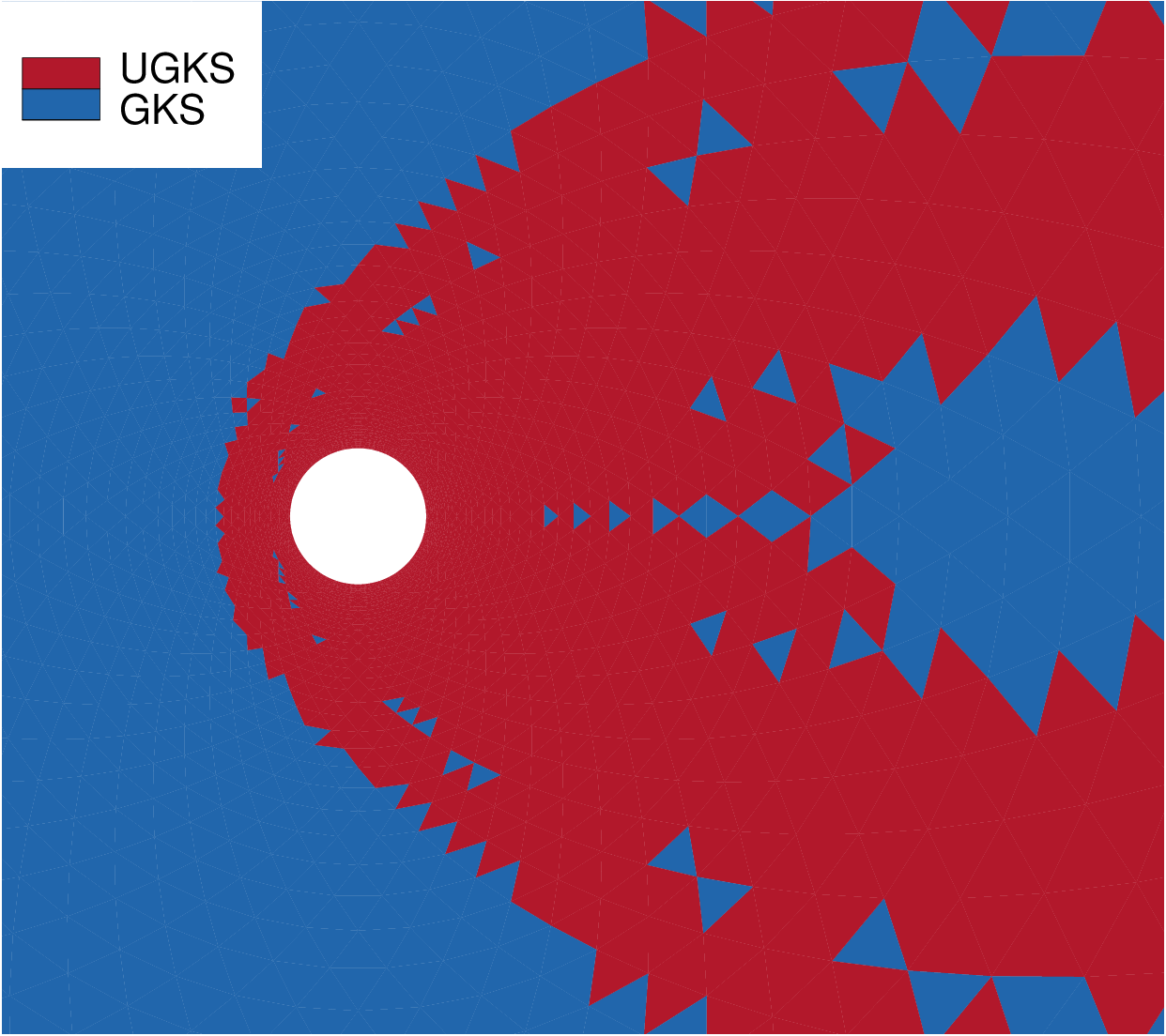}}~
	\subfloat[]{\includegraphics[width=0.3\textwidth]
		{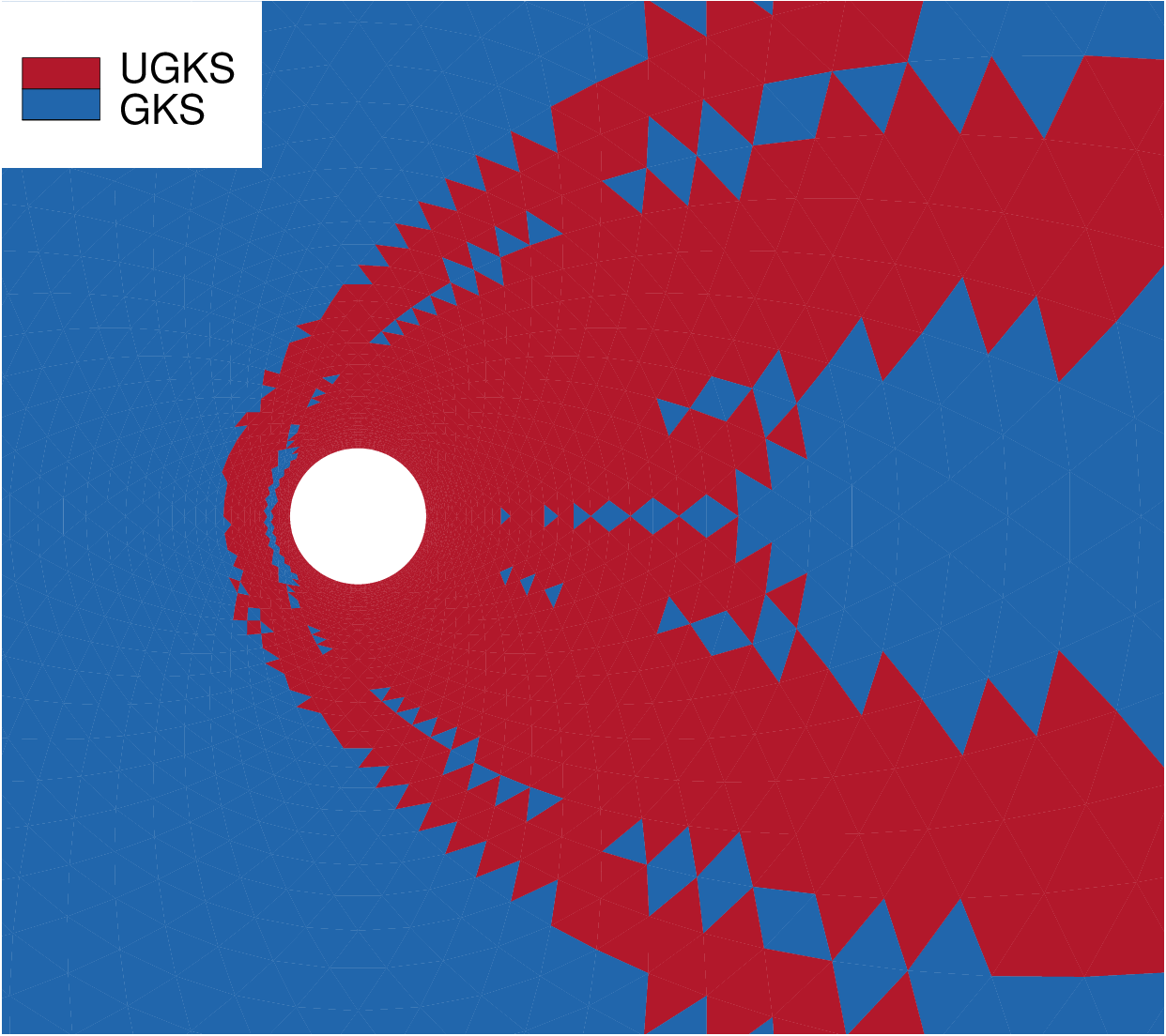}}~
	\subfloat[]{\includegraphics[width=0.3\textwidth]
		{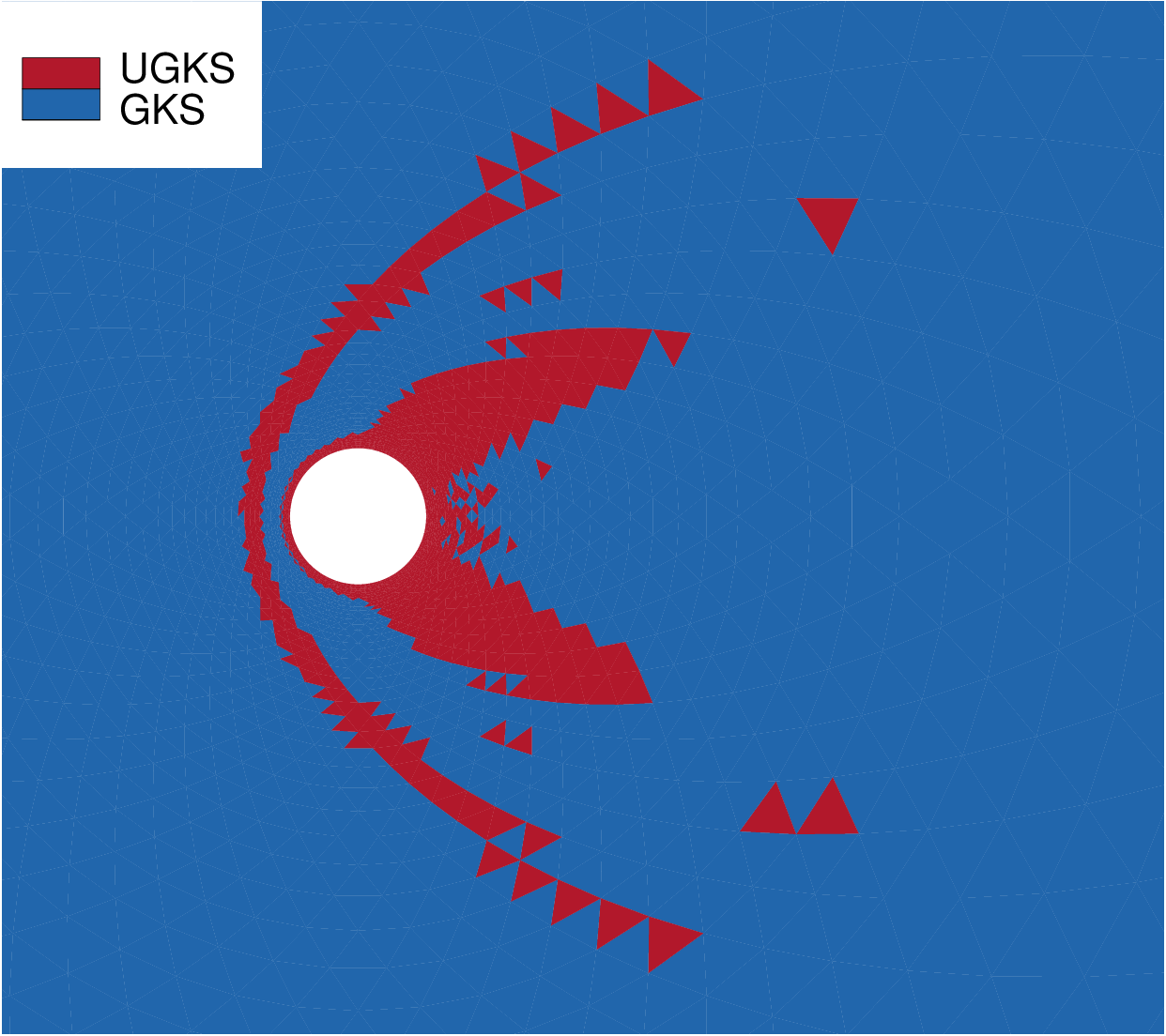}}\\
	\caption{Hypersonic flow at ${\rm Kn} = 0.1$ and ${\rm Ma} = 5$ passing over a circular cylinder by the IAUGKS. Distributions of velocity space adaptation with (a) $C_t = 0.005$, (b) $C_t = 0.01$, and (c)$C_t = 0.05$.}
	\label{fig:cylinder-Ma5-isDisc}
\end{figure}

Figure~\ref{fig:cylinder-Ma5-isDisc} depicts that 59.83\%, 55.01\%, and 29.71\% of the computational domain is covered by DVS (UGKS) for $C_t = 0.005$, $C_t = 0.01$, and $C_t = 0.05$ respectively. The physical CFL number is set to 0.4 and the numerical CFL number for implicit iteration is set to 400. As shown in Table~\ref{table:cylindertime}, the simulation time decreases as the proportion of the DVS region shrinks. Compared with the original explicit UGKS, the unstructured DVS helps to improve the efficiency by 3 times; by adding the implicit algorithm, the simulation time further diminishes by nearly 100 times. Moreover, with different decomposition criterion, the computation efficiency of the IAUGKS is raised to over 700 times higher than the original UGKS with structured DVS. All the IAUGKS simulations are conducted on a personal computer with a single core of intel 13700K @5.30 GHz.

The cases with different decomposition criteria give reasonable results with different acceleration rates. Considering the ability to capture non-equilibrium physics and computation efficiency in conjunction, $C_t = 0.01$ and $C_t = 0.05$ are suitable for non-equilibrium flow simulations.  

\begin{table}[H]
	\caption{The computational cost for simulations of hypersonic flow at ${\rm Kn}_\infty = 0.1$ and ${\rm Ma} = 5$ around a cylinder by the IAUGKS. The physical domain consists of 5,000 cells, and the unstructured DVS mesh is discretized by 2,060 cells.}
	\centering
	\begin{threeparttable}
		\begin{tabular}{ccccc}
			\hline
			Solver & \makecell[c]{Acceleration\\Method} & Steps & Simulation Time, h & Acceleration Rate  \\
			\hline			
			UGKS & Original UGKS & $40000$ & 190.7 & 1\\ \hline	
			UGKS & unstructured DVS & $45000$ & 63.3 & 3.01\\ \hline	
			UGKS & \makecell[c]{unstructured DVS\\implicit iteration} & $1000\tnote{1}+300$ & 0.665 & 286.8\\
			\hline	
			UGKS & \makecell[c]{unstructured DVS\\implicit iteration \\ Adaptation $C_t = 0.005$} & $1000\tnote{1}+300$ & 0.411 & 464.0\\
			\hline	
			UGKS & \makecell[c]{unstructured DVS\\implicit iteration \\ Adaptation $C_t = 0.01$} & $1000\tnote{1}+300$ & 0.379 & 503.2\\
			\hline	
			UGKS & \makecell[c]{unstructured DVS\\implicit iteration \\ Adaptation $C_t = 0.05$} & $1000\tnote{1}+300$ & 0.262 & 727.9\\
			\hline
		\end{tabular}
		
		\begin{tablenotes}
			\item[1] Steps of first-order GKS simulations.
		\end{tablenotes}
	\end{threeparttable}
	\label{table:cylindertime}
\end{table}

\begin{figure}[H]
	\centering
	\includegraphics[width=0.4\textwidth]
	{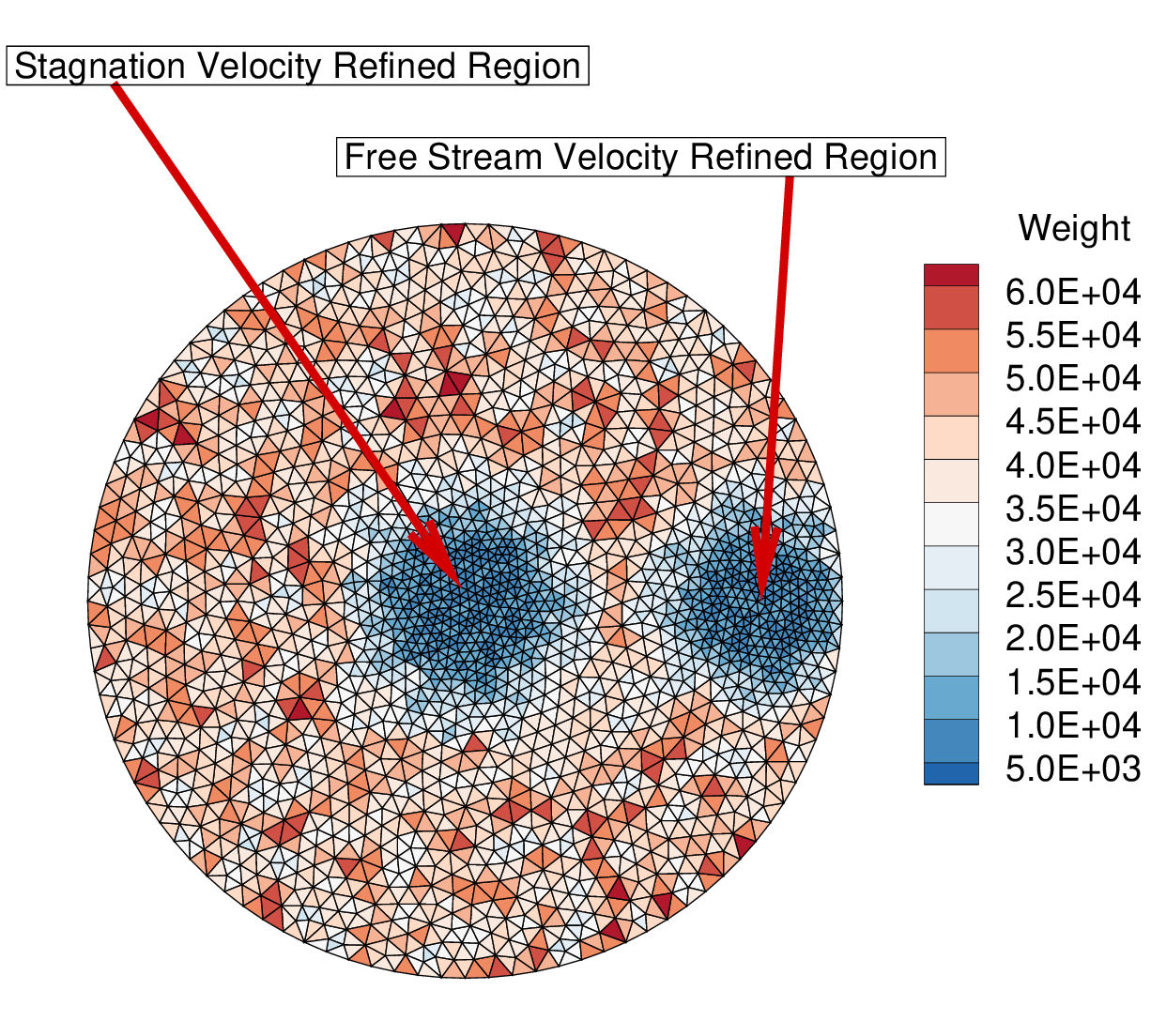}
	\caption{Unstructured DVS mesh used for hypersonic flow at ${\rm Kn} = 0.01$ and ${\rm Ma} = 15$ passing over a cylinder by the IAUGKS.}
	\label{fig:cylinder-Ma15-DVS}
\end{figure}

To further verify the computational accuracy and robustness of the current scheme, hypersonic flow passing over a circular cylinder at a very large Mach number ${\rm Ma}_\infty = 15$ and ${\rm Kn}_\infty = 0.01$ is simulated. The computation domain and geometry characteristics are the same as in the first case. The temperature in free stream flow is $T_\infty = 217.5$ K and the isothermal wall boundary is set to a fixed temperature of $T_w = 300$ K. Figure~\ref{fig:cylinder-Ma15-DVS} shows the unstructured DVS mesh consists of 3,420 cells. The DVS is discretized in a circle region with the center $0.4\times(U_\infty,V_\infty,W_\infty)$, and the radius is $5\sqrt{R T_s}$ where $T_s$ is the stagnation temperature of the free stream flow. The unstructured DVS mesh is refined at zero velocity point with a radius of $5\sqrt{R T_w}$, and the free stream velocity point with a radius of $5\sqrt{R T_\infty}$. 

\begin{figure}[H]
	\centering
	\subfloat[]{\includegraphics[width=0.3\textwidth]
		{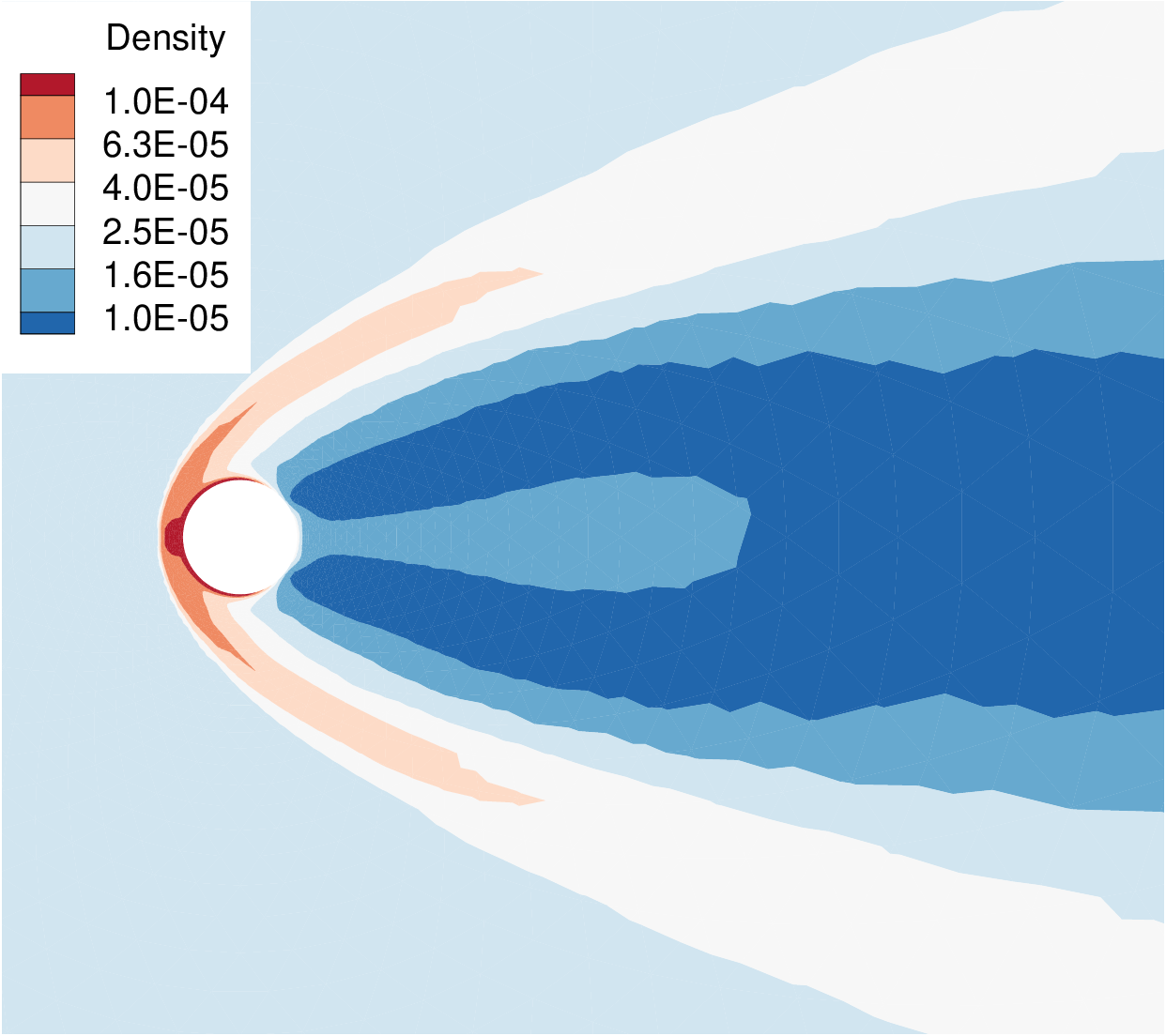}}~
	\subfloat[]{\includegraphics[width=0.3\textwidth]
		{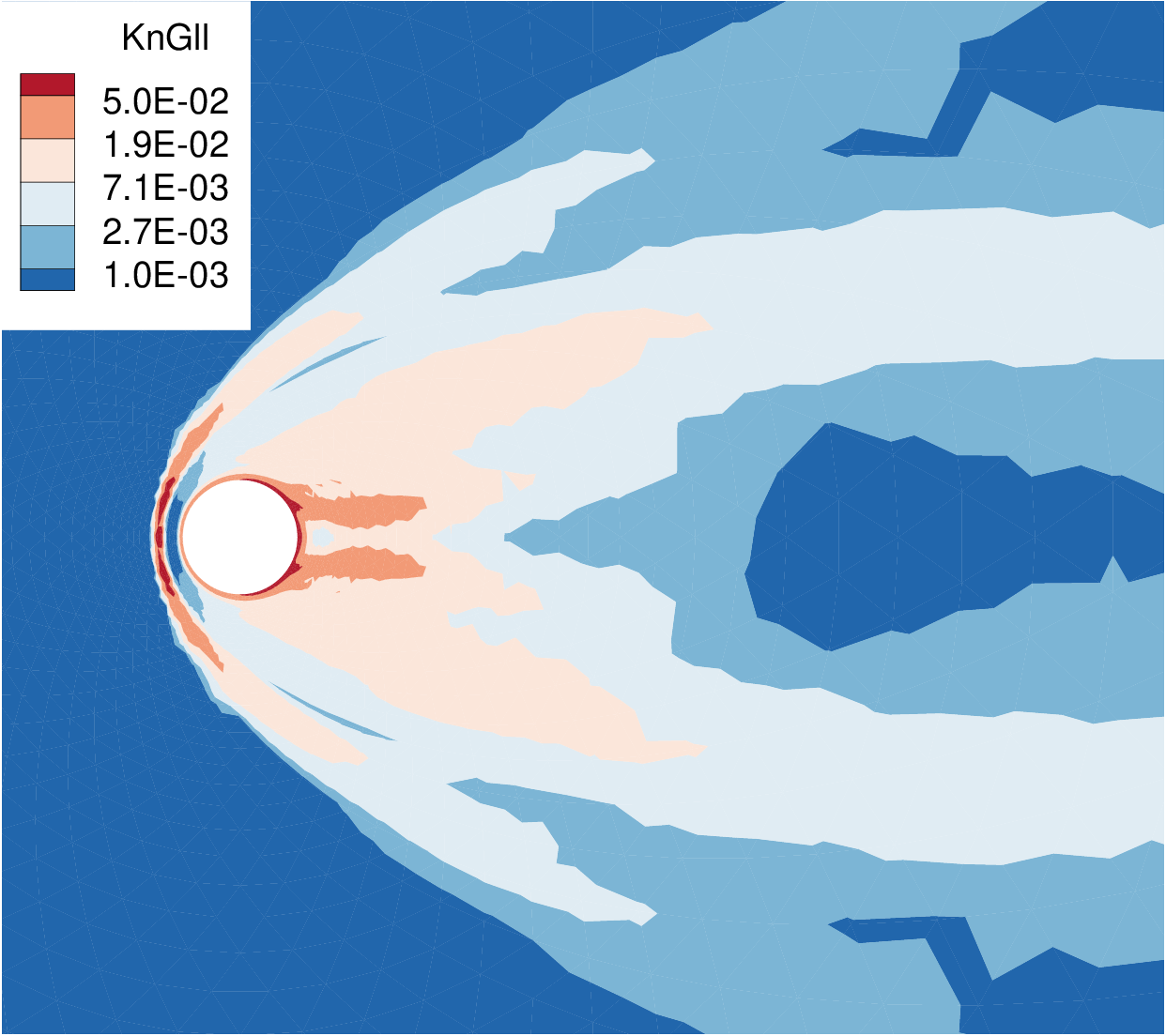}}\\
	\subfloat[]{\includegraphics[width=0.3\textwidth]
		{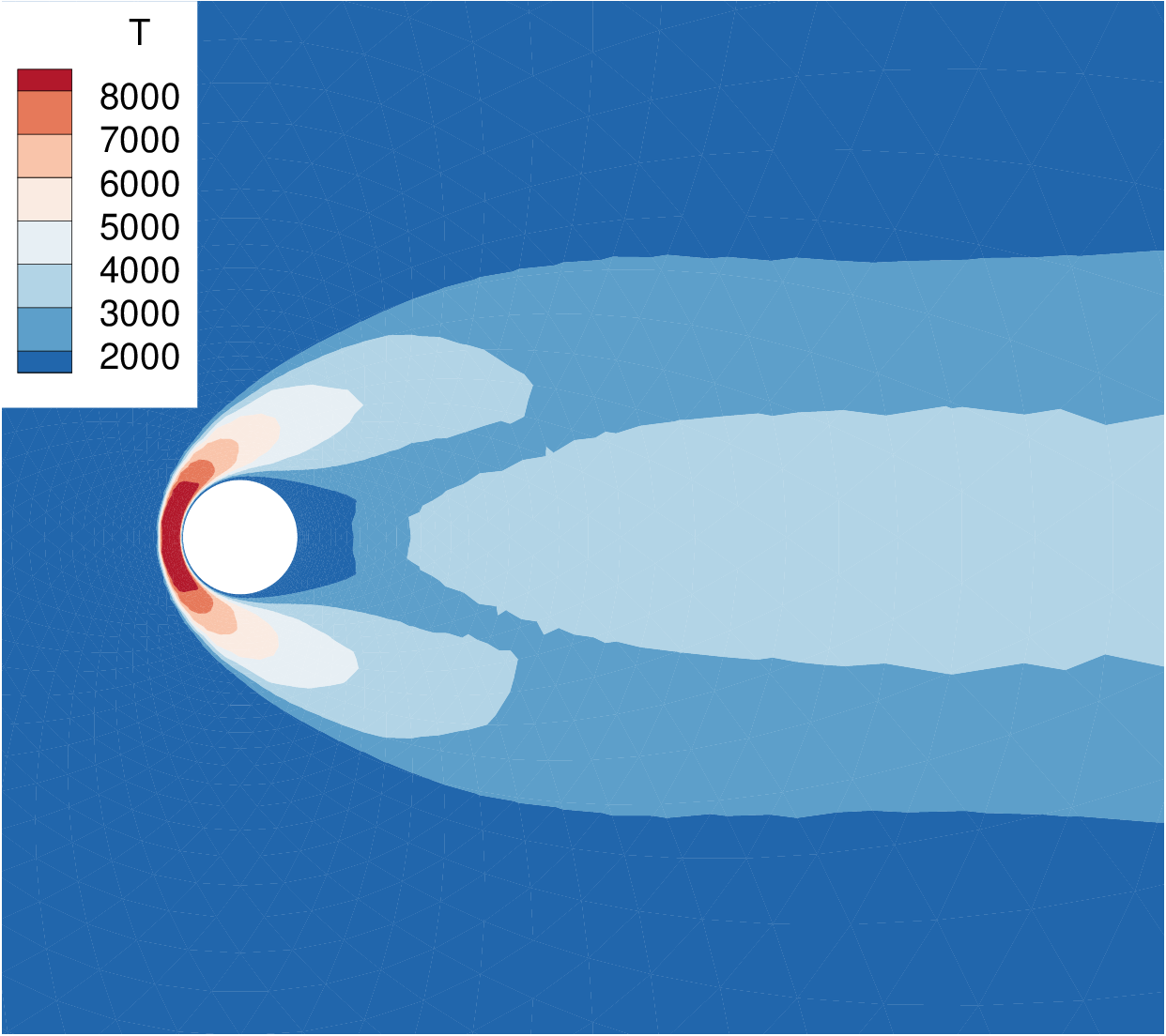}} ~
	\subfloat[]{\includegraphics[width=0.3\textwidth]
		{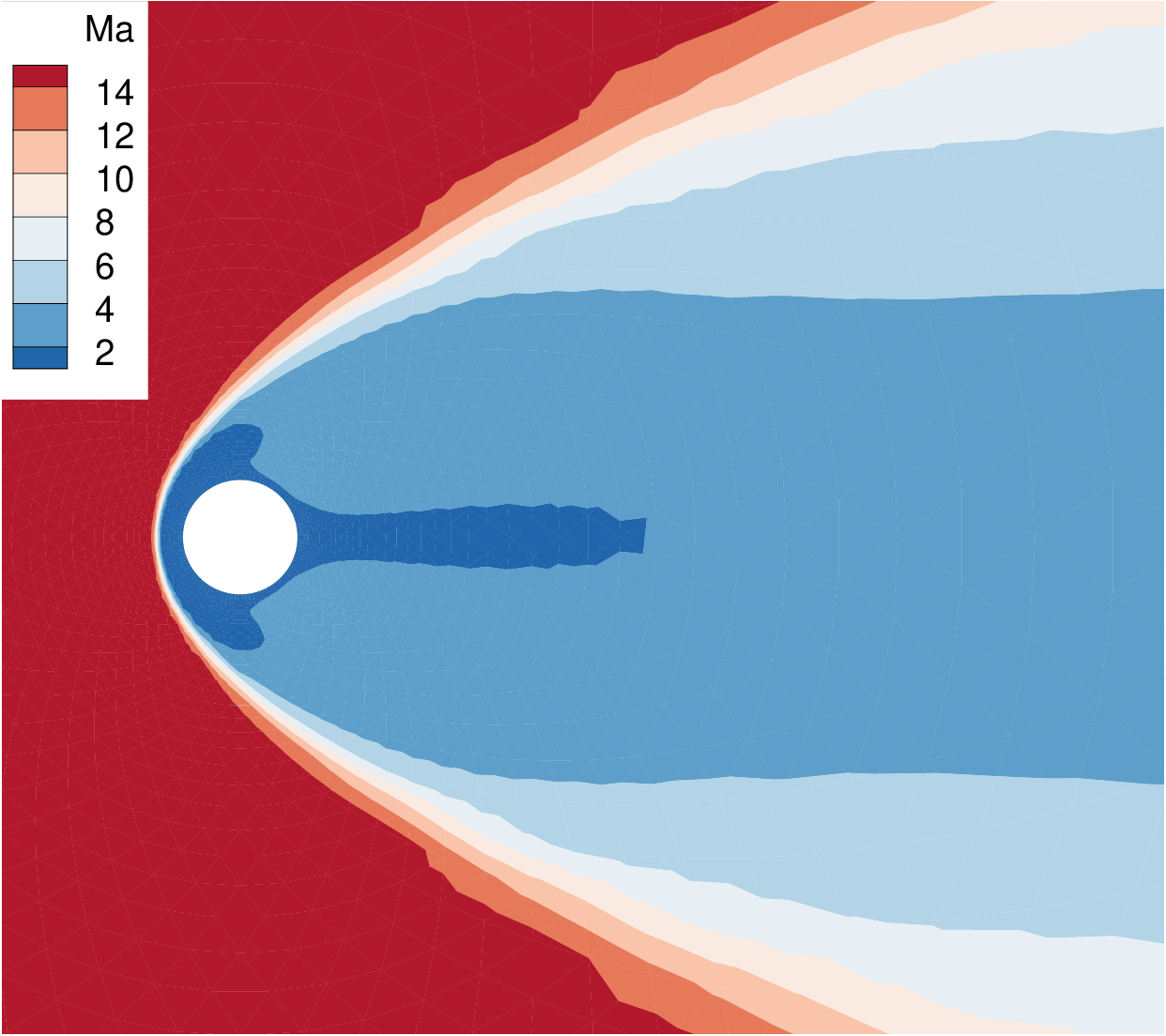}}\\
	\caption{Hypersonic flow at ${\rm Kn} = 0.01$ and ${\rm Ma} = 15$ passing over a circular cylinder with $C_t = 0.01$ by the IAUGKS. (a) Density, (b) $\rm{Kn}_{Gll}$,
		(c) temperature, (d) Mach number contours.}
	\label{fig:cylinder-Ma15-0.01}
\end{figure}

\begin{figure}[H]
	\centering
	\subfloat[]{\includegraphics[width=0.5\textwidth]
		{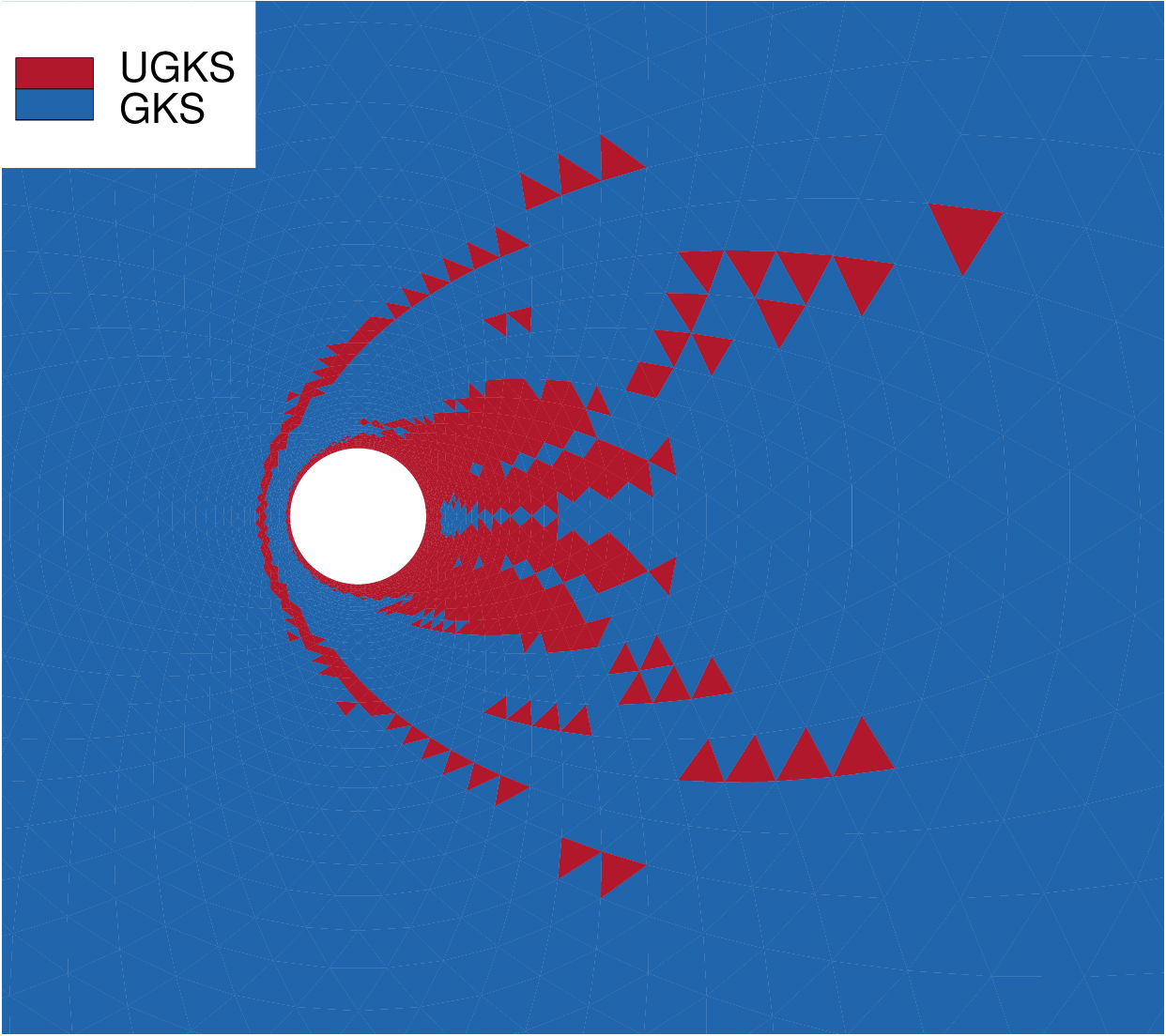}}\\
	\caption{Hypersonic flow at ${\rm Kn} = 0.01$ and ${\rm Ma} = 15$ passing over a circular cylinder by the IAUGKS. Distributions of velocity space adaptation with $C_t = 0.01$.}
	\label{fig:cylinder-Ma15-isDisc}
\end{figure}

The flow contours with $C_t = 0.01$ are plotted in Fig.~\ref{fig:cylinder-Ma15-0.01}. Figure~\ref{fig:cylinder-Ma15-isDisc} shows 25.12\% of the region is covered by DVS with the current criterion. The vicinity of shock and leeward regions are contained within the UGKS region, which is consistent with the highly non-equilibrium area shown in the $\rm{Kn}_{Gll}$ contour. Even though most of the computation domain is covered by the GKS, the IAUGKS gives accurate predictions of surface quantities compared with the explicit UGKS. 

The physical CFL number is set to 0.4 and the numerical CFL number for implicit iteration is set to 40. The computation efficiency is compared with the original UGKS with structured DVS and the IUGKS. All the IAUGKS simulations are conducted on a personal computer platform with a single core of intel 13700K @5.30 GHz. 

\begin{figure}[H]
	\centering
	\subfloat[]{\includegraphics[width=0.3\textwidth]
		{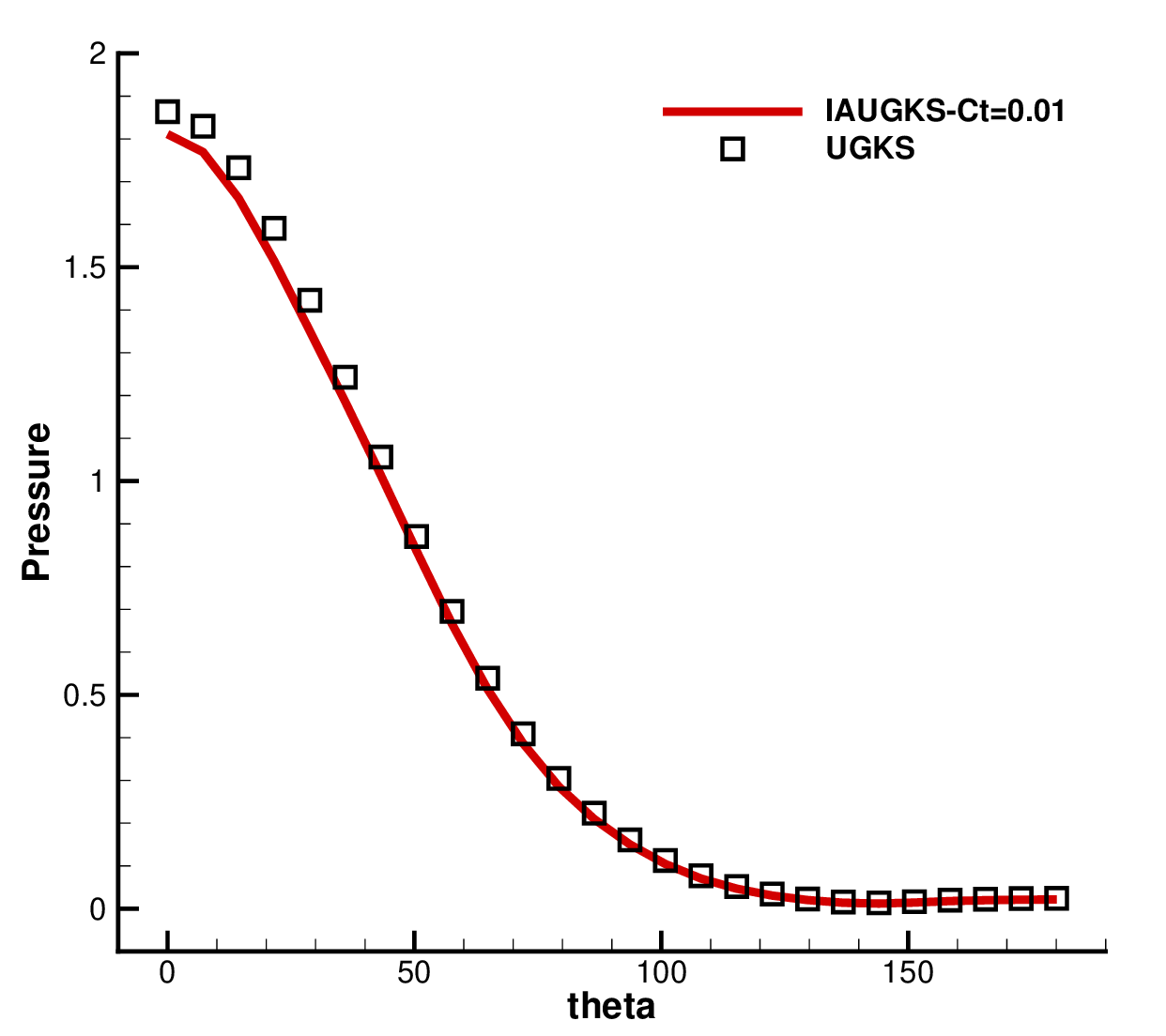}}
	\subfloat[]{\includegraphics[width=0.3\textwidth]
		{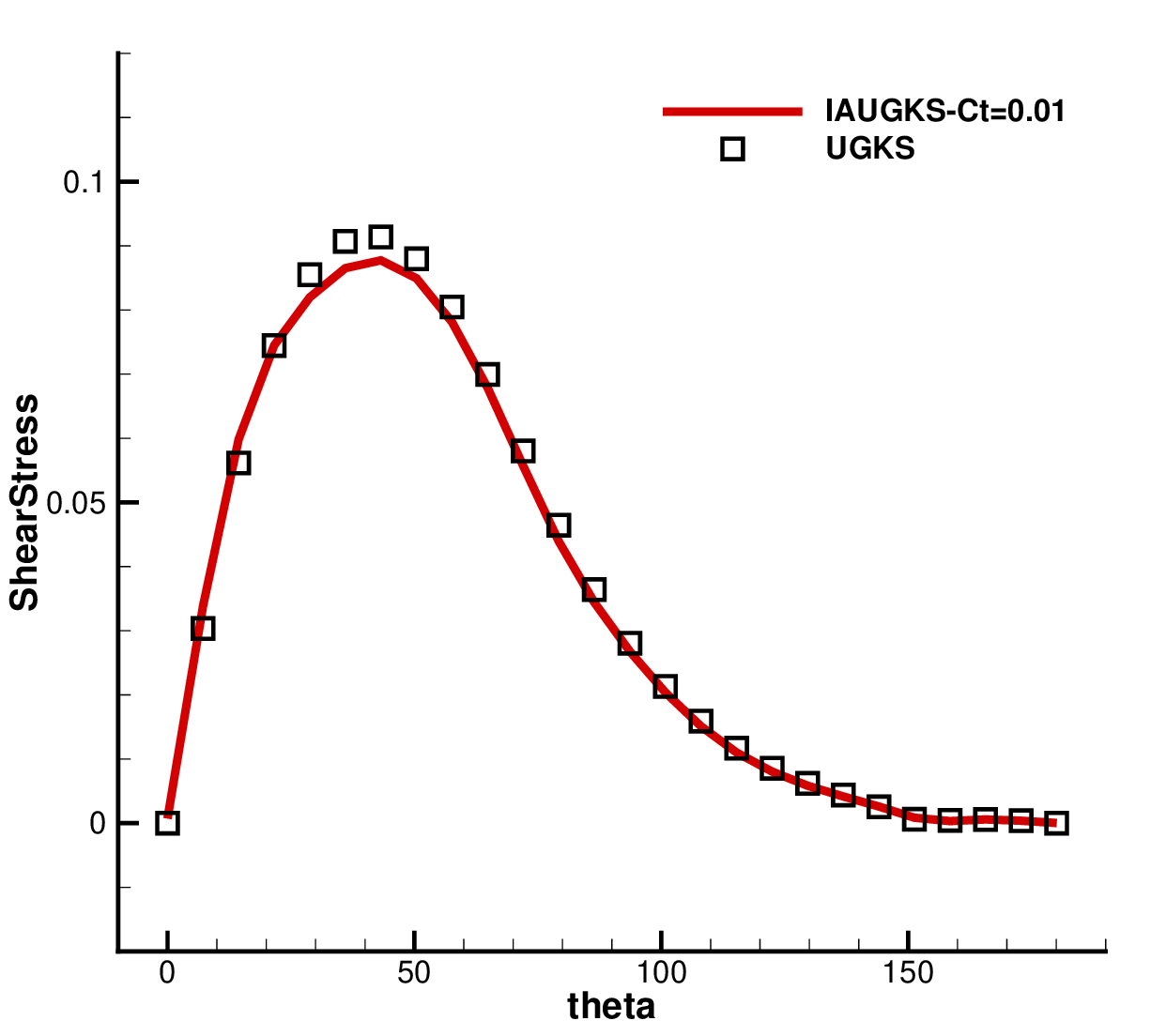}}
	\subfloat[]{\includegraphics[width=0.3\textwidth]
		{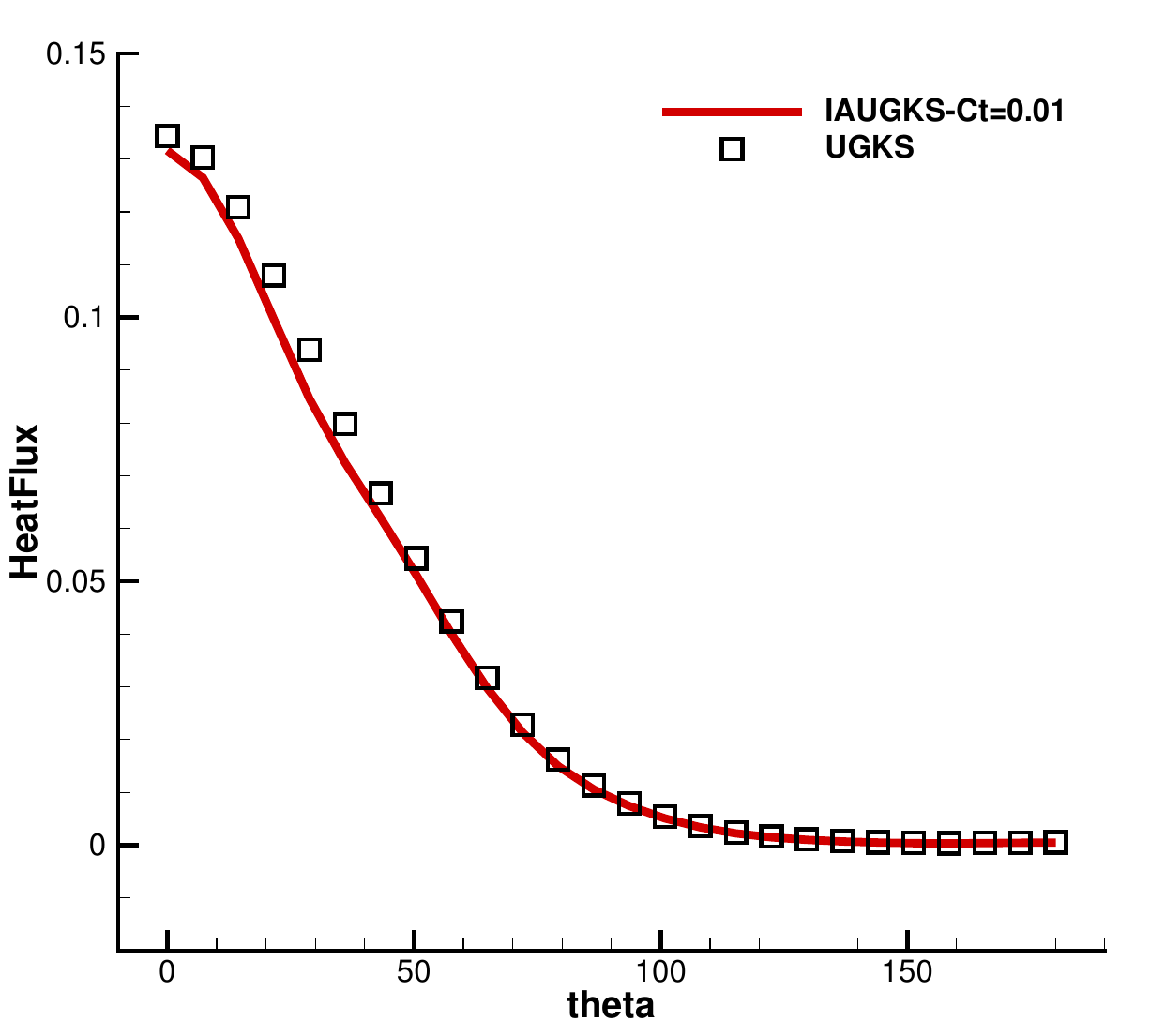}}\\
	\caption{Hypersonic flow at ${\rm Kn}_\infty = 0.01$ and ${\rm Ma} = 15$ passing over a circular cylinder by the IAUGKS. Surface quantities distributions: (a) pressure, (b) shear stress, and (c) heat flux.}
	\label{fig:cylinder-Ma15-surface}
\end{figure}

\begin{table}[H]
	\caption{The computational cost for simulations of hypersonic flow at ${\rm Kn}_\infty = 0.01$ and ${\rm Ma} = 15$ around a cylinder by the IAUGKS. The physical domain consists of 5,000 cells, and the unstructured DVS mesh is discretized by 3,420 cells.}
	\centering
	\begin{threeparttable}
		\begin{tabular}{ccccc}
			\hline
			Solver & \makecell[c]{Acceleration\\Method} & Steps & Simulation Time, h & Acceleration Rate  \\
			\hline			
			UGKS & Original UGKS & $2600\tnote{1}+12000$ & 340.1 & 1\\ \hline		
			UGKS & unstructured DVS & $2600\tnote{1}+10000$ & 95.47 & 3.6\\ \hline	
			UGKS & \makecell[c]{unstructured DVS\\implicit iteration} & $2600\tnote{1}+300$ & 1.462 & 232.5\\
			\hline	
			UGKS & \makecell[c]{unstructured DVS\\implicit iteration \\ Adaptation $C_t = 0.01$} & $2600\tnote{1}+400$ & 0.432 & 786.4\\
			\hline
		\end{tabular}
		
		\begin{tablenotes}
			\item[1] Step of first-order GKS simulations.
		\end{tablenotes}
	\end{threeparttable}
	\label{table:cylindertime2}
\end{table}
\subsection{Supersonic flow around a sphere}

\begin{figure}[H]
 	\centering
 	 	\subfloat[]{\includegraphics[width=0.4\textwidth]
 		{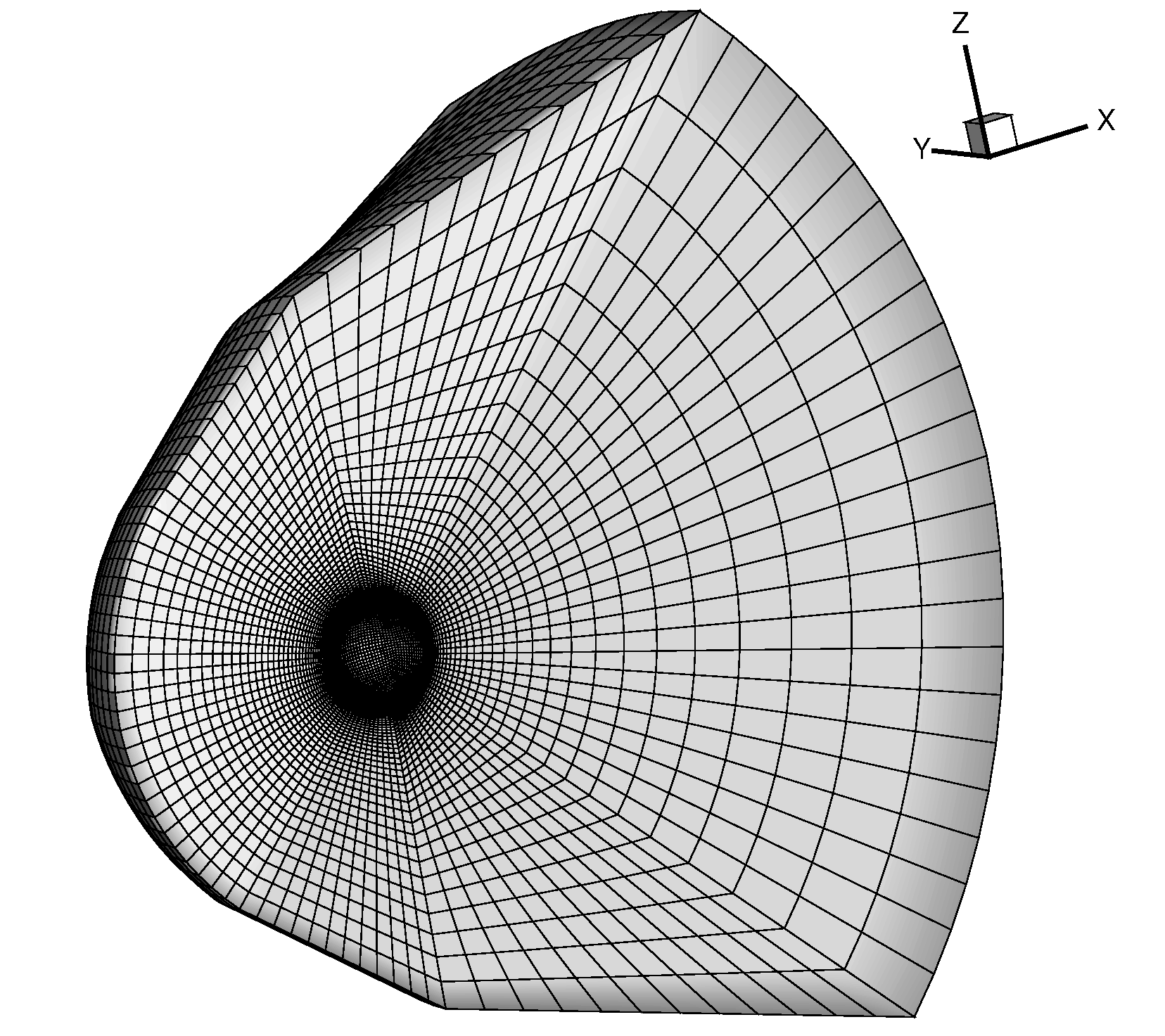}}
 	\subfloat[]{\includegraphics[width=0.4\textwidth]
 		{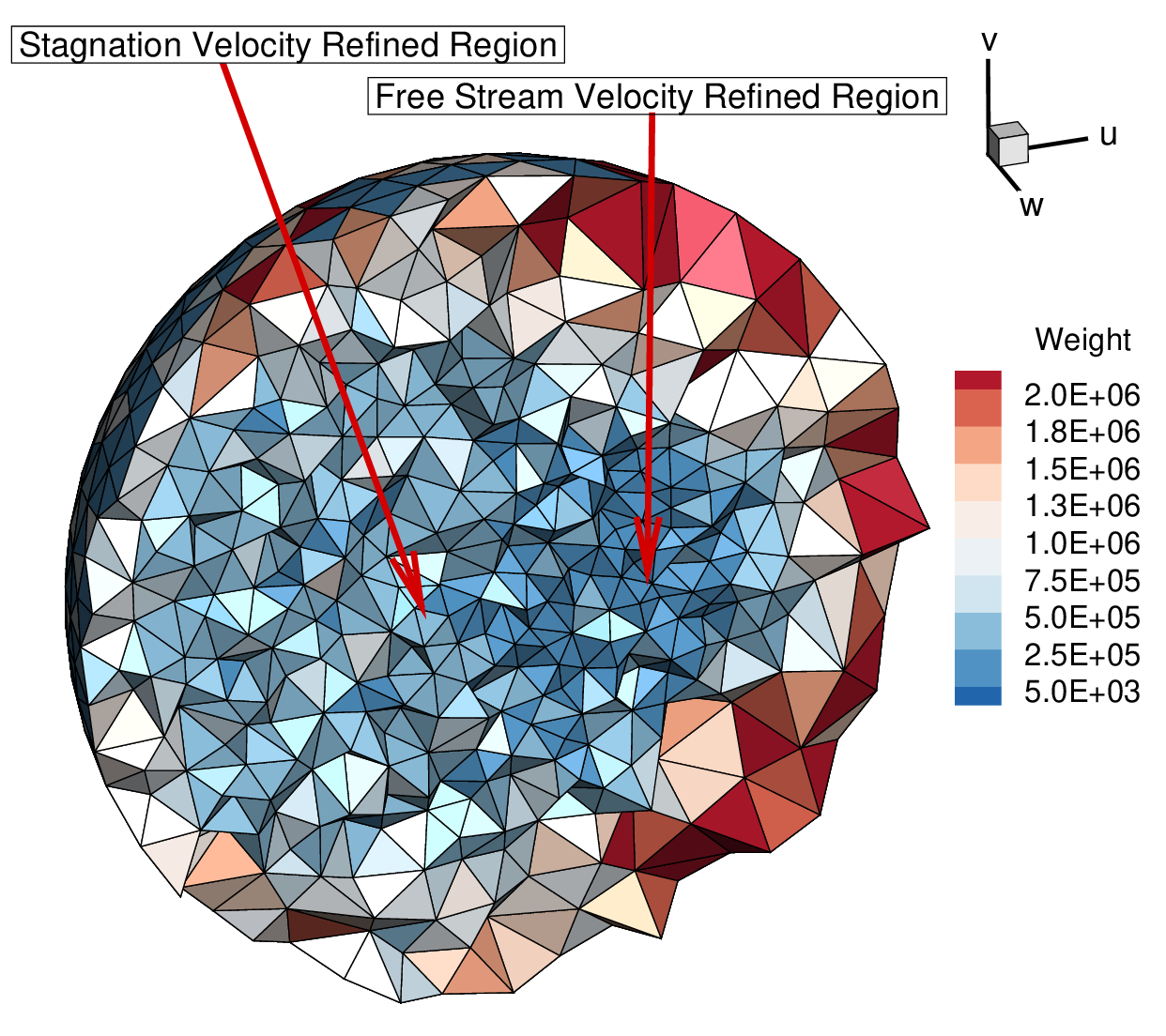}}
 	\caption{Supersonic flow at ${\rm Ma}_\infty = 4.25$ passing over sphere by the IAUGKS. (a) Physical mesh consisting of 138,240 cells, and (b) unstructured DVS mesh consisting of 18,802 cells.}
 	\label{fig:sphere-mesh}
\end{figure}

To further verify the effectiveness and efficiency of the IAUGKS in three-dimensional cases, the supersonic flow passing over a sphere at Mach number $4.25$ for ${\rm Kn}_\infty = 0.031$ and ${\rm Kn}_\infty = 0.0031$ are simulated for nitrogen gas. The characteristic length is chosen as the sphere diameter $D = 0.002$ m to define the Knudsen number. The physical domain contains $3,456 \times 40$ hexahedron cells, and the surface mesh is divided into 6 domains with $24 \times 24$ cells in each domain. Figure~\ref{fig:sphere-mesh} illustrates the section view of unstructured DVS mesh with 18,802 cells. The DVS is discretized into a sphere mesh with a radius of $6\sqrt{R T_s}$. The sphere center is located at $0.4\times(U_\infty,V_\infty,W_\infty)$, and the velocity space near the zero and free stream velocity point are refined within a spherical region of radius $r=3\sqrt{R T_w}$ and $r=3\sqrt{R T_\infty}$ respectively. Table ~\ref{table:spherecondition} gives the initial conditions of free stream flow, i.e., $\rho_\infty$ and $T_\infty$ for different ${\rm Kn}_\infty$, and the temperature of isothermal wall boundary condition $T_w$. 

\begin{table}[H]
	\caption{Free stream flow parameters of supersonic flow at ${\rm Ma}_\infty=4.25$ around a sphere.}
	\centering
	\begin{tabular}{ccccc}
		\hline
		 ${\rm Kn}_\infty$ & Altitude, km & $\rho_\infty$, kg/($\rm{m}^3 \cdot \rm{s}$) & $T_\infty$, K & $T_w$, K \\
		\hline
		0.031 & 63.2 & $6.879\times 10^{-4}$ & $65.04$ & $302$  \\
		0.0031& 41.5 & $6.879\times 10^{-3}$ & $65.04$ & $302$  \\
		\hline
	\end{tabular}
	\label{table:spherecondition}
\end{table}

\begin{figure}[H]
	\centering
	\subfloat[]{\includegraphics[width=0.3\textwidth]
		{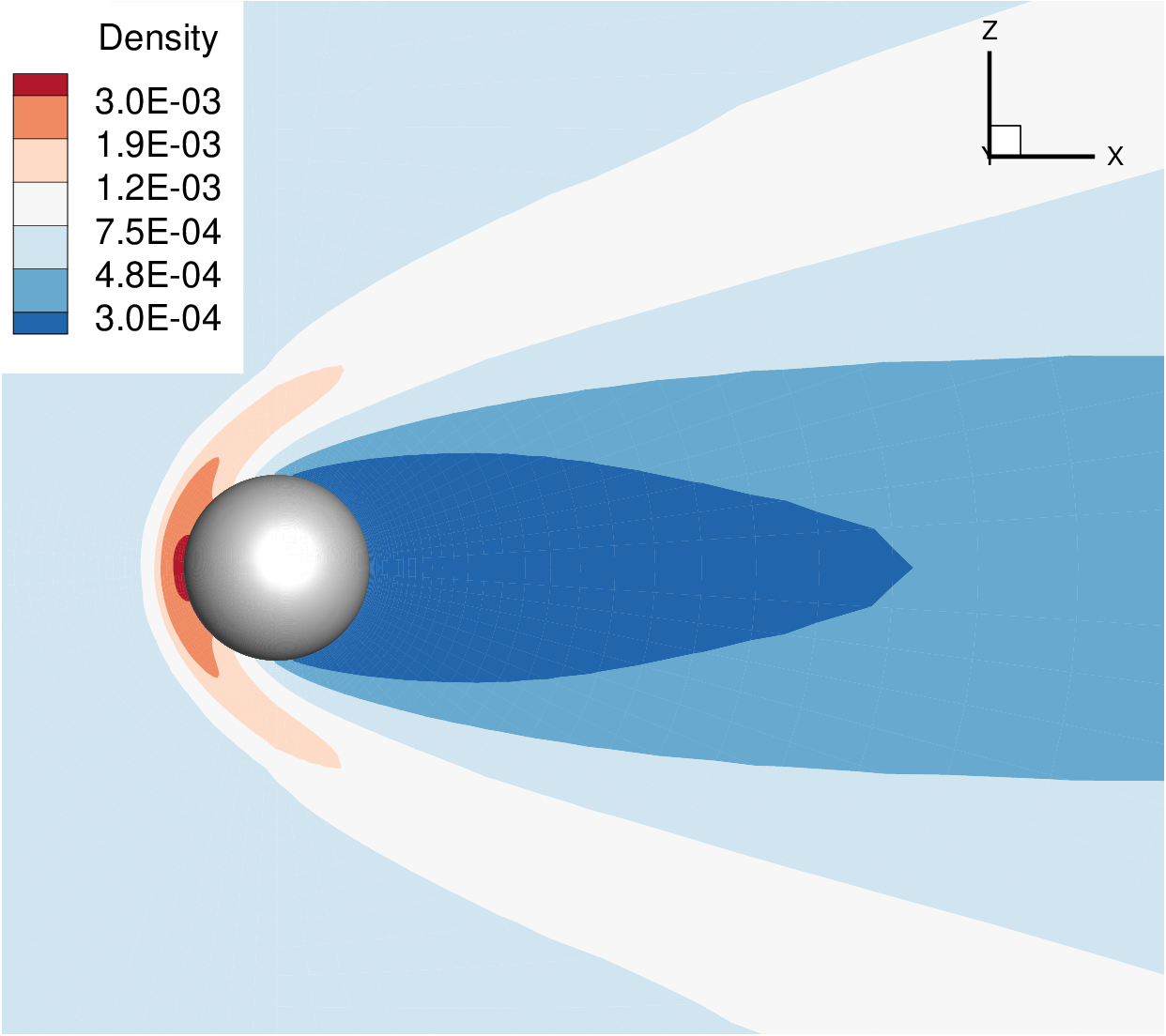}}~
	\subfloat[]{\includegraphics[width=0.3\textwidth]
		{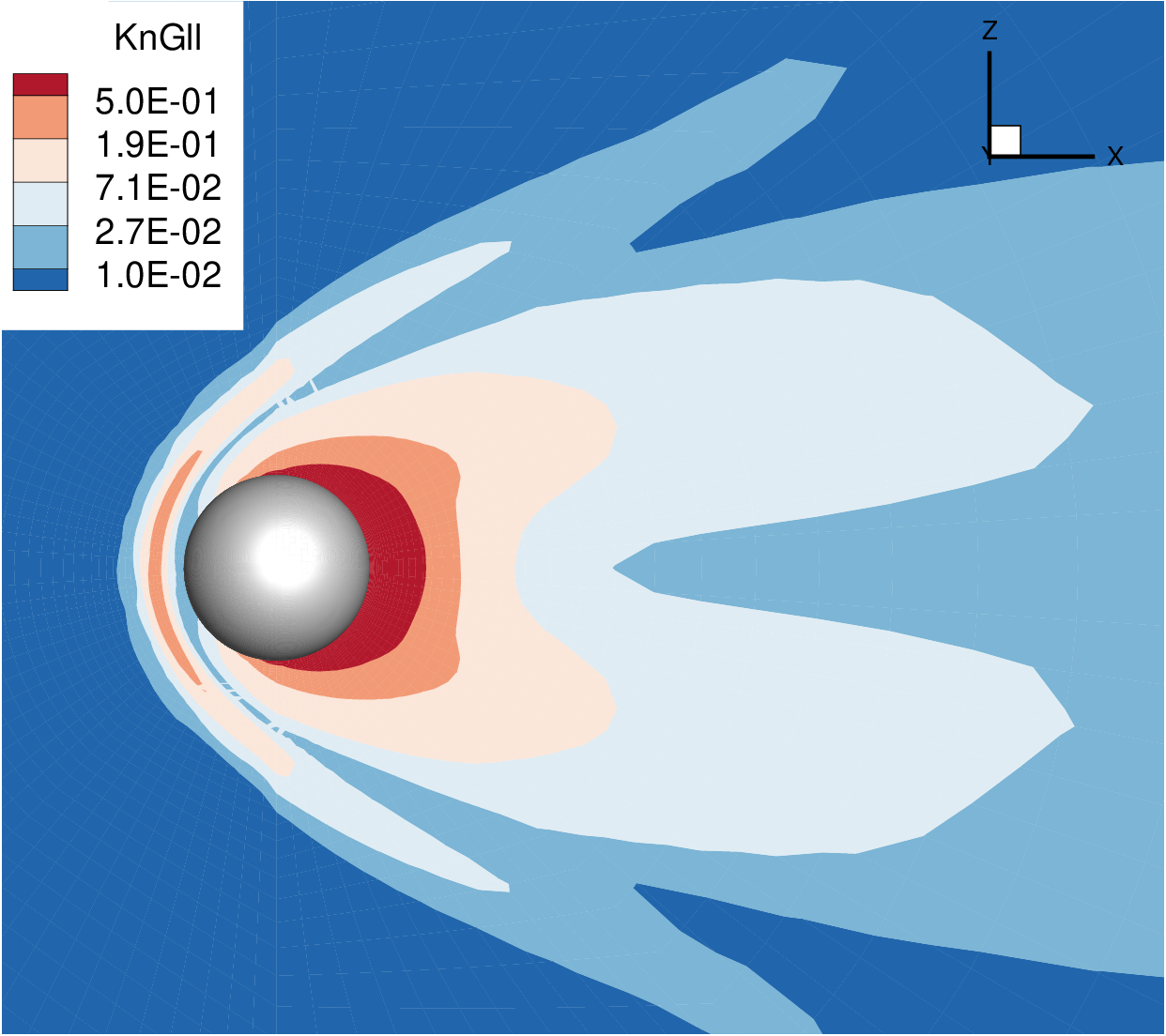}}\\
	\subfloat[]{\includegraphics[width=0.3\textwidth]
		{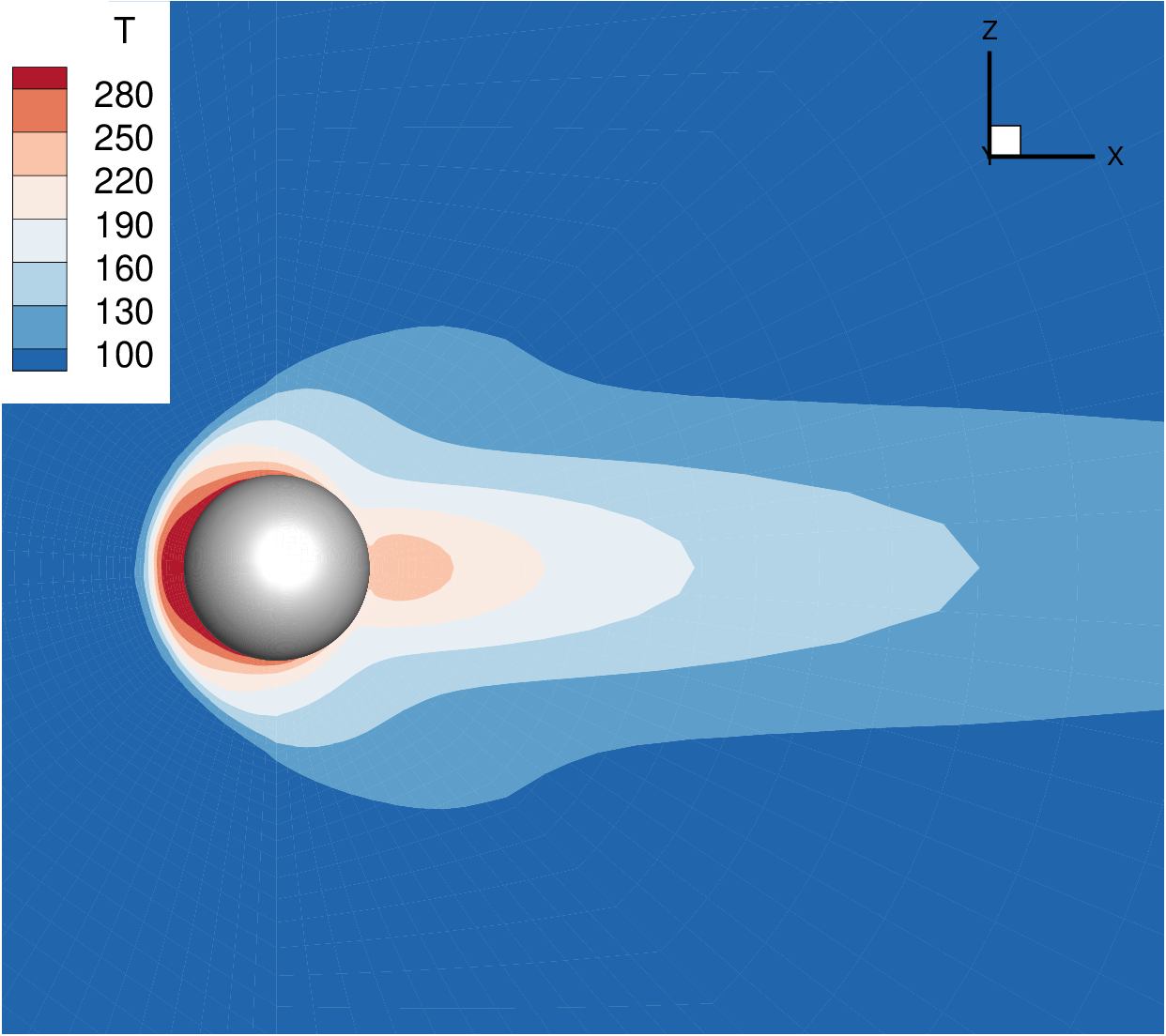}}~
	\subfloat[]{\includegraphics[width=0.3\textwidth]
		{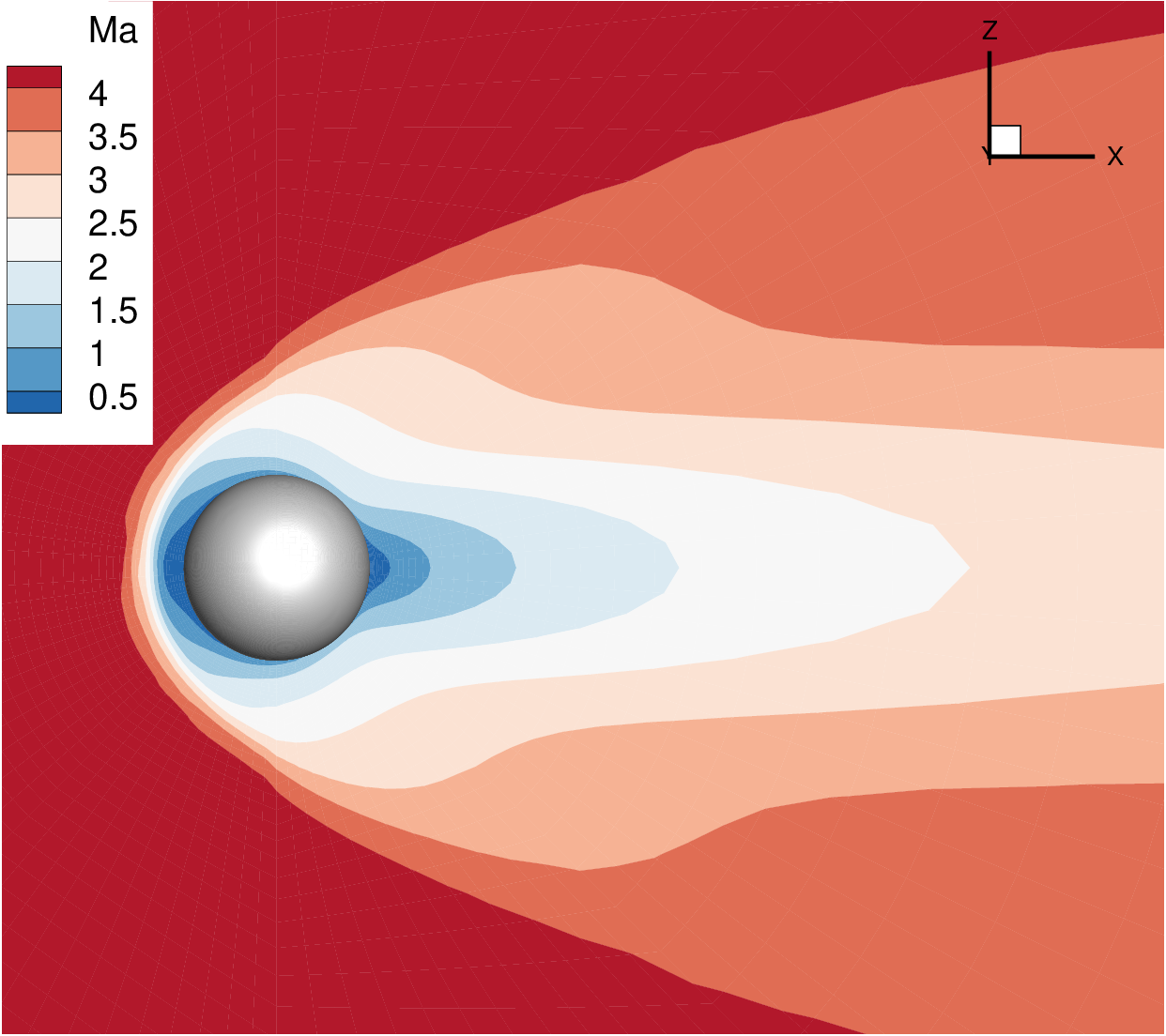}}\\
	\caption{Supersonic flow
		around a sphere at ${\rm Ma}_\infty = 4.25$ for ${\rm Kn}_\infty = 0.031$ by the IAUGKS. (a) Density, (b) $\rm{Kn}_{Gll}$,
		(c) temperature, and (d) Mach number contours.}
	\label{fig:sphere-Ma4.25-0.031}
\end{figure}

\begin{figure}[H]
	\centering
	\subfloat[]{\includegraphics[width=0.3\textwidth]
		{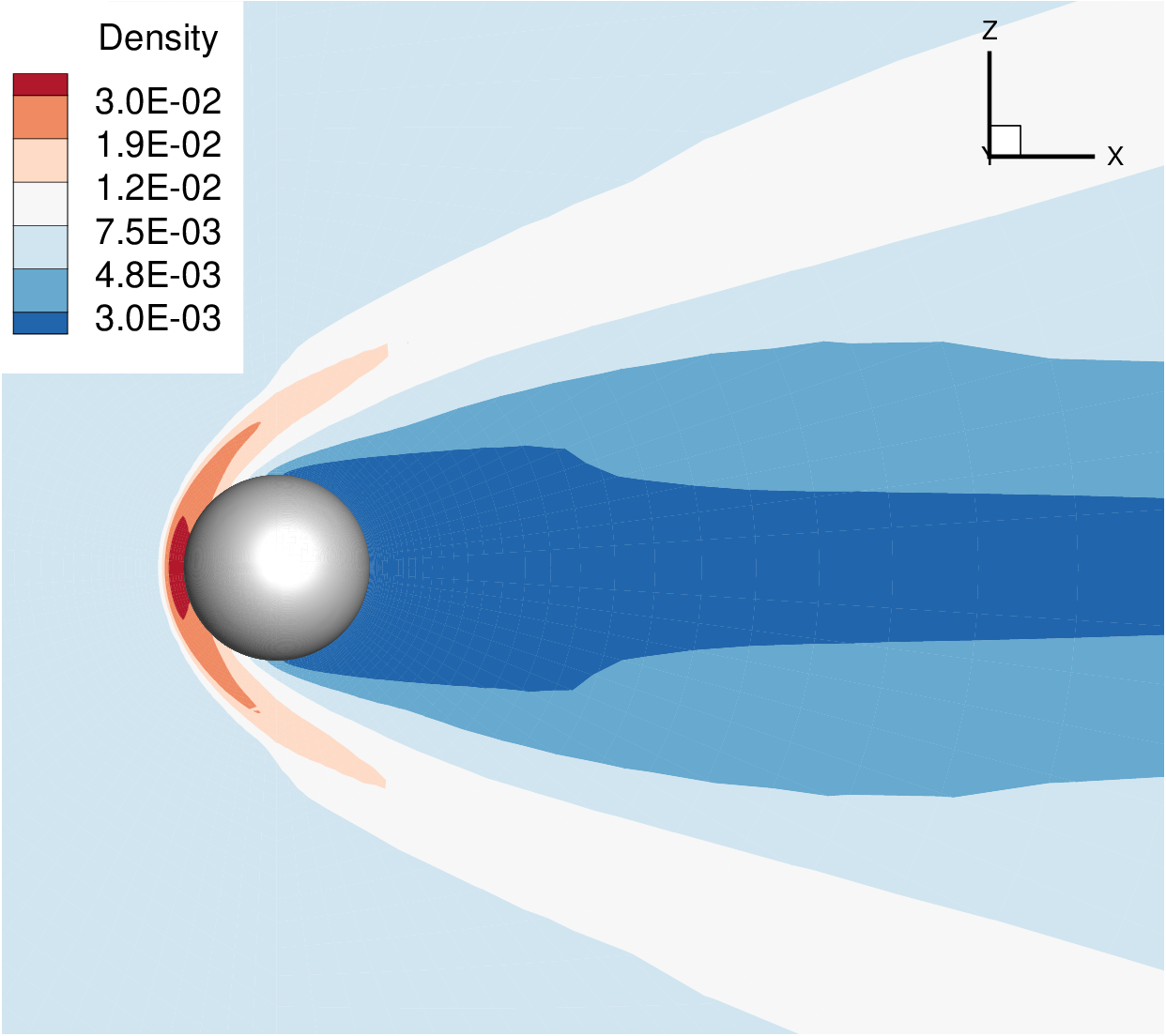}}~
	\subfloat[]{\includegraphics[width=0.3\textwidth]
		{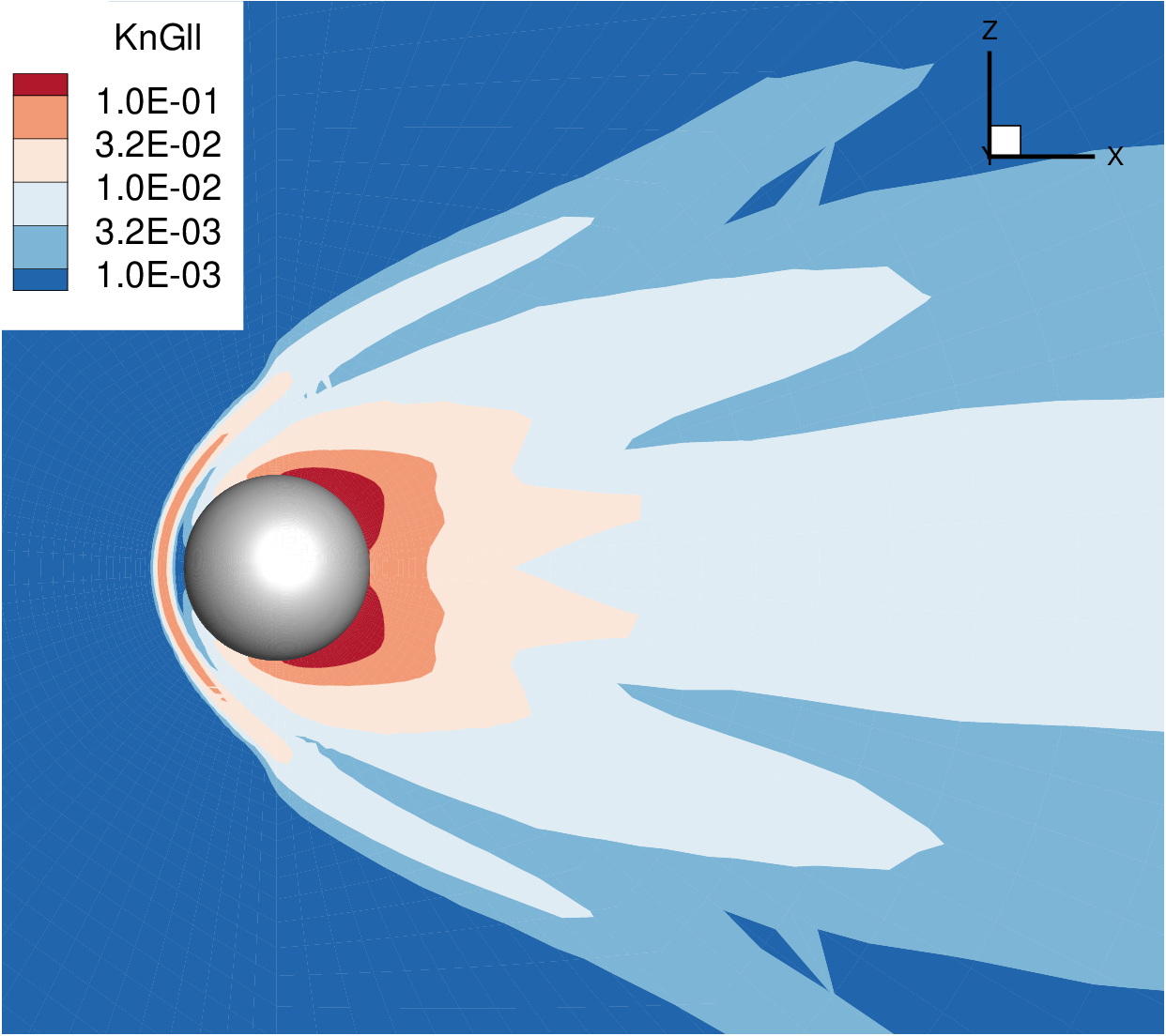}}\\
	\subfloat[]{\includegraphics[width=0.3\textwidth]
		{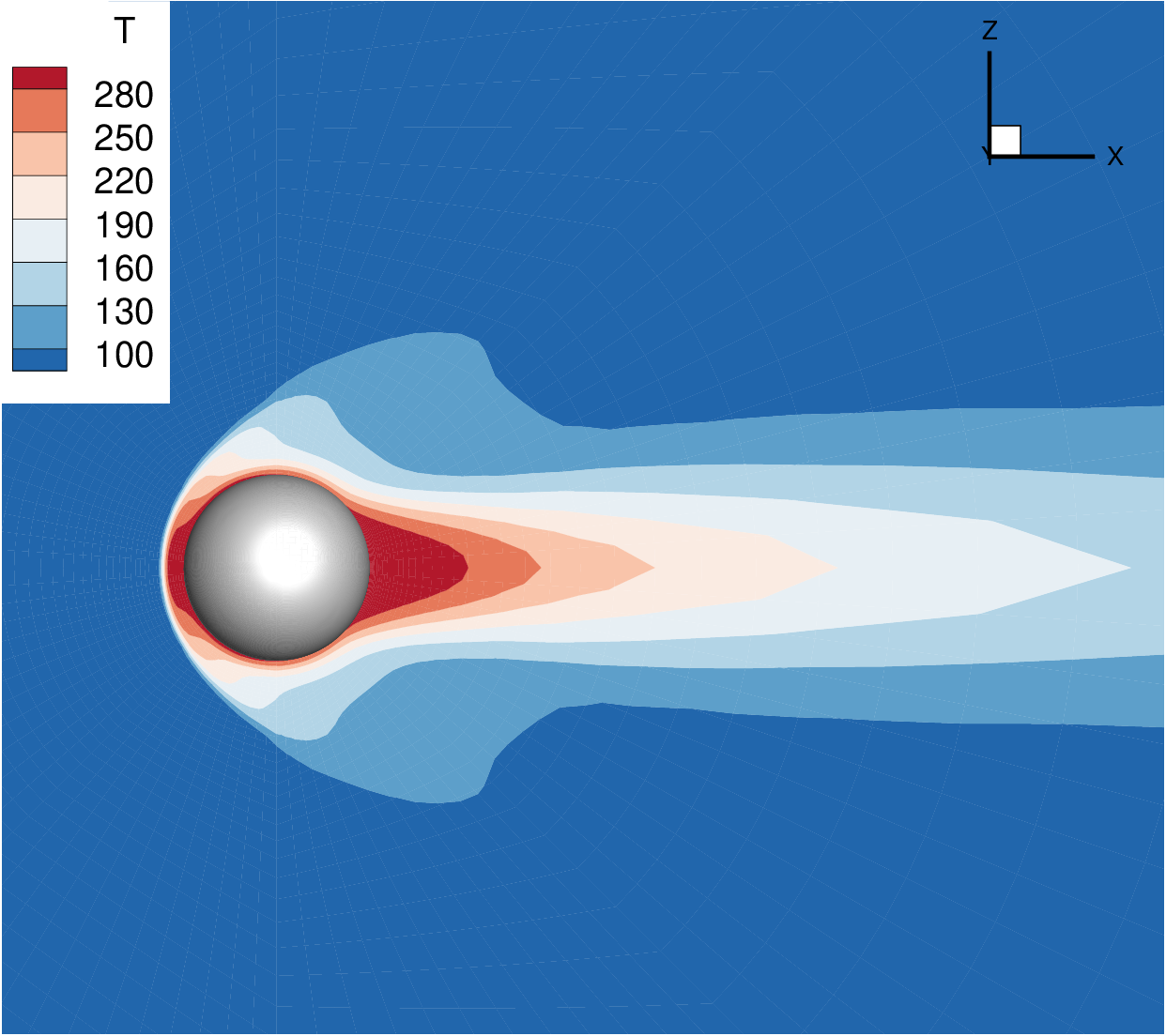}}~
	\subfloat[]{\includegraphics[width=0.3\textwidth]
		{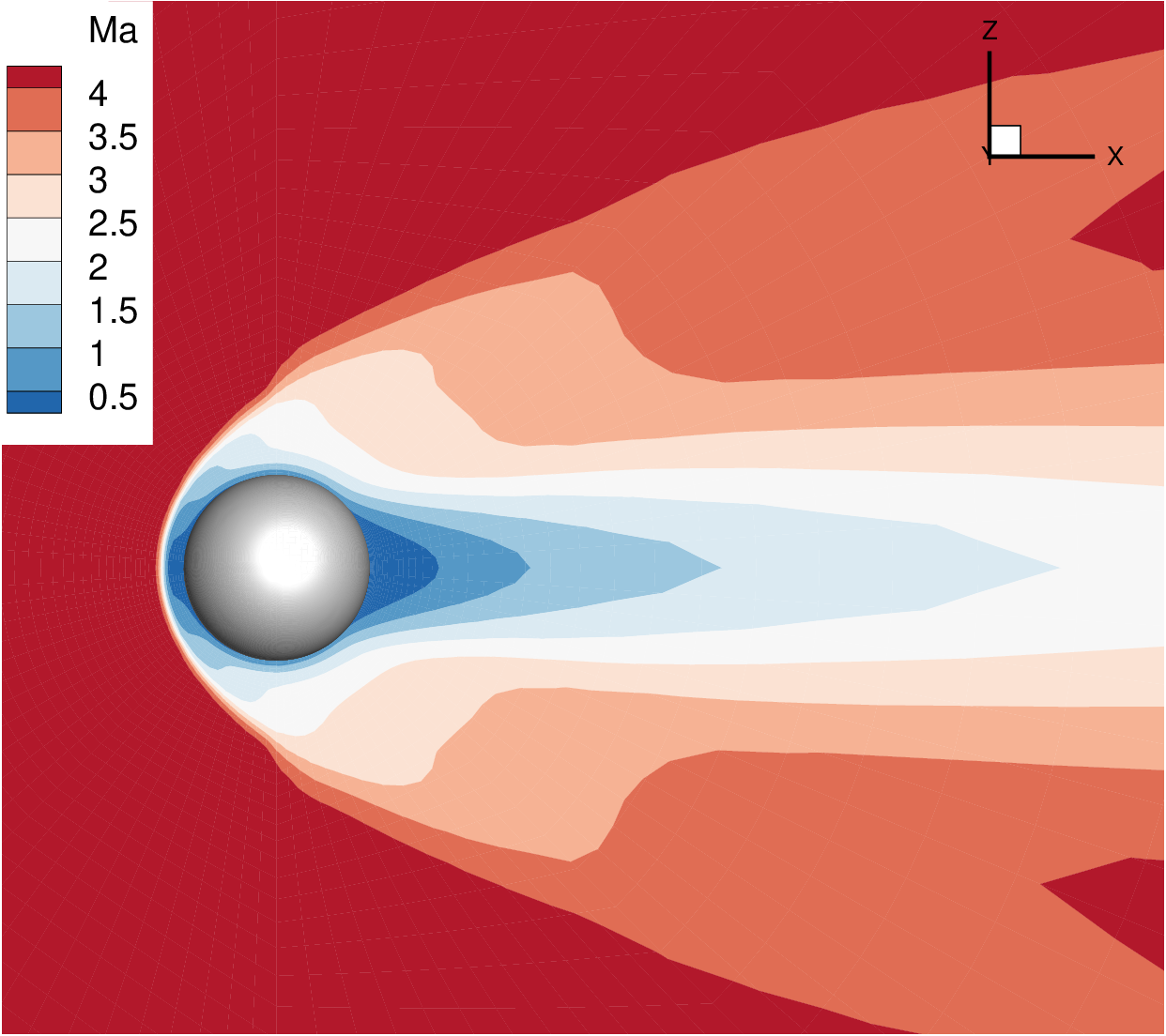}}\\
	\caption{Supersonic flow
		around a sphere at ${\rm Ma}_\infty = 4.25$ for ${\rm Kn}_\infty = 0.0031$ by the IAUGKS. (a) Density, (b) $\rm{Kn}_{Gll}$,
		(c) temperature, and (d) Mach number contours.}
	\label{fig:sphere-Ma4.25-0.0031}
\end{figure}

\begin{figure}[H]
	\centering
	\subfloat[]{\includegraphics[width=0.3\textwidth]
		{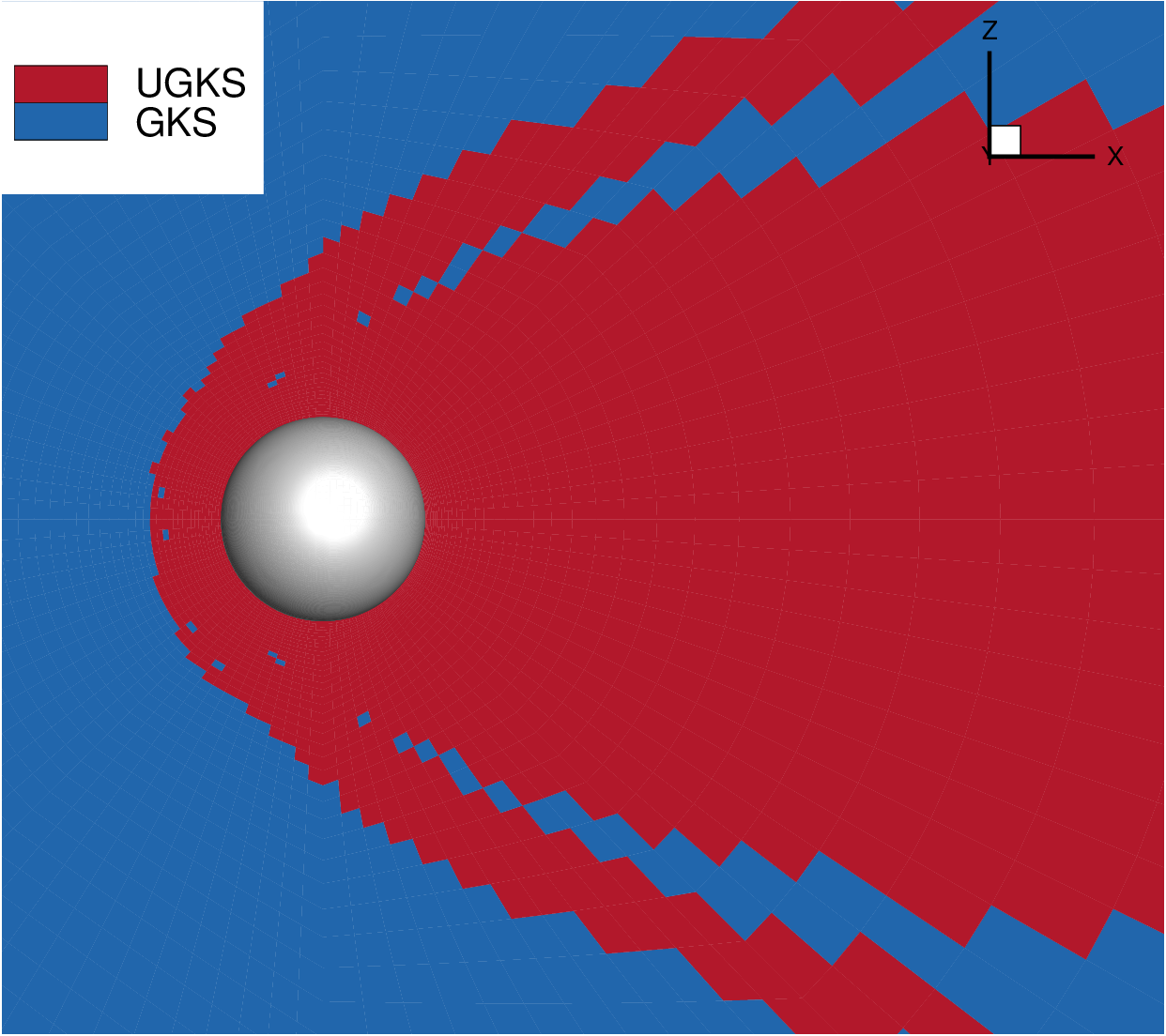}}~
	\subfloat[]{\includegraphics[width=0.3\textwidth]
		{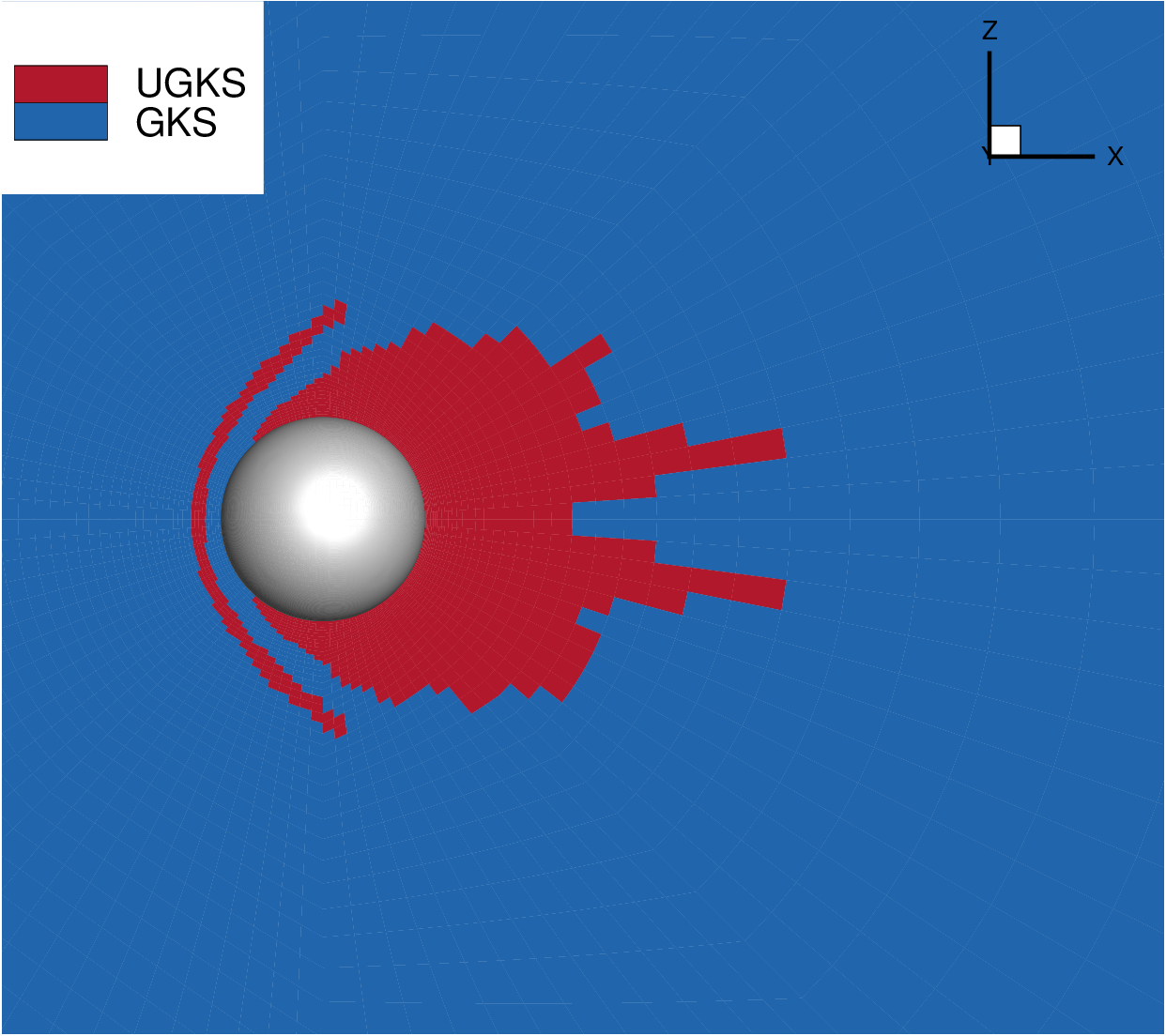}}
	\caption{Supersonic flow around a sphere at ${\rm Ma}_\infty = 4.25$ by the IAUGKS. Proportions of DVS with $C_t = 0.01$ for (a) ${\rm Kn}_\infty = 0.031$ is $63.87\%$, and for (b) ${\rm Kn}_\infty = 0.0031$ is $36.96\%$.}
	\label{fig:sphere-isDisc}
\end{figure}

Figures~\ref{fig:sphere-Ma4.25-0.031} and~\ref{fig:sphere-Ma4.25-0.0031} depict the contours of density, gradient length local Knudsen number, Mach number, and temperature by the IAUGKS for different ${\rm Kn}_\infty$. Figure~\ref{fig:sphere-isDisc} illustrates the domain decomposition with $C_t = 0.01$, where 63.87\% and 36.96\% region are covered by DVS for ${\rm Kn}_\infty = 0.031$ and ${\rm Kn}_\infty = 0.0031$ respectively.

The drag coefficients predicted by the IAUGKS are shown in Table~\ref{table:spherecd}. Compared with experiment data for Air~~\cite{wendtJF} and UGKS data for nitrogen~\cite{jiang2019implicit}, the IAUGKS gives accurate results with errors as small as 1.91\%.
To further verify the ability to accurately predict surface quantities, the pressure, shear stress, and heat flux coefficient distribution are compared with the explicit UGKS for ${\rm Kn}_\infty = 0.031$. The non-dimensionalized surface coefficients are given by
\begin{equation*}
	C_p     = \dfrac{p_s-p_\infty} {\frac12 \rho_\infty U_\infty^2},~
	C_\tau  = \dfrac{f_s}{\frac12 \rho_\infty U_\infty^2},~
	C_h     = \dfrac{h_s}{\frac12 \rho_\infty U_\infty^3},
\end{equation*}
where velocity $U_\infty$ can be calculated by free-stream Mach number ${\rm Ma}_\infty$, $p_s$ is the surface pressure, $p_{\infty}$ is the pressure in free stream flow, $f_s$ is the surface friction and $h_s$ is the surface heat flux. For $C_t = 0.01$, the simulation results of the IAUGKS fit well with the explicit UGKS; for $C_t = 0.05$, the heat flux coefficient at the windward region deviates from the reference. For three-dimensional flows, the validity of the criterion is consistent with that of the two-dimensional flows.

\begin{table}[H]
	\caption{Drag coefficients of supersonic flow around a sphere at ${\rm Ma_\infty} = 4.25$ by the IAUGKS.}
	\centering
	\begin{tabular}{ccccc}
		\toprule
		\multirow{2}{*}{${\rm Ma}_\infty$} & \multirow{2}{*}{${\rm Kn}_\infty$} & \multicolumn{3}{c}{Drag Coefficient (Error)} \\
		\cline{3-5} &  &
		\begin{tabular}[c]{@{}c@{}}Experiment (Air)\end{tabular} &
		\begin{tabular}[c]{@{}c@{}}UGKS      (${\rm N}_2$)\end{tabular} &
		\begin{tabular}[c]{@{}c@{}}IAUGKS (${\rm N}_2$)\end{tabular}
		\\ \midrule
		4.25 & 0.031 & 1.35 & 1.355 (0.39\%) & 1.376 (1.91\%)\\
		4.25 & 0.0031 & - &  {1.162} (-) & {1.166} (-)\\
		\bottomrule
	\end{tabular}
	\label{table:spherecd}
\end{table}

\begin{figure}[H]
	\centering
	\subfloat[]{\includegraphics[width=0.3\textwidth]
		{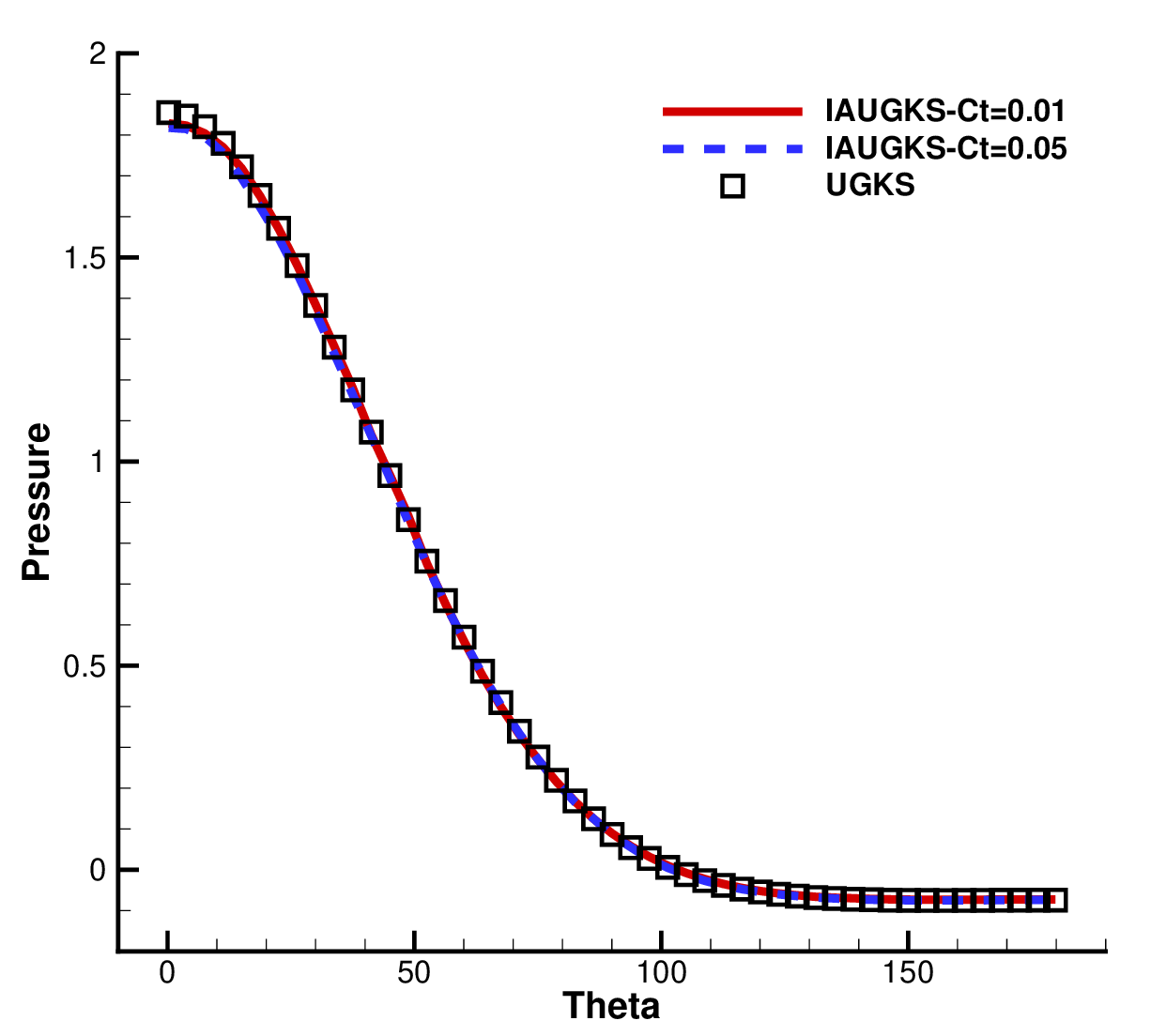}}~
	\subfloat[]{\includegraphics[width=0.3\textwidth]
		{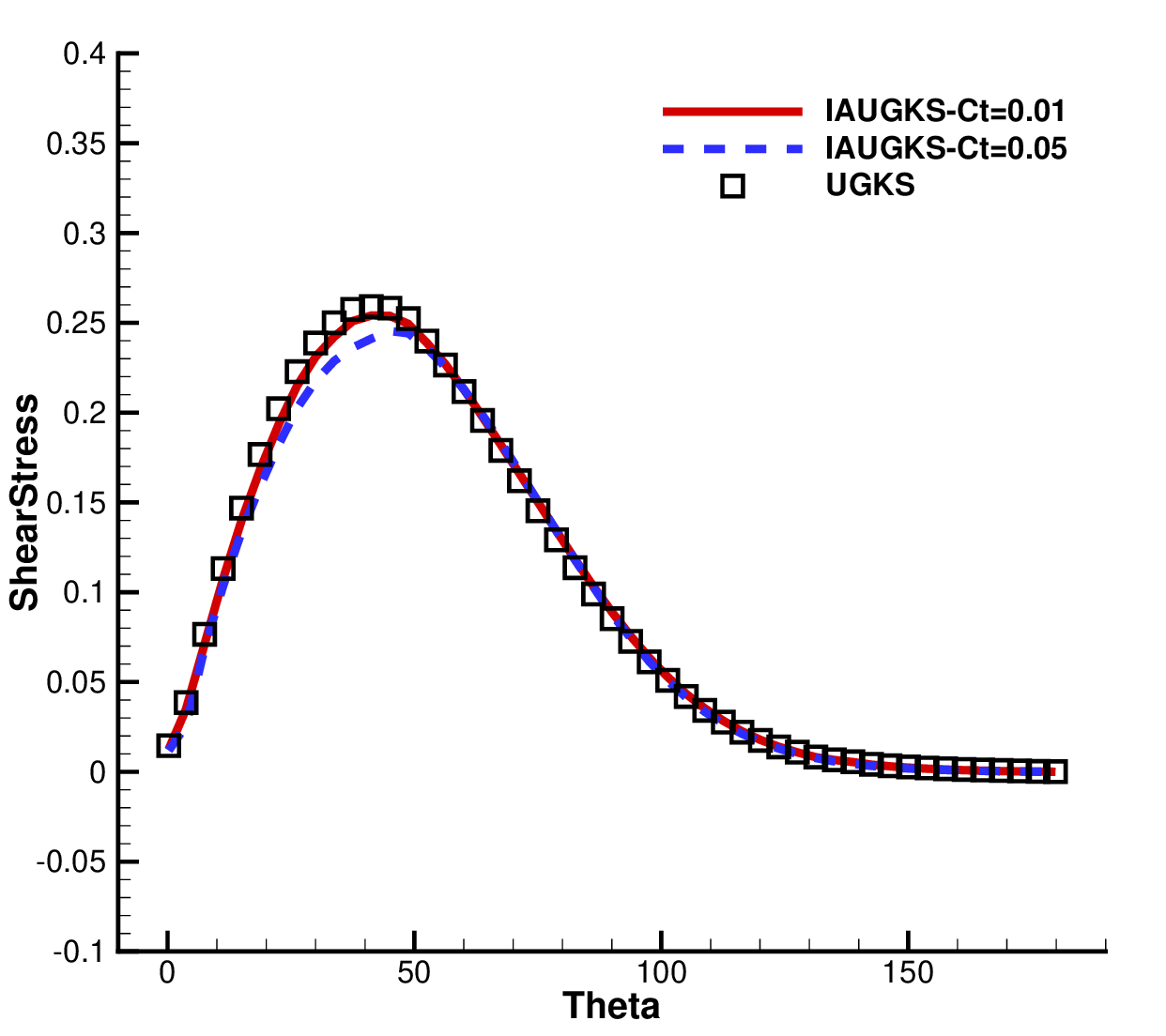}}~
	\subfloat[]{\includegraphics[width=0.3\textwidth]
		{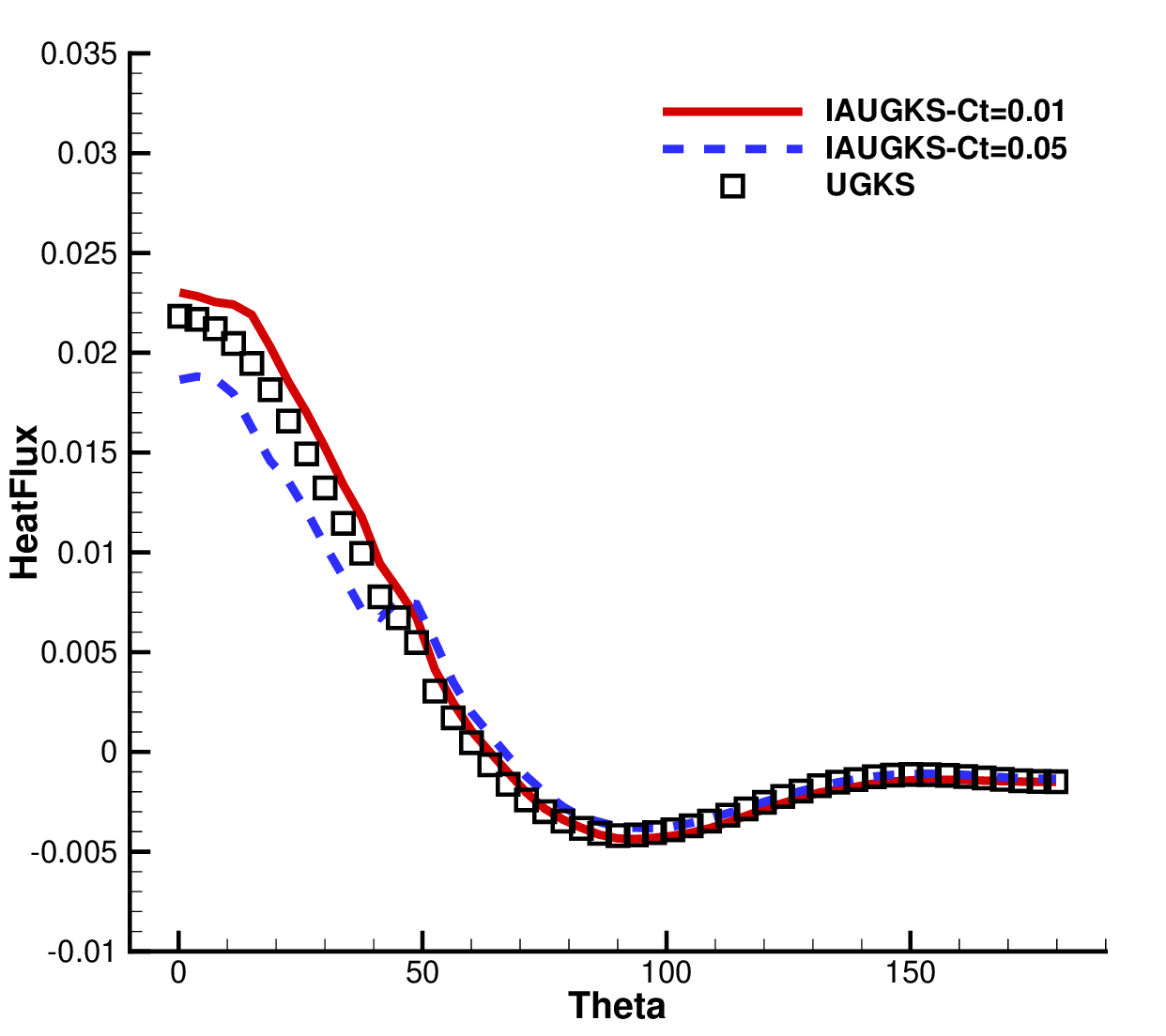}}\\
\caption{Surface quantities of supersonic flow
	around a sphere at ${\rm Ma}_\infty = 4.25$ for ${\rm Kn}_\infty = 0.031$ by the IAUGKS. (a) Pressure coefficient, (b) shear stress coefficient,
	and (c) heat flux coefficient.}
\label{fig:spheresurface}
\end{figure}

\begin{figure}[H]
	\centering
	\subfloat[]{\includegraphics[width=0.3\textwidth]
		{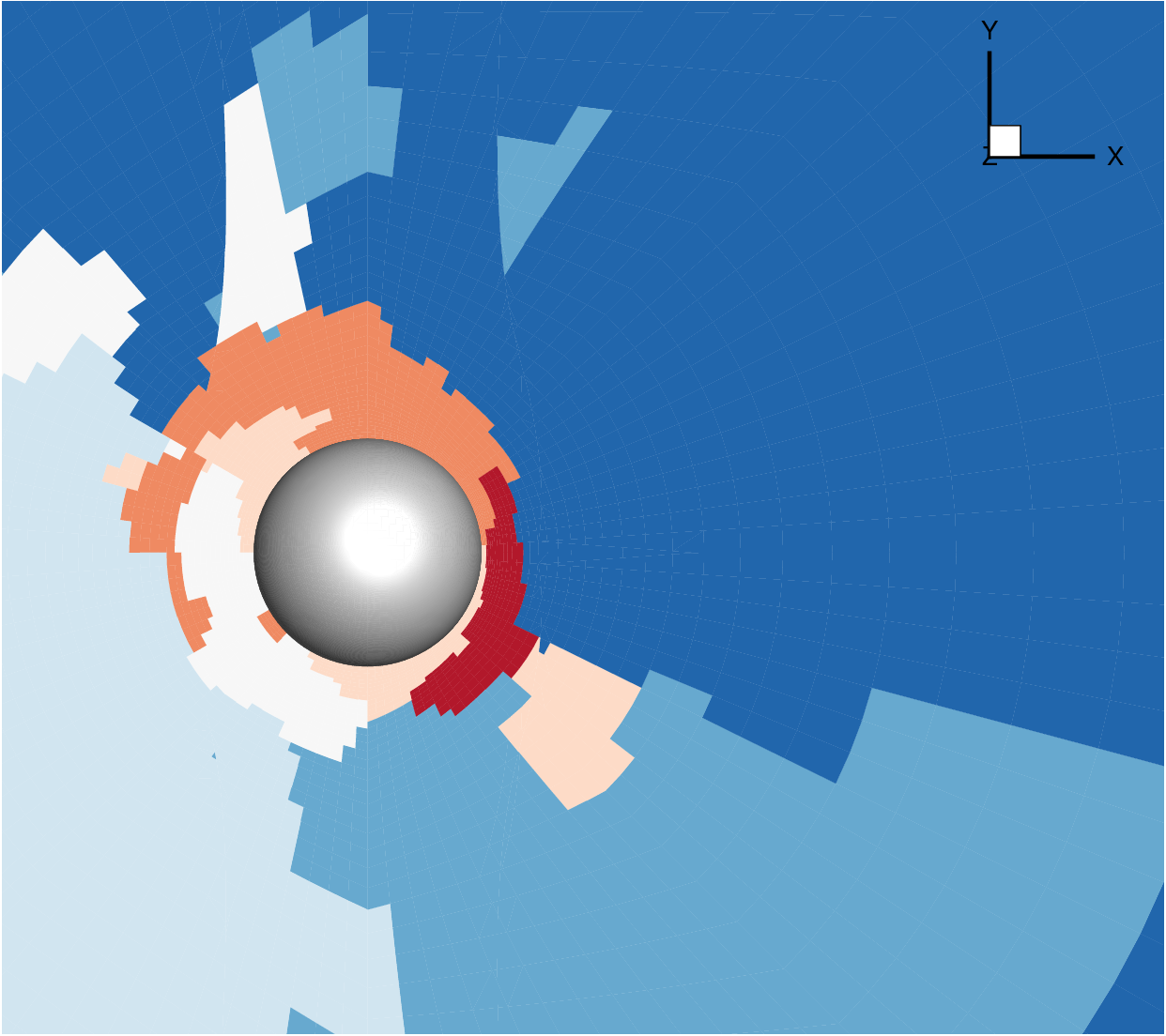}}~
	\subfloat[]{\includegraphics[width=0.3\textwidth]
		{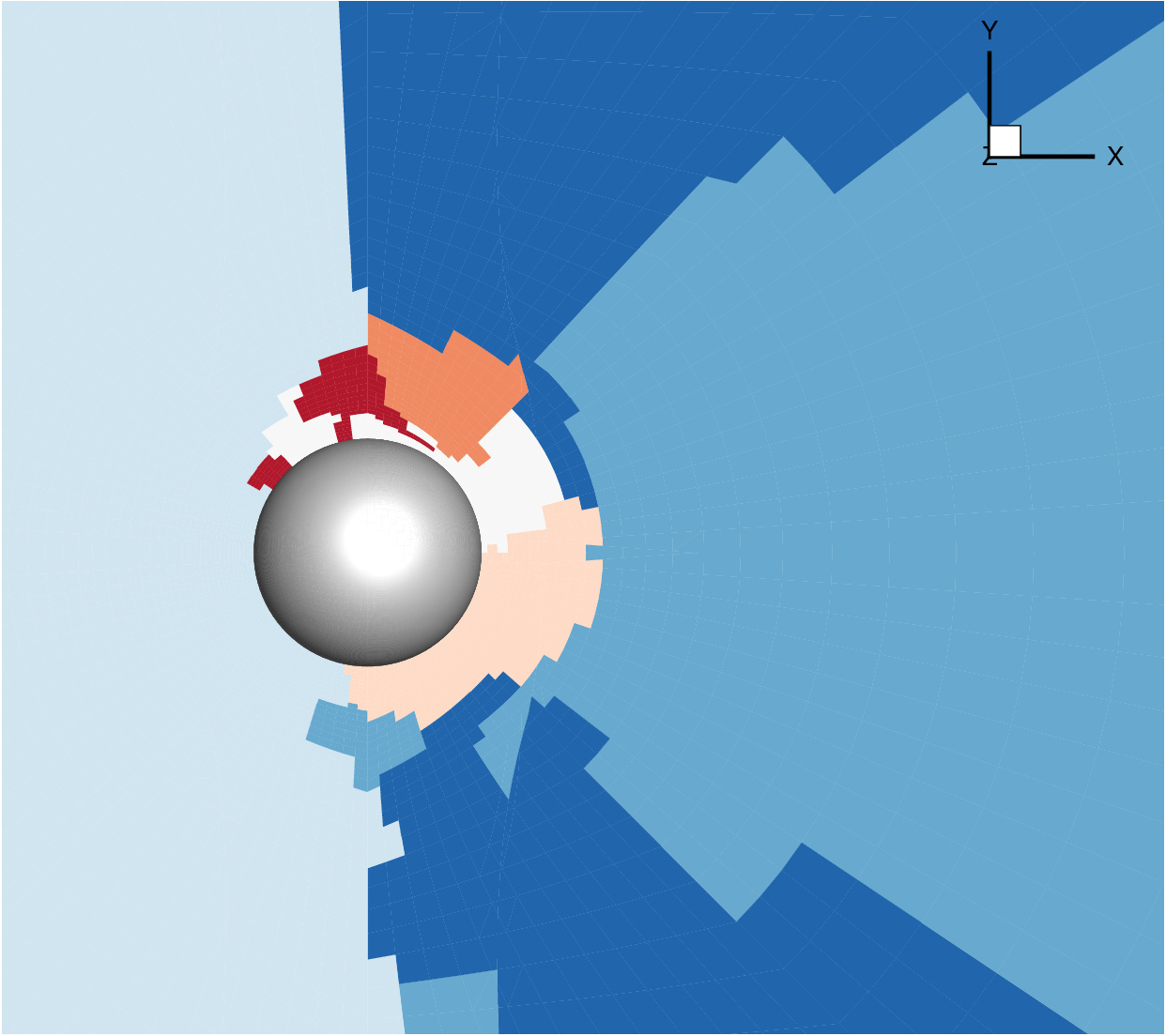}}
	\caption{Supersonic flow around a sphere at ${\rm Ma}_\infty = 4.25$ and ${\rm Kn}_\infty = 0.031$ by the IAUGKS. The partitioning of the computation domain (a) before the load balancing, and (b) after the load balancing.}
	\label{fig:sphere-partition}
\end{figure}

The physical CFL number is set to 0.4 and the numerical CFL number for implicit iteration is set to 400. Table~\ref{table:spheretime2} shows the computational efficiency and resource consumption for ${\rm Kn}_\infty = 0.031$ and ${\rm Kn}_\infty = 0.0031$. In three-dimensional cases, the implicit algorithm dramatically enhances the computational efficiency of the explicit AUGKS by nearly 26 times, achieving the same level of efficiency as the UGKWP method. A simple static load balancing method is adopted. Fig.~\ref{fig:sphere-partition} shows that after the load balancing process, regions using DVS in Fig.~\ref{fig:sphere-isDisc} (a) are decomposed into more partitions. All the simulations are conducted on the SUGON computation platform with a CPU model of 7285 32C 2.0GHz.

\begin{table}[H]
	\caption{The computational cost for simulations of the supersonic flow around a sphere at ${\rm Ma_\infty} = 4.25$ by the IAUGKS. The physical domain consists of 138,240 cells, and the unstructured DVS mesh is discretized by 18,802 cells. The criterion of the velocity space adaptation is $C_t = 0.01$ for the IAUGKS and the AUGKS simulations.}
	\centering
	\begin{threeparttable}
		\begin{tabular}{ccccc}
			\hline
			${\rm Kn}_\infty$ & Method & Cores & Steps & Wall Clock Time, h   \\
			\hline		
			$0.031$ & IAUGKS  & 640 & $3200\tnote{1}+90$ & 1.51 \\		
			$0.031$ & AUGKS & 1920 & $4000\tnote{1}+3500$ & 9.97 \\ 
			$0.031$ & UGKWP & 128 & $12000+13000\tnote{2}$ & 3.46 \\ 
			$0.0031$ & IAUGKS & 640 & $3400\tnote{1}+100$ & 0.416\\ 
			$0.0031$ & AUGKS & 1920 & $2500\tnote{1}+3100$ & 7.96 \\
			\hline
		\end{tabular}
		
		\begin{tablenotes}
			\item[1] Steps of first-order GKS simulations.
			\item[2] Average steps of UGKWP simulations.
		\end{tablenotes}
	\end{threeparttable}
	\label{table:spheretime2}
\end{table}

\subsection{Hypersonic flow around an X38-like space vehicle}
In this section, hypersonic flow at ${\rm Ma}_\infty = 8.0$ passing over an X38-like space vehicle for ${\rm Kn}_\infty = 0.00275$ at angles of attack of ${\rm AoA} = 20^\circ$ is simulated. The reference length to define the Knudsen number is $L_{ref} = 0.28$ m. In this case, due to the hypersonic free stream flow in the transition regime and the complex geometric shape, all flow regimes are involved, posing a great challenge to the numerical solver. The free stream temperature is $T_\infty = 56$ K, and an isothermal wall is applied to the vehicle surface with $T_w = 302$ K.

\begin{figure}[H]
	\centering
	\subfloat[]{\includegraphics[width=0.33\textwidth]
		{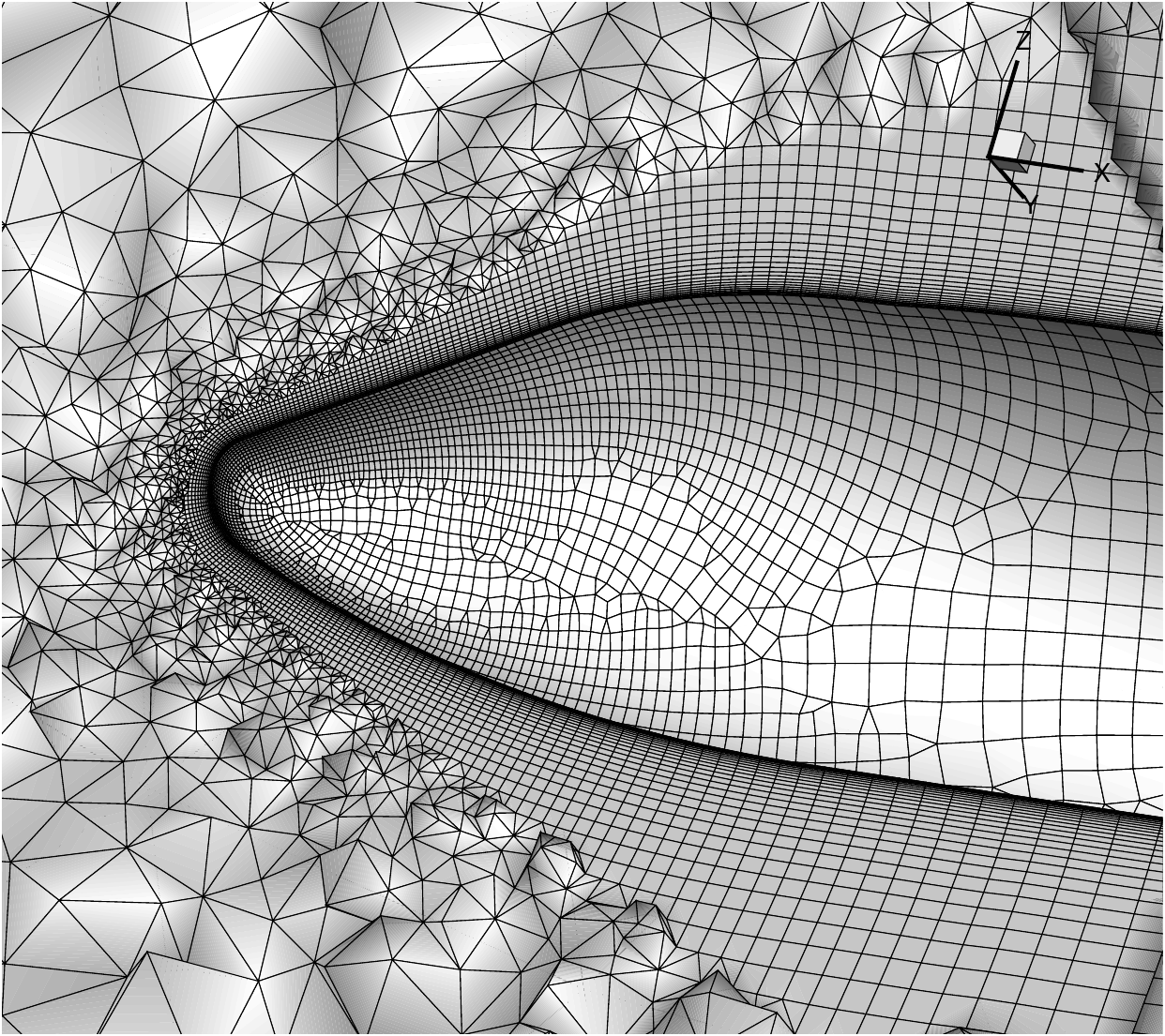}}
	\subfloat[]{\includegraphics[width=0.33\textwidth]
		{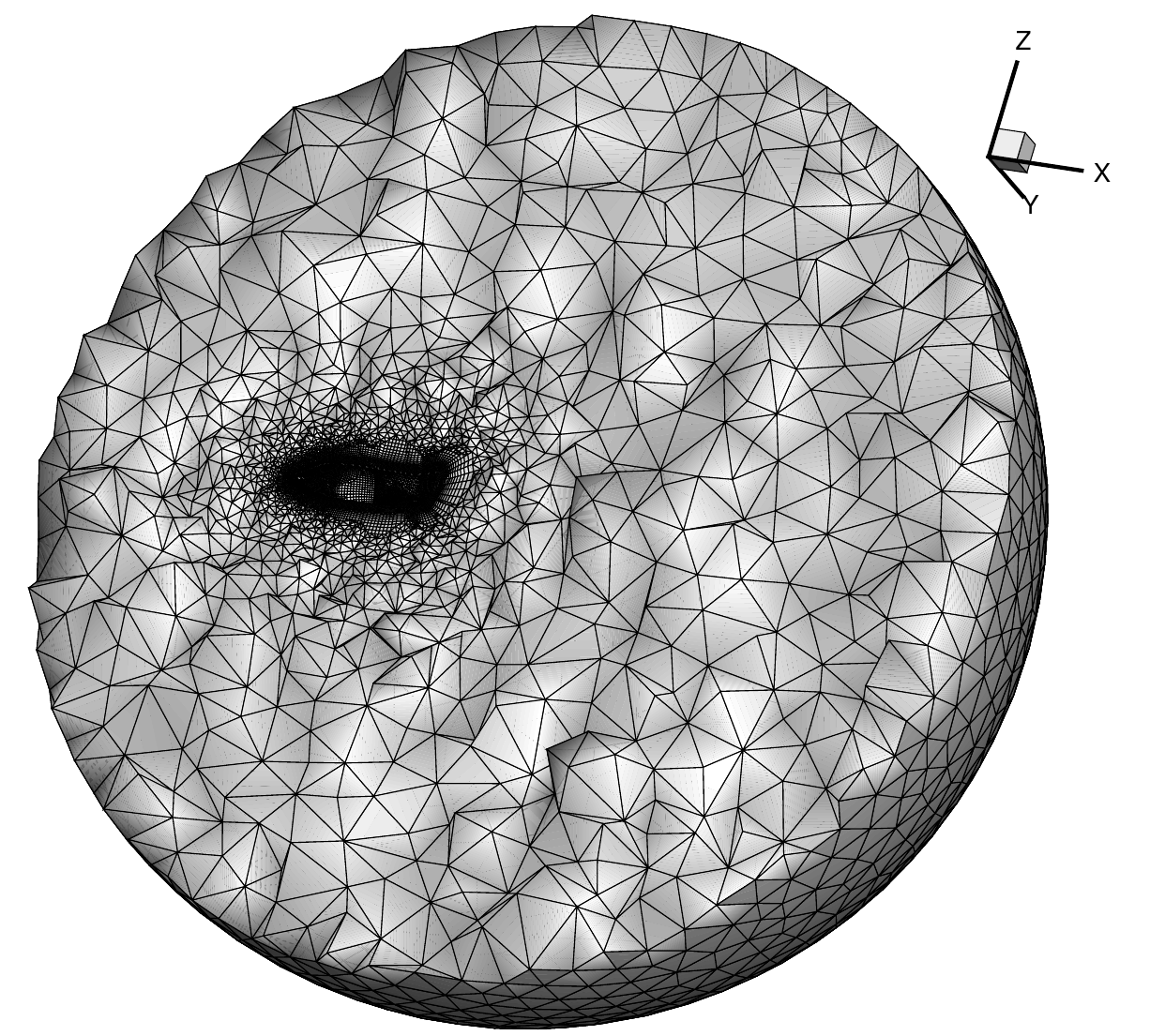}} \\	

	\caption{Section view of the physical mesh of the X38-like space vehicle with 600,078 cells: (a) close view of the vehicle nose, and (b) half of the domain. }
	\label{fig:x38-mesh}
\end{figure}

\begin{figure}[H]
	\centering
	\includegraphics[width=0.4\textwidth]
	{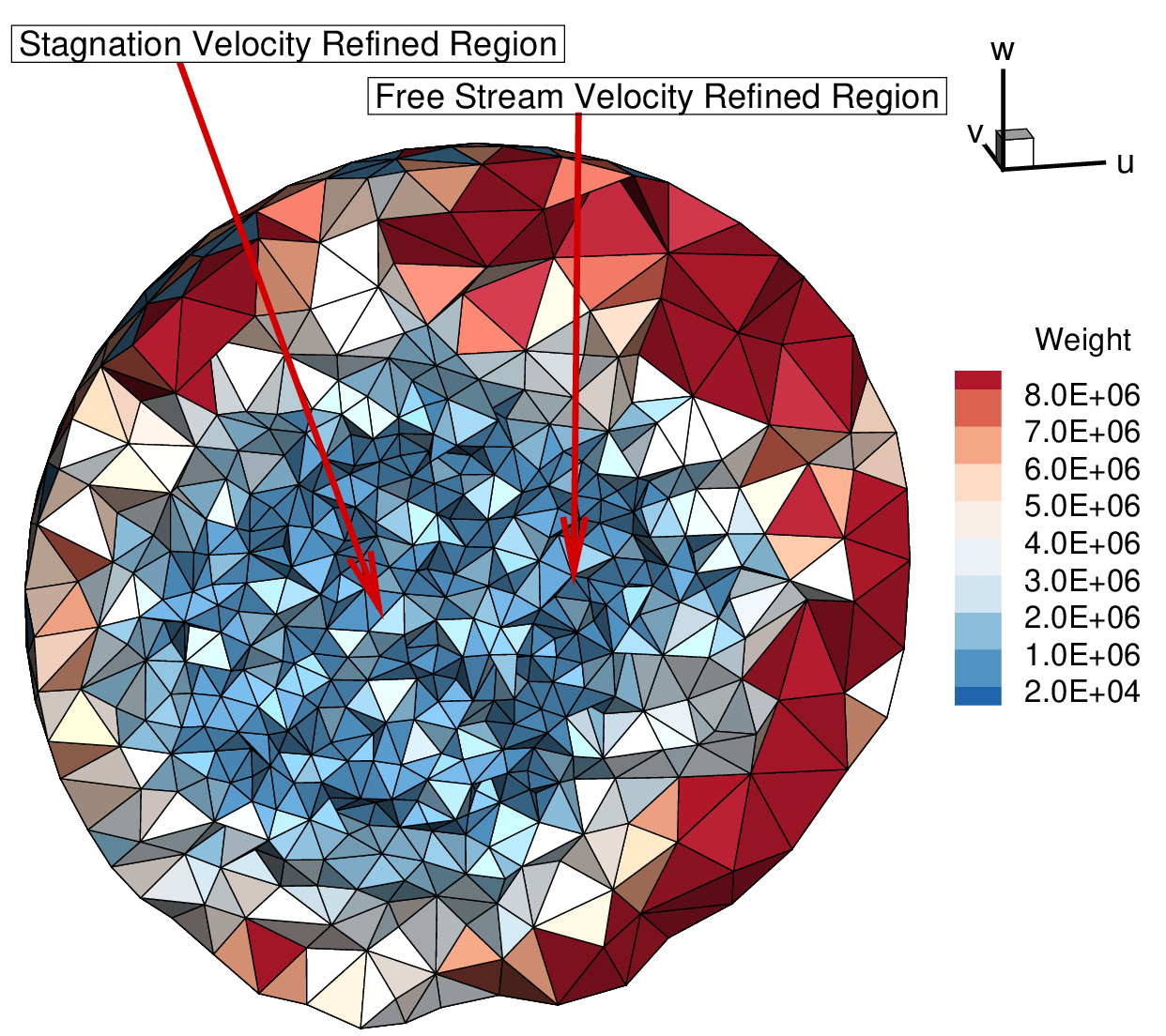}
	\caption{Unstructured DVS mesh with 23,520 cells used for hypersonic flow at ${\rm Ma}_\infty = 8.0$ and ${\rm Kn}_\infty = 0.00275$ passing over an X38-like space vehicle by the IAUGKS.}
	\label{fig:x38-DVS}
\end{figure}

The physical mesh consists of 600,078 all kinds of cells with the cell height of the first layer mesh $h=1.5\times 10^{-4}$ m, as shown in Fig.~\ref{fig:x38-mesh}. The unstructured DVS mesh is depicted in Fig.~\ref{fig:x38-DVS}. The DVS is discretized into 23,520 cells in a sphere mesh with a radius of $5\sqrt{R T_s}$. The sphere center is located at $0.4\times(U_\infty,V_\infty,W_\infty)$, the velocity space near the zero and free stream velocity point are refined within a spherical region of radius $r=5\sqrt{R T_w}$ and $r=5\sqrt{R T_\infty}$ respectively. 

\begin{figure}[H]
	\centering
	\subfloat[]{\includegraphics[width=0.3\textwidth]
		{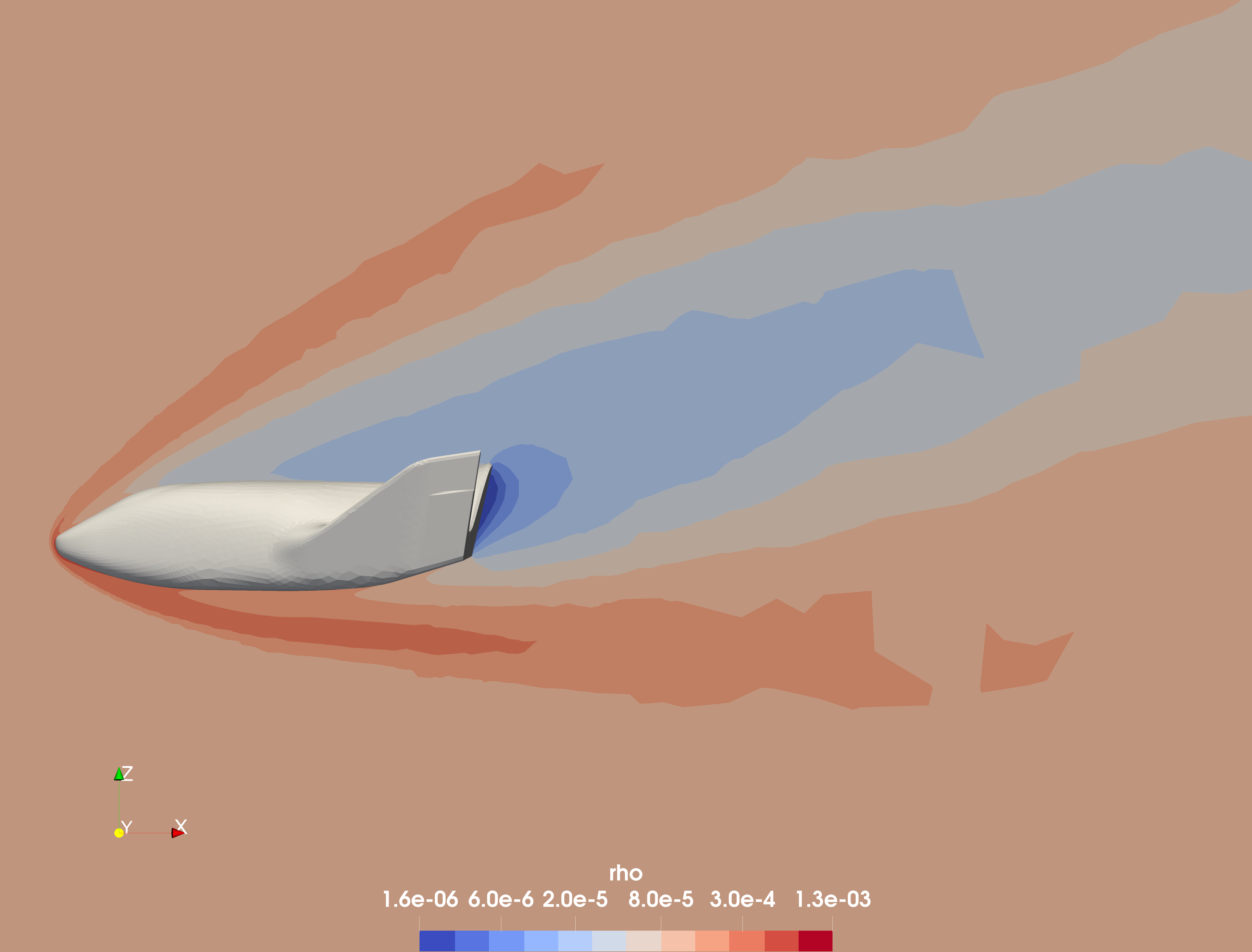}}~
	\subfloat[]{\includegraphics[width=0.3\textwidth]
		{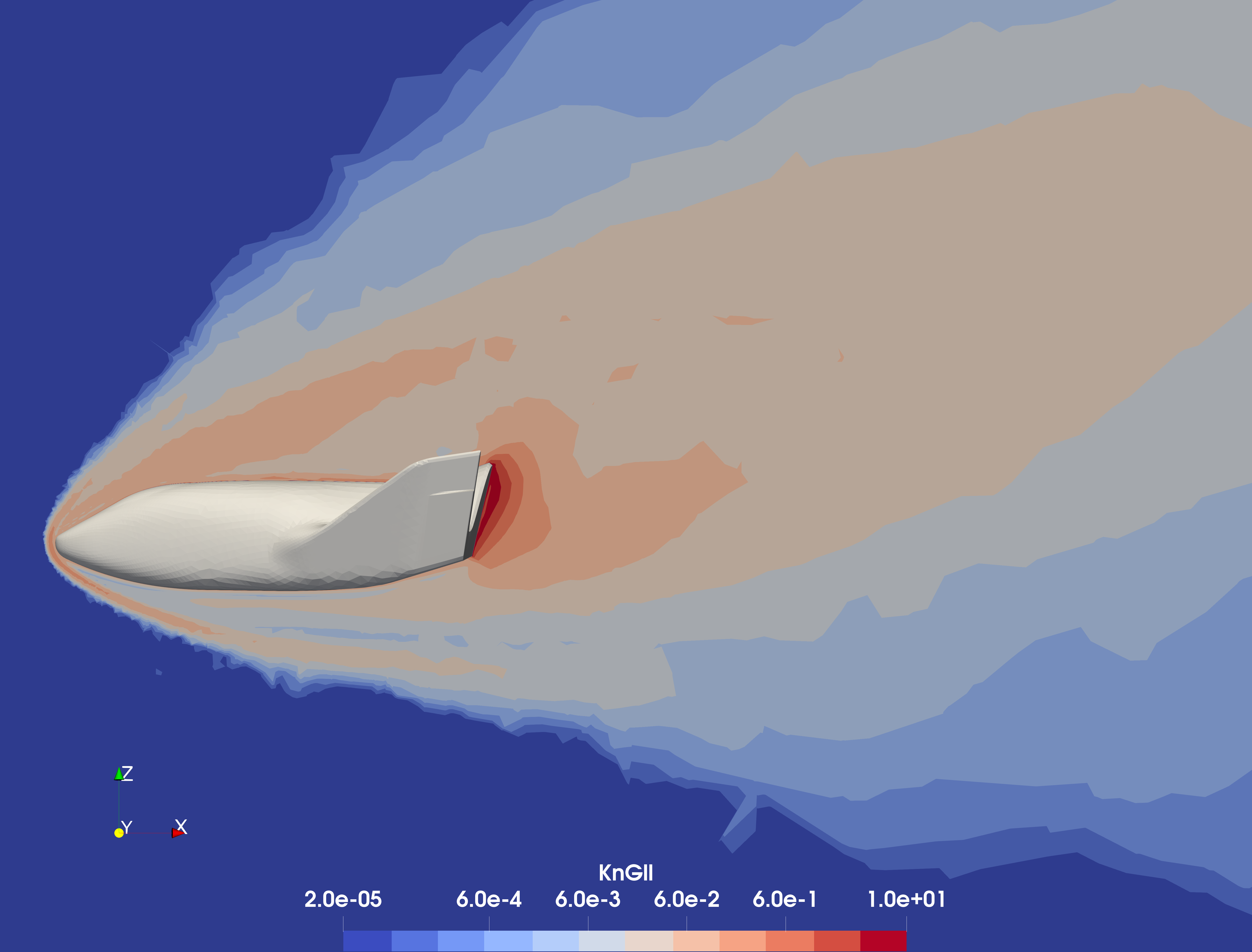}}\\
	\subfloat[]{\includegraphics[width=0.3\textwidth]
		{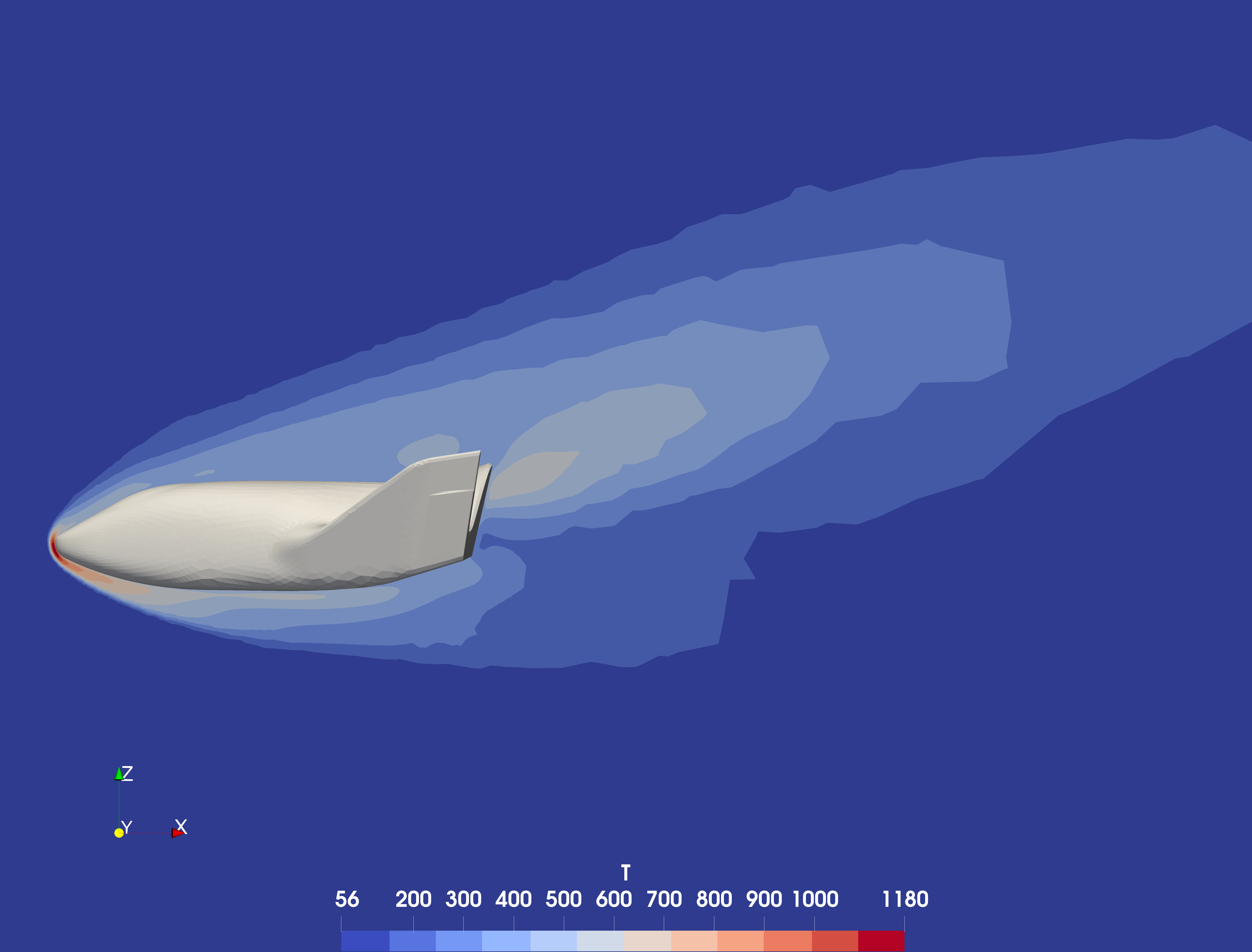}}~
	\subfloat[]{\includegraphics[width=0.3\textwidth]
		{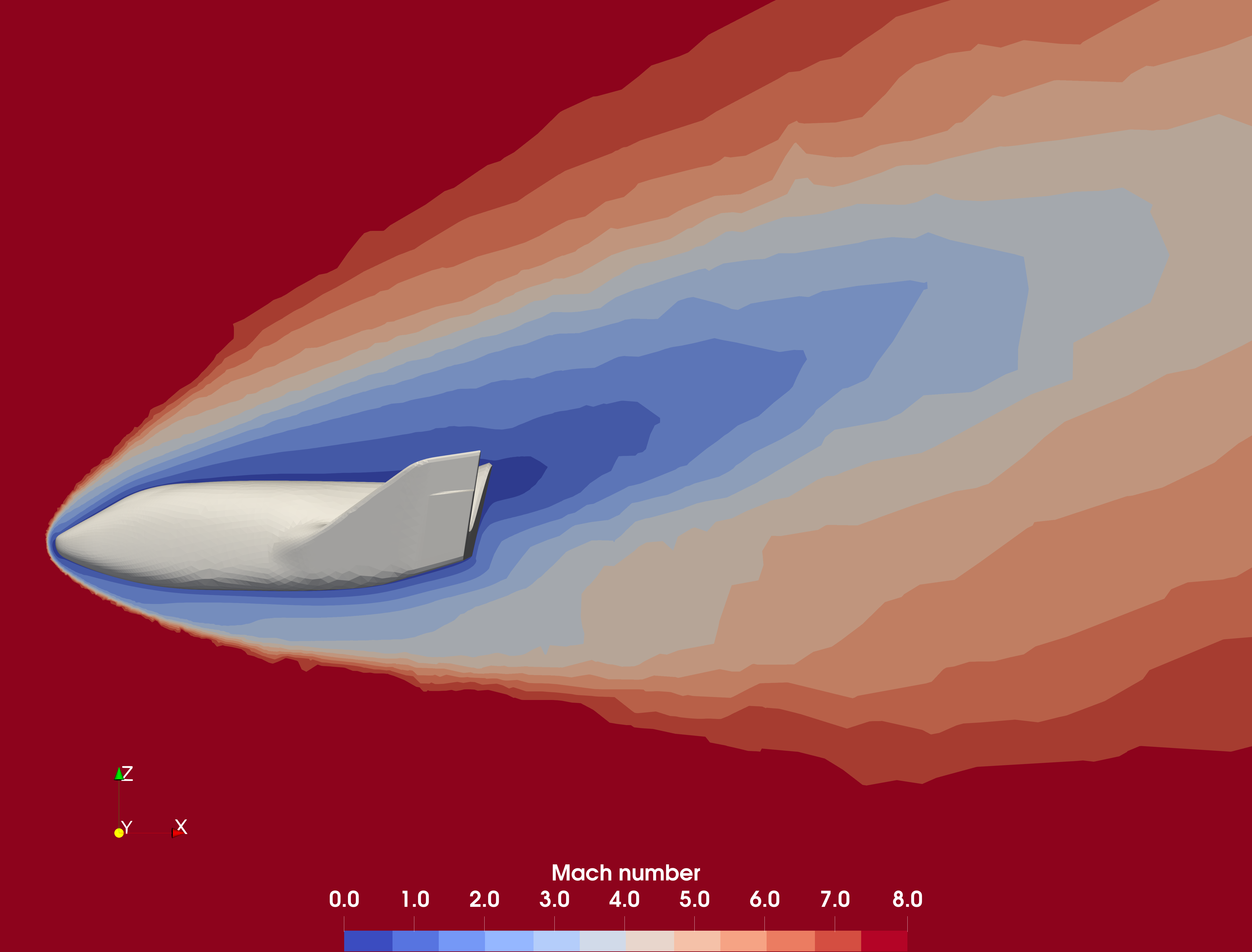}}\\
	\caption{Hypersonic flow around an X38-like space vehicle at ${\rm Ma}_\infty = 8$ for ${\rm Kn}_\infty = 0.00275$ by the IAUGKS. Distributions of (a) Density, (b) $\rm{Kn}_{Gll}$,
		(c) temperature, and (d) Mach number contours.}
	\label{fig:x38-contour}
\end{figure}

\begin{figure}[H]
	\centering
	\includegraphics[width=0.4\textwidth]
	{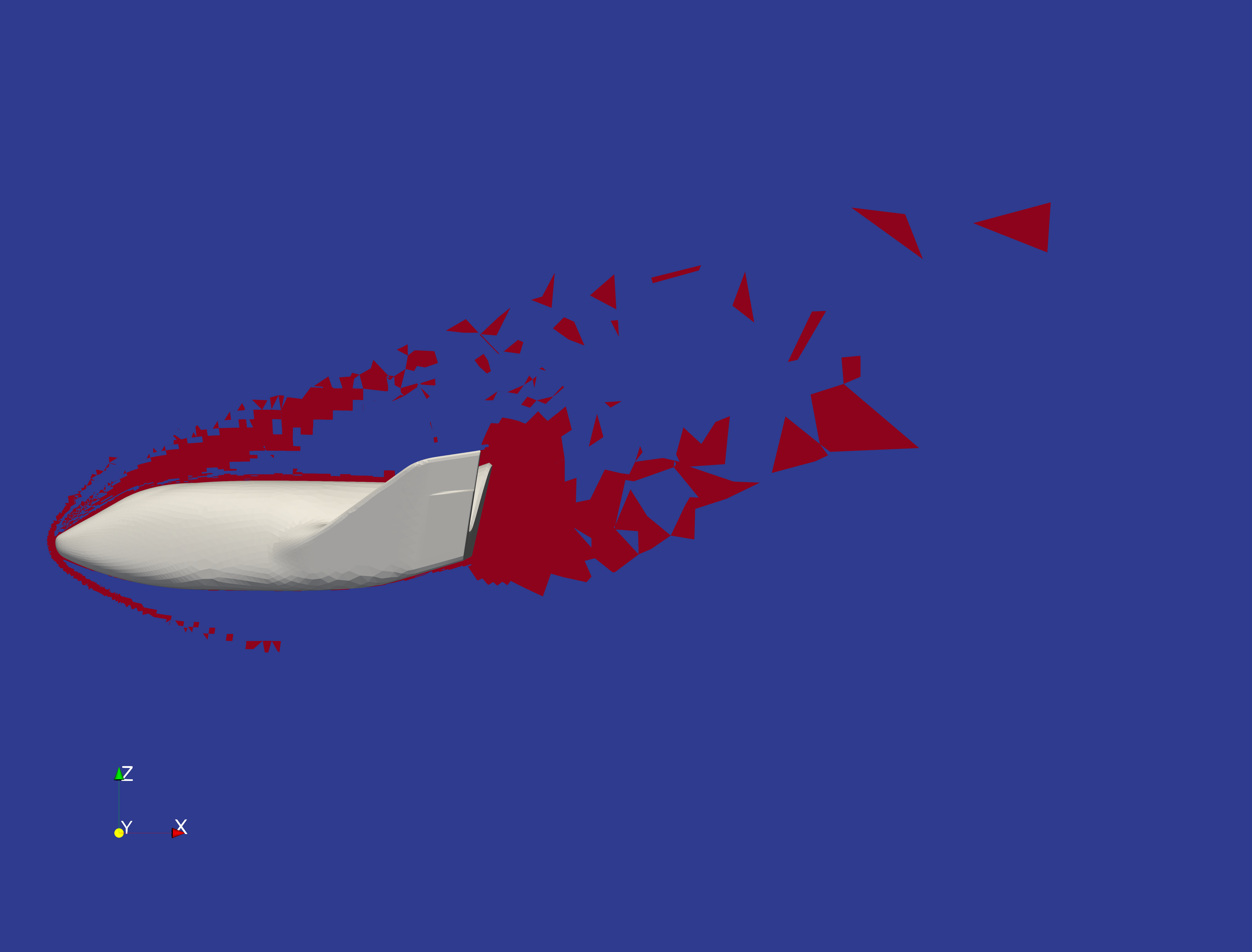}
	\caption{Hypersonic flow around an X38-like space vehicle at ${\rm Ma}_\infty = 8$ for ${\rm Kn}_\infty = 0.00275$ by the IAUGKS. Distributions of velocity space adaptation with $C_t = 0.05$ where the DVS (UGKS) is used in 46.37\% of ths physical domain.}
	\label{fig:x38-isDisc}
\end{figure}

Figure~\ref{fig:x38-contour} illustrates contours of the density, gradient-length local Knudsen number, temperature, and Mach number. Even though the free stream flow is in near continuum regime, the gradient-length local contour denotes strong non-equilibrium flow in the leeward region, where the $\rm{Kn}_{Gll}$ is 4 to 5 orders of magnitude higher than in the free stream flow. Figure~\ref{fig:x38-isDisc} shows that the velocity space adaptation mechanism successfully detects the non-equilibrium region and uses DVS for 46.37\% of the whole computation domain.

\begin{figure}[H]
	\centering
	\subfloat[]{\includegraphics[width=0.33\textwidth]
		{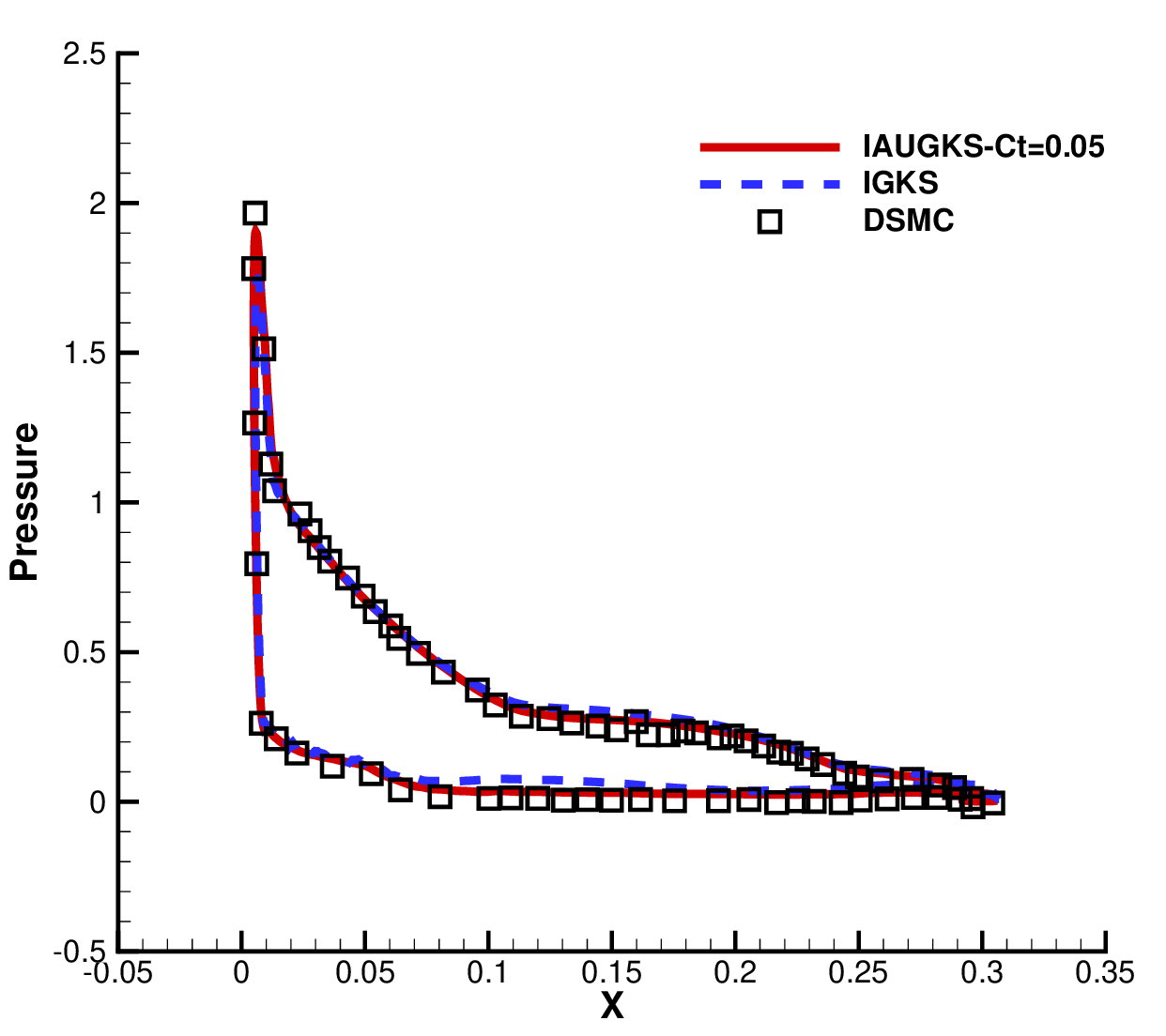}}
	\subfloat[]{\includegraphics[width=0.33\textwidth]
		{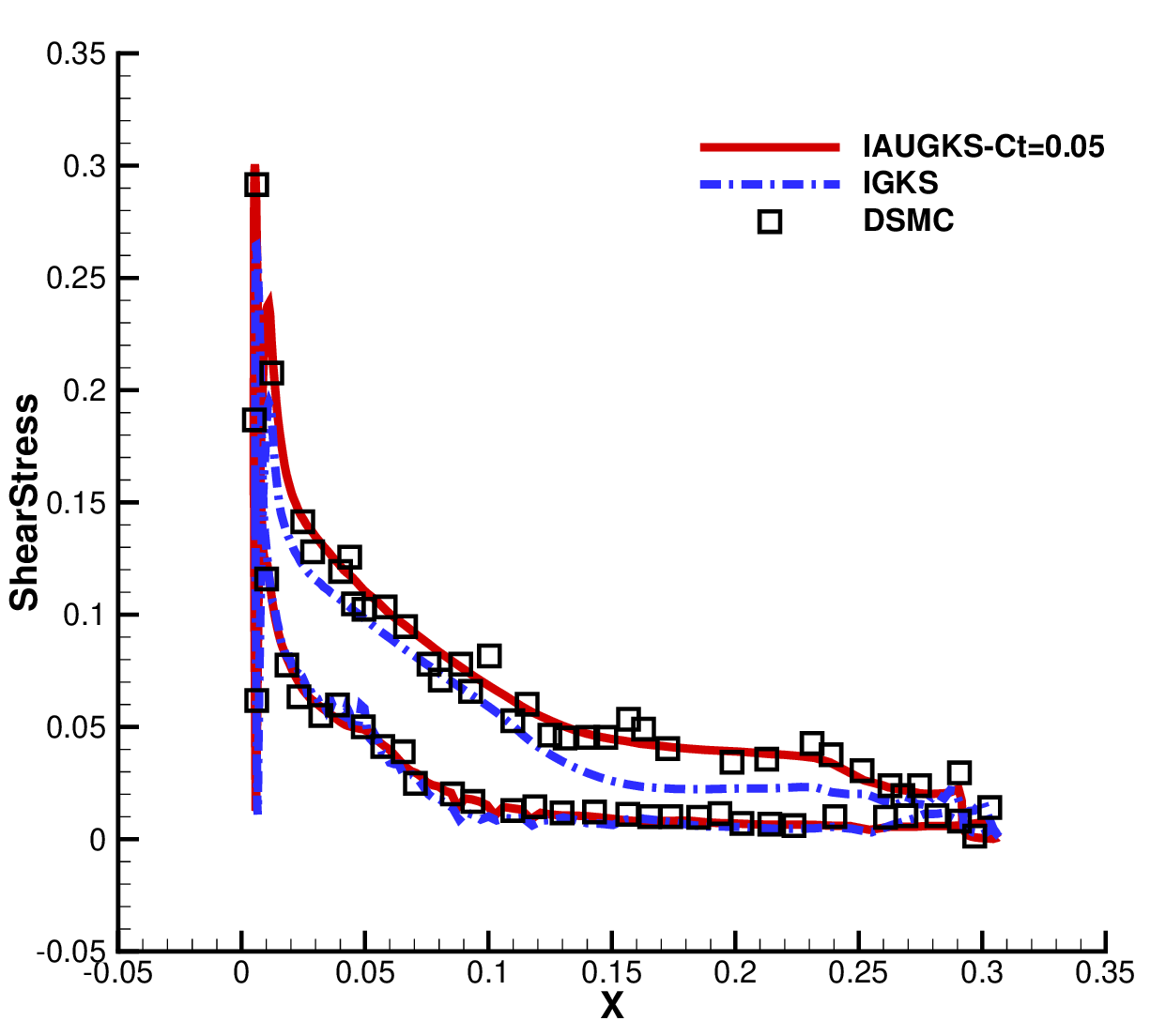}}
	\subfloat[]{\includegraphics[width=0.33\textwidth]
		{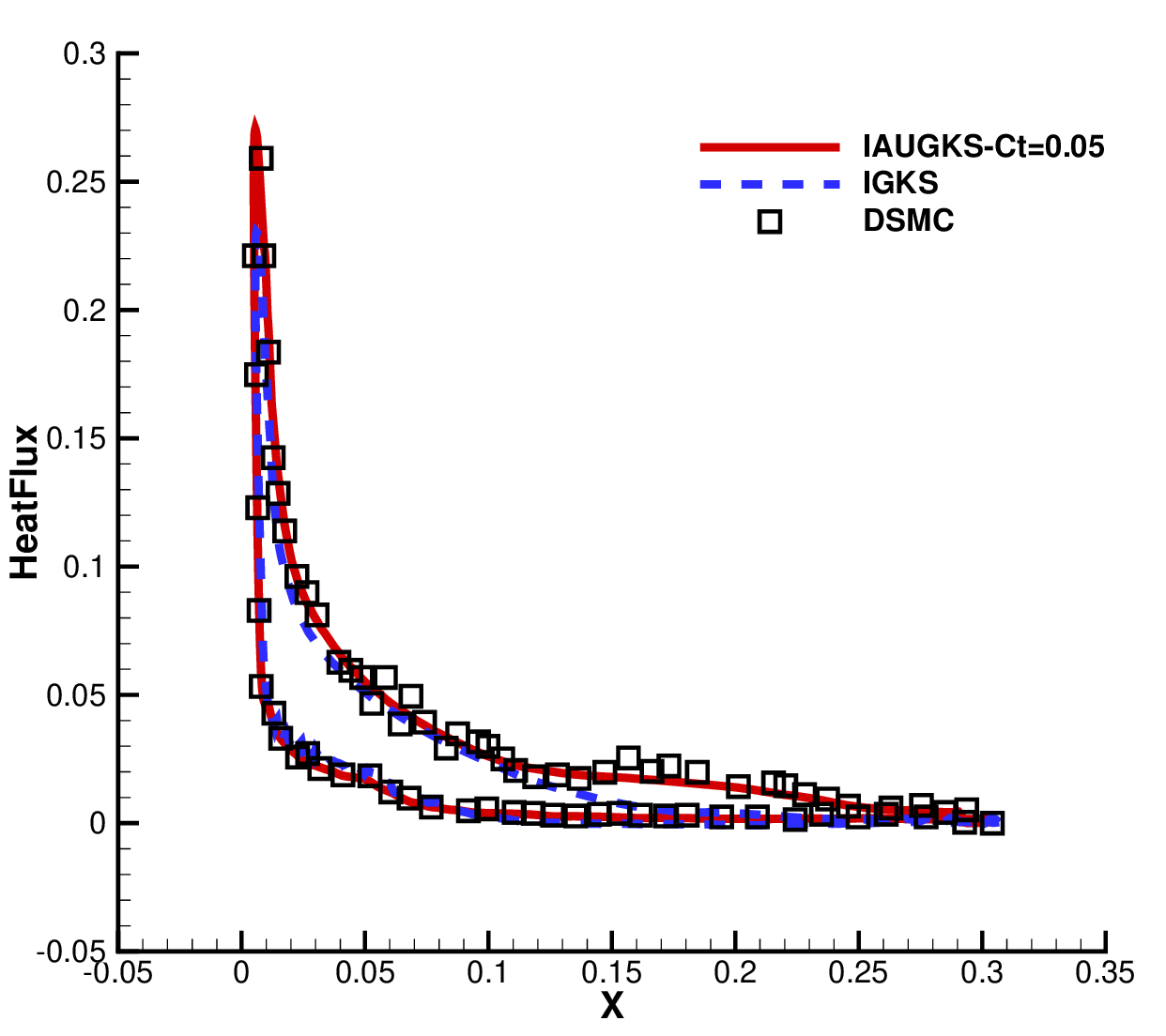}}\\
	\caption{Surface quantities of hypersonic flow
		around an X38-like space vehicle at ${\rm Ma}_\infty = 8.0$ for ${\rm Kn}_\infty = 0.00275$ by the IAUGKS for argon gas compared with the DSMC method and the IGKS. (a) Pressure coefficient, (b) shear stress coefficient,
		and (c) heat flux coefficient.}
	\label{fig:x38-surface-line}
\end{figure}

Figure~\ref{fig:x38-surface-line} depicts the surface quantities on the symmetric cross-section perpendicular to the y-axis and comparisons with the DSMC data. All coefficients predicted by the IAUGKS with $C_t = 0.05$ align well with the DSMC reference. However, if continuous velocity space is applied for the whole computation domain, the IGKS gives false result in terms of shear stress and heat flux at the leeward region and the forebody of the vehicle. This also implies strong non-equilibrium effect exists within these regions, which cannot be captured by the traditional continuum solver. This shows the importance of the velocity space adaptation. 

The simulation takes 35,000 steps of the GKS and 250 steps of the IAUGKS, and the physical CFL number is set to 0.4 and the numerical CFL number for implicit iteration is set to 100. A simple static load balancing method is adopted. The simulation is conducted on the SUGON computation platform which takes 4.86 hours of wall clock time on 640 cores of CPU 7285 32C 2.0GHz.

\subsection{Hypersonic flow over a space station}
Hypersonic flow at ${\rm Ma}_\infty = 25.0$ passing over a space station for ${\rm Kn}_\infty = 0.01$ is simulated. The characteristic length to define the Knudsen number is selected as the scale of solar panel $L = 0.01$ m. The free stream temperature is $T_\infty = 142.2$ K, and the isothermal wall is applied with $T_w = 500$ K.

\begin{figure}[H]
	\centering
	\includegraphics[width=0.4\textwidth]
	{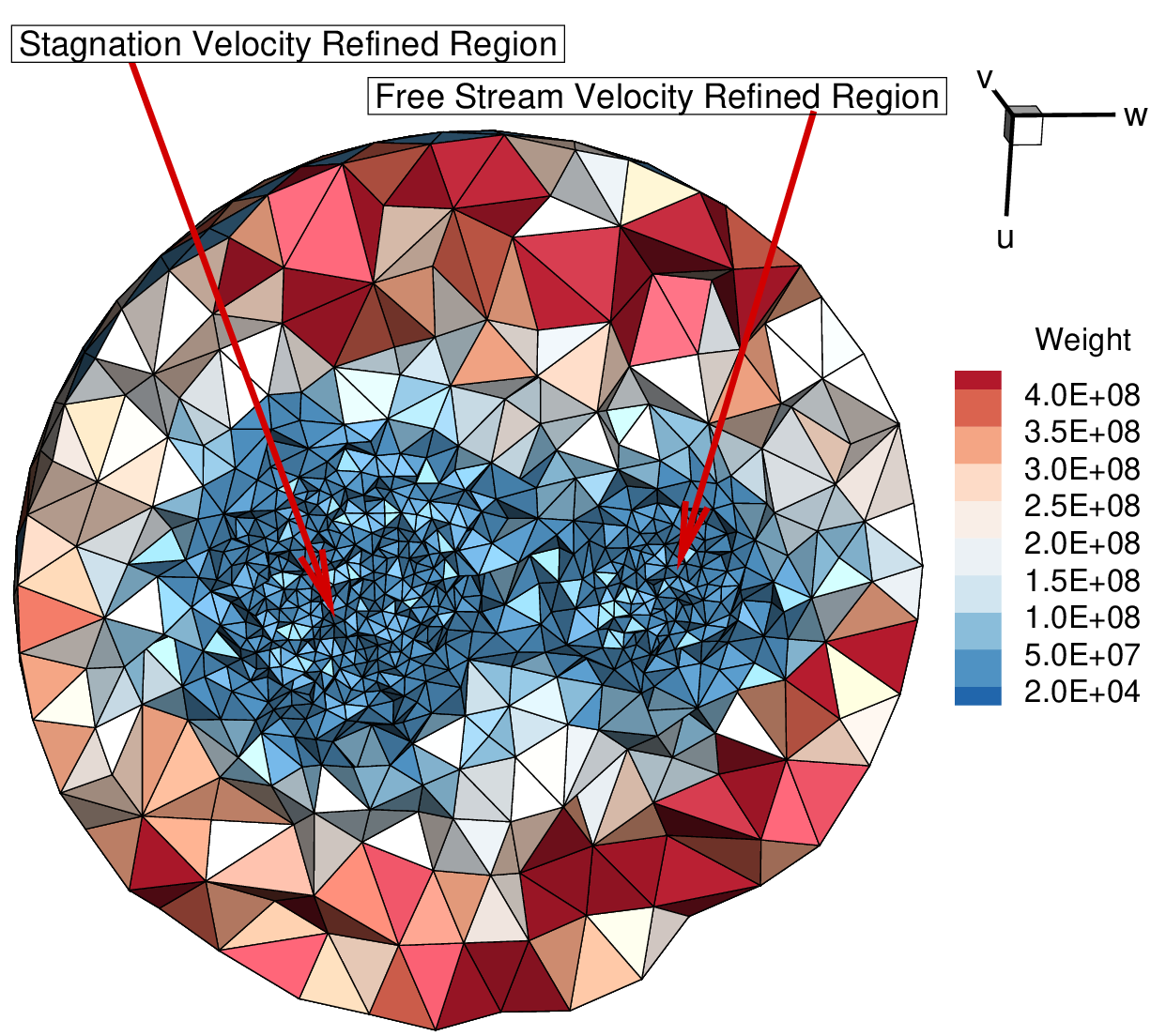}
	\caption{Unstructured DVS mesh with 32,646 cells used for hypersonic flow at ${\rm Ma}_\infty = 25.0$ and ${\rm Kn}_\infty = 0.01$ passing over a space station by the IAUGKS.}
	\label{fig:station-DVS}
\end{figure}

The whole physical domain consists of 5,640,776 cells, and the unstructured DVS mesh is discretized into 32,646 cells. In Fig.~\ref{fig:station-DVS}, the DVS mesh is a sphere centered at $0.4\times(U_\infty,V_\infty,W_\infty)$ with a radius of $42\sqrt{R T_{\infty}}$.The velocity space is refined in a spherical region near zero velocity point and free stream velocity point with radius $r=5\sqrt{R T_w}$ and $r=5\sqrt{R T_\infty}$ respectively. 

\begin{figure}[H]
	\centering
	\subfloat[]{\includegraphics[width=0.6\textwidth]
		{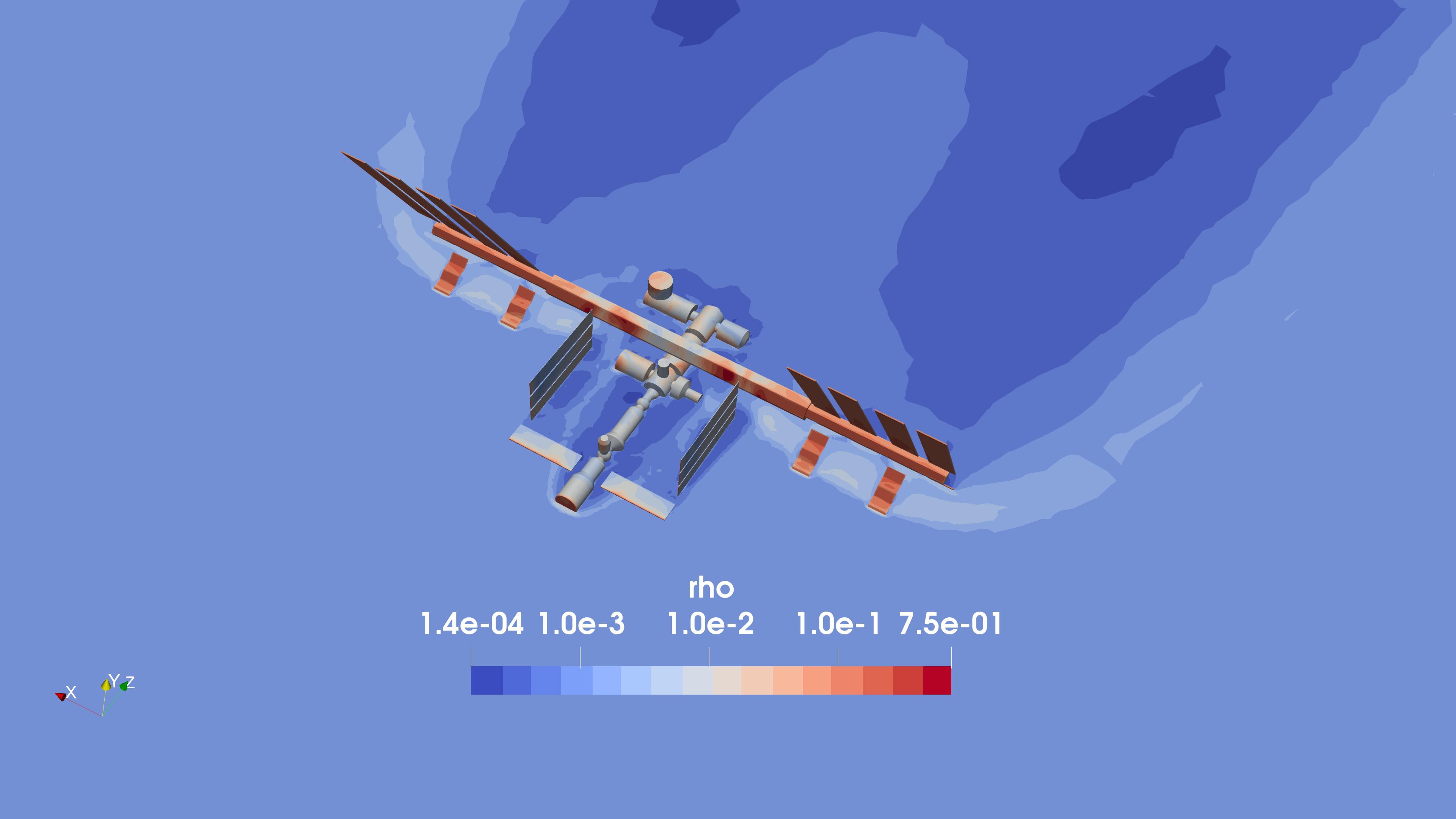}}
	\\
	\subfloat[]{\includegraphics[width=0.6\textwidth]
		{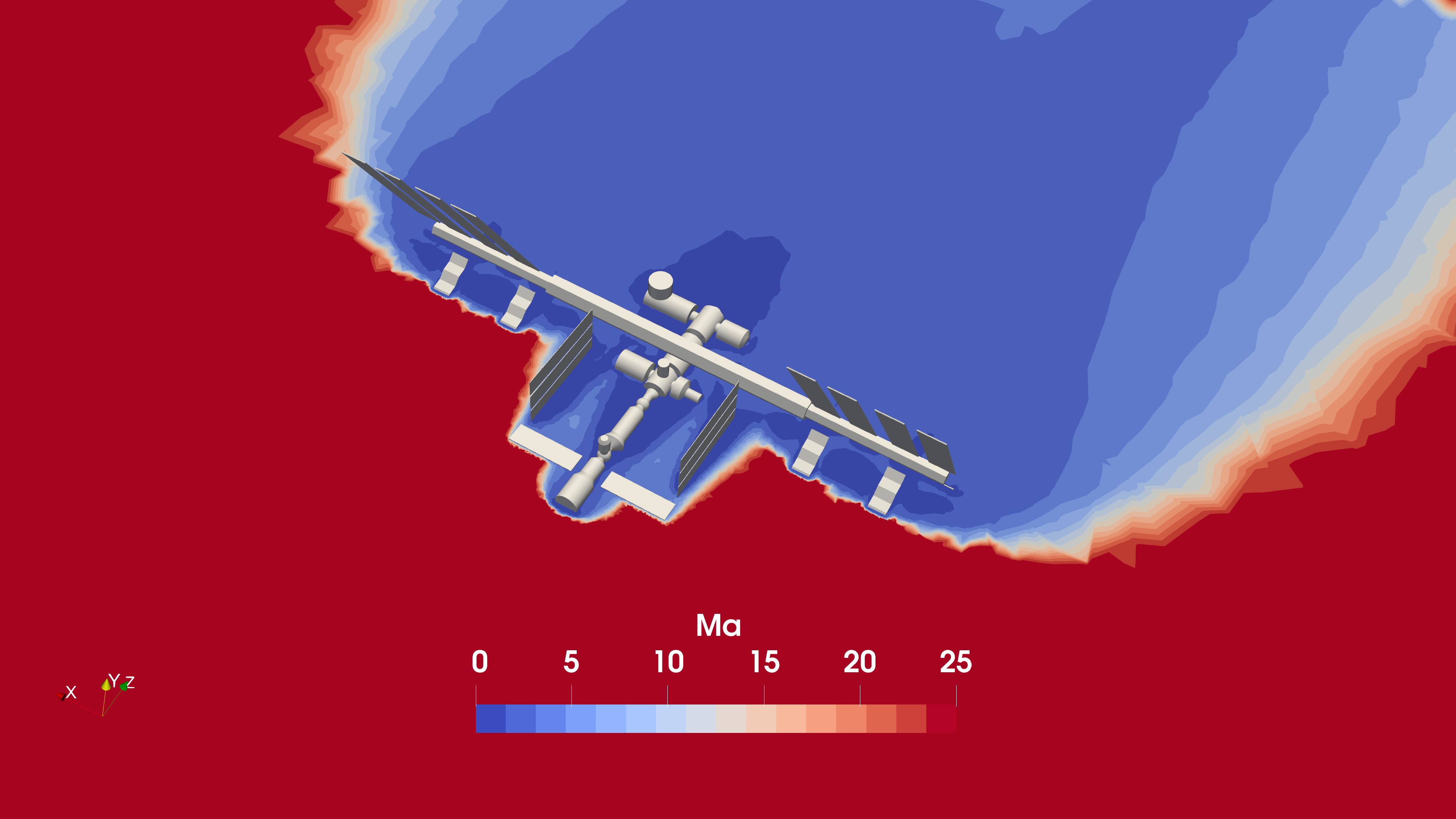}}
	\\
	\subfloat[]{\includegraphics[width=0.6\textwidth]
		{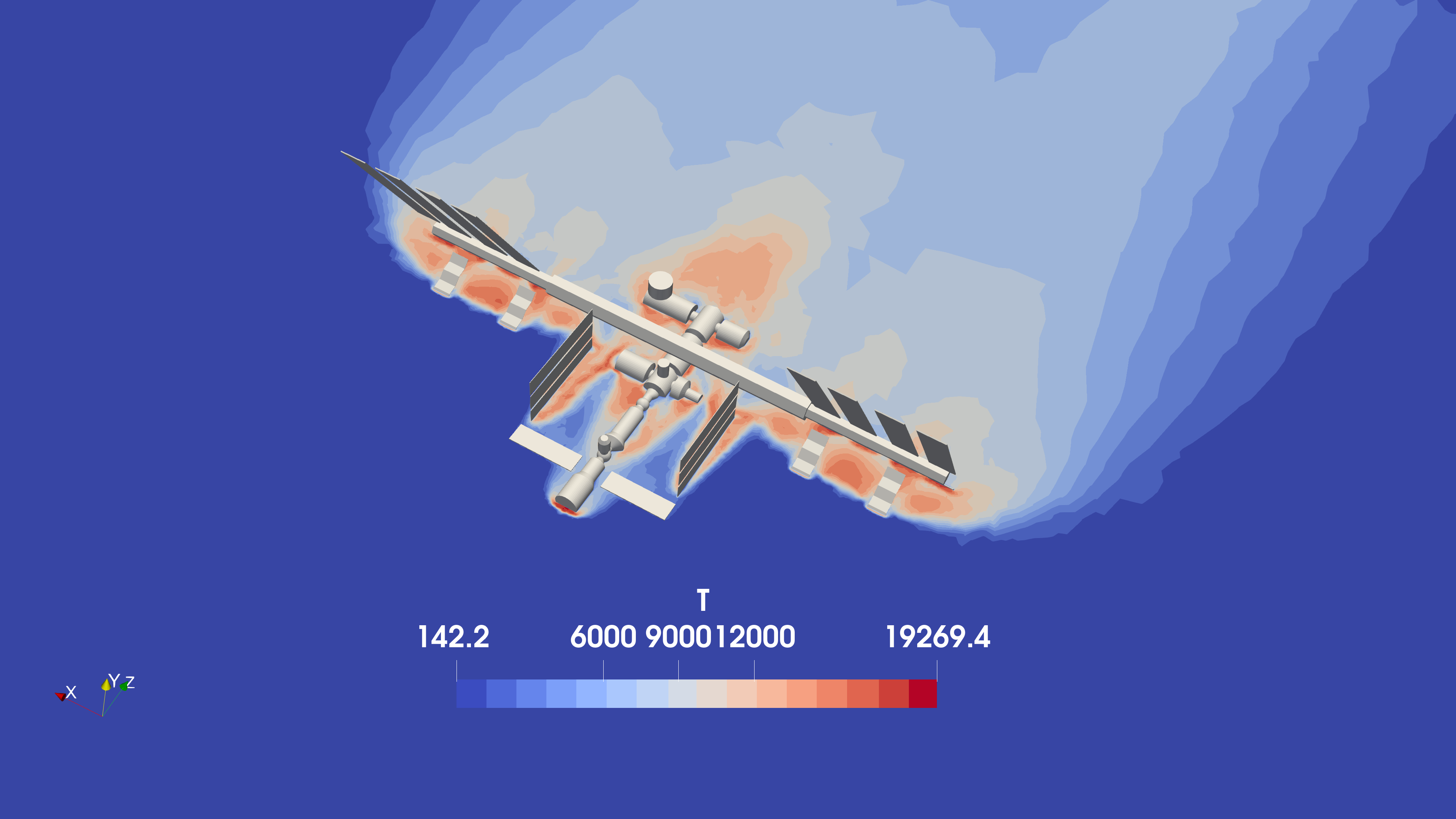}}
	\\
	\caption{Hypersonic flow around a space station at ${\rm Ma}_\infty = 25$ for ${\rm Kn}_\infty = 0.01$. Distributions of (a) density, (b) Mach number, and (c) temperature contours.}
	\label{fig:station-contour}
\end{figure}

\begin{figure}[H]
	\centering
	\includegraphics[width=0.6\textwidth]
		{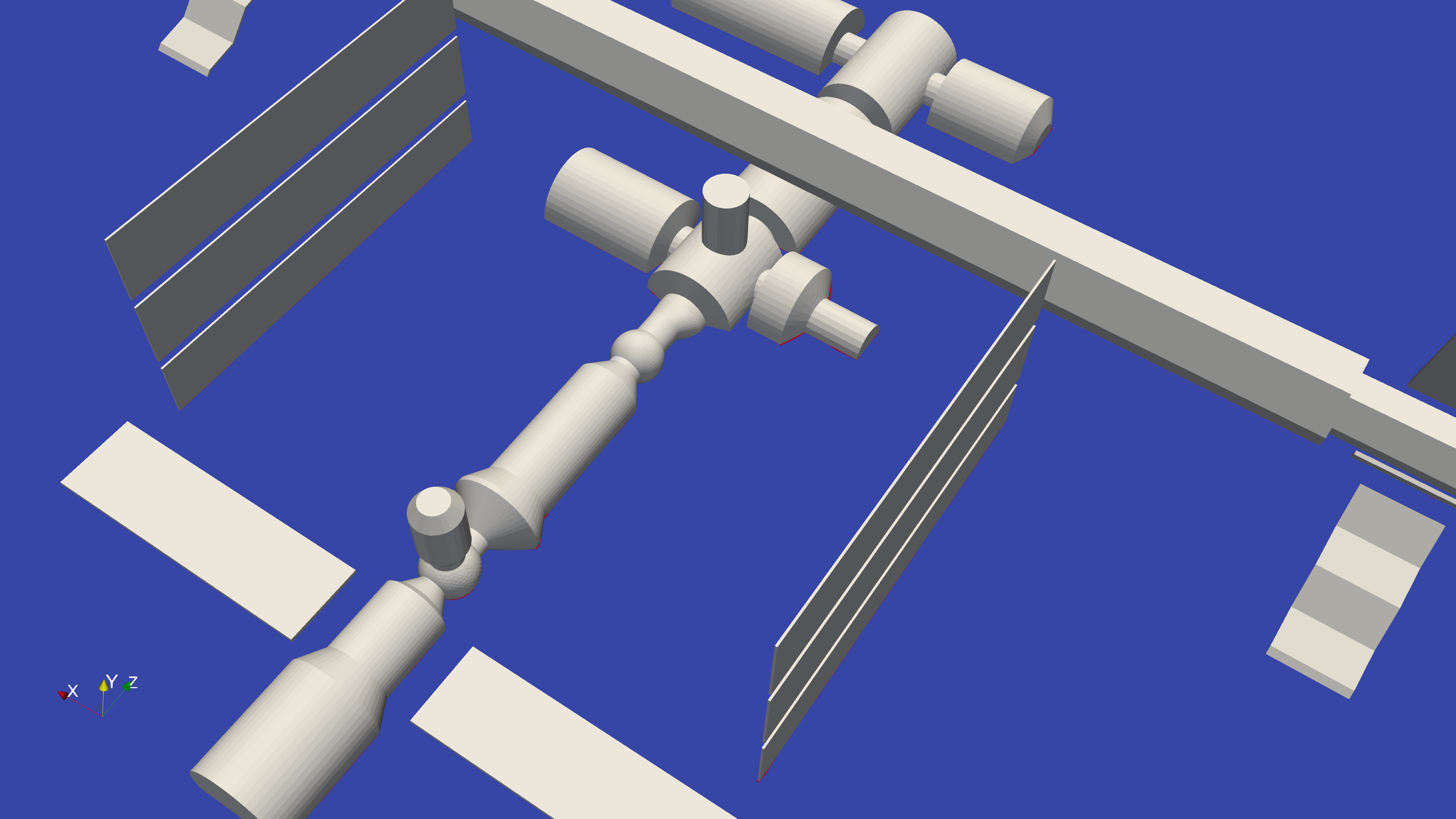}
	\caption{Hypersonic flow at ${\rm Kn} = 0.01$ and ${\rm Ma} = 25$ passing over a space station by the IAUGKS. Distributions of velocity space adaptation with $C_t = 0.05$, where the computation domain using DVS takes up $1.53\%$.}
	\label{fig:station-isDisc}
\end{figure}

Figure~\ref{fig:station-contour} shows the density, Mach number, and temperature contour of hypersonic flow around a space station. As depicted in Fig.~\ref{fig:station-isDisc}, the computation region adopting DVS only exists in the close vicinity of the space station, and only occupies 1.53\% of the whole domain. As a result, the demand for CPU cores and memory is greatly reduced. As shown in Table~\ref{table:stationtime}, the GSIS, which applies DVS to the whole domain, has to use as many as 9216 cores and 21.5TB to simulate the case, while the IAUGKS only needs 320 cores and less than 2.5TB memory. It is worth noting that the current scheme does not yet account for thermal non-equilibrium, while the GSIS~\cite{zhang2023efficient} needs handling an additional velocity distribution function, resulting in approximately double the memory consumption and computational cost. If linearly estimate the acceleration rate by considering the core hours, the IAUGKS is more efficient than the GSIS in terms of the computational cost. By implementing better load balancing techniques, it is possible to further reduce both memory consumption and computation time. The physical CFL number is set to 0.4 and the numerical CFL number for implicit iteration is set to 100, and the simulation is conducted on the SUGON computation platform with CPU 7285 32C 2.0GHz.
\begin{table}[H]
	\caption{Hypersonic flow at ${\rm Ma}_\infty = 25.0$ passing over a space station for ${\rm Kn}_\infty = 0.01$ with $C_t = 0.05$. The simulation time and cost are compared with the GSIS.}
	\centering
	\begin{threeparttable}
		\begin{tabular}{cccccc}
			\hline
			Method & Cores & Steps & Wall Clock Time, h& Core hours, h  & Acceleration Rate   \\
			\hline
			IAUGKS & 320 & $3500\tnote{1}+450$ & 11.87 & 3798.4 & 2.10 \\
			GSIS & 9216 & $4000+52$ & 0.867 & 7990.3 & 1.0\\

			\hline
		\end{tabular}
		
		\begin{tablenotes}
			\item[1] Steps of first-order GKS simulations.
		\end{tablenotes}
	\end{threeparttable}
	\label{table:stationtime}
\end{table}


\section{Conclusion}\label{sec:conclusion}
In this paper, an implicit adaptive unified gas-kinetic scheme (IAUGKS) for non-equilibrium flow simulation is constructed. Based on the UGKS, a combination of macroscopic implicit prediction and microscopic implicit iteration is adopted, achieving fully implicit and accelerating convergence by 1 to 2 orders of magnitude in all flow regimes. The velocity space
adaptation dynamically switches to a continuous distribution function instead of discrete velocity space in the near-continuum flow regime. This automatic dynamic interface between continuous and discrete distribution functions eliminates the need for a buffer zone. Compared to the original UGKS, the IAUGKS demonstrates faster convergence and reduced memory consumption in non-equilibrium flow simulations. In two-dimensional cases, the current scheme exhibits computational efficiency one order of magnitude greater than the UGKWP method, while performing comparably in three-dimensional cases. The velocity space adaptation can save half of the memory in hypersonic flows, utilizing fewer computational resources and simulation time. In summary, the IAUGKS serves as a valuable tool for simulating non-equilibrium flows. The current scheme can be further accelerated by employing techniques such as the multigrid method, GMRES method, and others. Additionally, the IAUGKS provides a solid foundation for multiphysics models and other non-equilibrium flows.

\section*{Author's contributions}

All authors contributed equally to this work.

\section*{Acknowledgments}

This work was supported by National Key R$\&$D Program of China (Grant Nos. 2022YFA1004500), National Natural Science Foundation of China (12172316), Hong Kong research grant council (16208021,16301222).

\section*{Data Availability}

The data that support the findings of this study are available from the corresponding author upon reasonable request.

\appendix
	




\bibliographystyle{elsarticle-num}
\bibliography{iaugks.bib}







\end{document}